\documentclass[range]{ar2e}

\usepackage{epsfig}     
\def\nodata{ ~$\cdots$~ }%
\def\nhi{\noindent \hangindent=0.3cm}
\def\etal{{\it et al.}}
\def\arcmin   {$^{\prime}$}
\def\arcsec   {$^{\prime\prime}$}
\def\la{\mathrel{\mathpalette\fun <}}
\def\ga{\mathrel{\mathpalette\fun >}}
\def\fun#1#2{\lower3.6pt\vbox{\baselineskip0pt\lineskip.9pt
  \ialign{$\mathsurround=0pt#1\hfil##\hfil$\crcr#2\crcr\sim\crcr}}}

\newcommand\epsscale[1]{\gdef\eps@scaling{#1}}%
\newcommand\plotone[1]{%
 \typeout{Plotone included the file #1}
 \centering
 \leavevmode
 \includegraphics[width={\eps@scaling\columnwidth}]{#1}%
}
\newcommand\plottwo[2]{{%
 \typeout{Plottwo included the files #1 #2}
 \centering
 \leavevmode
 \columnwidth=.45\columnwidth
 \includegraphics[width={\eps@scaling\columnwidth}]{#1}%
 \hfil
 \includegraphics[width={\eps@scaling\columnwidth}]{#2}%
}}%

\begin{document}

\input epsf.tex    
\input epsf.def   

\input psfig.sty

\jname{Annu. Rev. Astron. Astrophys.}
\jyear{2006}
\jvol{}
\ARinfo{1056-8700/97/0610-00}

\title{Extragalactic Globular Clusters and Galaxy Formation}

\markboth{Brodie \& Strader}{Extragalactic Globular Clusters}

\author{Jean P. Brodie and Jay Strader
\affiliation{UCO/Lick Observatory}}


\begin{abstract}

Globular cluster (GC) systems have now been studied in galaxies ranging from 
dwarfs to giants and spanning the full Hubble sequence of morphological 
types. Imaging and spectroscopy with the Hubble Space Telescope and large 
ground-based telescopes have together established that most galaxies have bimodal 
color distributions that reflect two subpopulations of old GCs: metal-poor and 
metal-rich. The characteristics of both subpopulations are correlated with those 
of their parent galaxies. We argue that metal-poor GCs formed in low-mass dark 
matter halos in the early universe and that their properties reflect biased galaxy 
assembly. The metal-rich GCs were born in the subsequent dissipational buildup of 
their parent galaxies and their ages and abundances indicate that most massive 
early-type galaxies formed the bulk of their stars at early times. Detailed 
studies of both subpopulations offer some of the strongest constraints on 
hierarchical galaxy formation that can be obtained in the near-field.

\end{abstract}

\maketitle

\section{Introduction}

Globular star clusters (GCs) are among the oldest radiant objects in the universe. With typical
masses $\sim 10^4-10^6$ M$_\odot$ (corresponding to luminosities of $\sim M_V = -5$ to $-10$) 
and compact sizes (half-light radii of a few pc), they are readily observable in external
galaxies. The 15 years since the Annual Review by Harris (1991, ``Globular Cluster Systems in Galaxies Beyond the 
Local Group") have seen a revolution in the field of extragalactic GCs. It is becoming 
increasingly apparent that GCs provide uniquely powerful diagnostics of fundamental 
parameters in a wide range of astrophysical processes. Observations of GCs are being used to constrain the 
star formation and assembly histories of galaxies, nucleosynthetic processes governing chemical evolution, 
the epoch and homogeneity of cosmic reionization, the role of dark matter in the formation of structure in 
the early universe, and the distribution of dark matter in present-day galaxies. GCs are valuable tools for 
theoretical and observational astronomy across a wide range of disciplines from cosmology to stellar 
spectroscopy.

It is not yet widely recognized outside the GC community that recent advances in GC research provide 
important constraints on galaxy formation that are complementary to in situ studies of galaxies at medium to 
high redshift. The theme of this article is the role of GC systems as tracers of galaxy formation and 
assembly, and one of our primary aims is to emphasize the current and potential links with results from 
galaxy surveys at high redshift and interpretations from stellar population synthesis, numerical 
simulations, and semi-analytical modeling. In what follows we will attempt to chronicle the observations 
that mark recent milestones of achievement and to place them in the wider theoretical and observational 
context. We will focus most closely on work carried out since about 2000. The preceding period is 
well-covered by the book of Ashman \& Zepf (1998) and the Saas-Fee lectures of Harris 
(2001).\footnote{Available online at http://physwww.mcmaster.ca/$\sim$harris/Publications/saasfee.ps.} Among 
the significant topics not directly covered are young massive clusters (potential ``proto-GCs"), X-ray 
sources in extragalactic GCs, and ultra-compact dwarf galaxies. Neither do we include a comprehensive 
discussion of the Galactic GC system.

The fundamental premise in what follows is that GCs are good tracers of the star formation histories of 
spheroids (early-type galaxies, spiral bulges, and halos), in the sense that major star-forming episodes are 
typically accompanied by significant GC formation. Low-level star formation (e.g., in quiescent galactic 
disks) tends to produce few, if any, GCs. Since most of the stellar mass in the local universe is in 
spheroids ($\sim 75$\%; Fukugita, Hogan, \& Peebles 1998), GCs trace the bulk of the star formation history 
of the universe. Although the relationships between star formation, GC formation and GC survival are complex 
and do not necessarily maintain relative proportions under all conditions, this underlying assumption is 
supported by a number of lines of argument. Massive star clusters appear to form during all major 
star-forming events, such as those accompanying galaxy-galaxy interactions (e.g., Schweizer 2001). In these 
situations, the number of new clusters formed scales with the amount of gas involved in the interaction 
(e.g., Kissler-Patig, Forbes, \& Minniti 1998). The cluster formation efficiency (the fraction of star 
formation in clusters) scales with the star formation rate, at least in spiral galaxies where it can be 
directly measured at the present epoch (Larsen \& Richtler 2000). This may suggest that massive clusters 
form whenever the star-formation rate is high enough, and that this occurs principally during spheroid 
formation. Perhaps most importantly, the properties of GCs (especially their metallicities) are correlated 
with the properties of their host galaxies.

\section{Color Bimodality: Globular Cluster Subpopulations}

Perhaps the most significant development of the decade in the field of extragalactic GCs was the 
discovery that the color distributions of GC systems are typically bimodal. Indeed, color bimodality is 
the basic paradigm of modern GC studies. Nearly every massive galaxy studied to date with sufficiently 
accurate photometry has been shown to have a bimodal GC color distribution, indicating two 
subpopulations of GCs.  In principle, these color differences can be due to age or metallicity 
differences or some combination of the two. Due to the well-known degeneracy between age and metallicity 
(e.g., Worthey 1994), the cause of this bimodality is not readily deduced from optical colors alone. 
Nonetheless, the significance of the finding was immediately recognized. The presence of bimodality 
indicates that there have been at least two major star-forming epochs (or mechanisms) in the histories of 
most---and possibly all---massive galaxies. Subsequent spectroscopic studies (see \S 4) have shown that 
color bimodality is due principally to a metallicity difference between two old subpopulations.

With our ``bimodality-trained" modern eyes, we can see evidence of the phenomenon in the $B-I$ 
CFHT imaging of NGC 4472 in Couture, Harris \& Allwright (1991) and $C-T_{1}$ CTIO imaging of 
NGC 5128 by Harris \etal~(1992). However, the first groups to propose bimodality (or 
``multimodality") were Zepf \& Ashman (1993) for NGC 4472 and NGC 5128 and Ostrov, Geisler, \& 
Forte (1993) for NGC 1399 (in fact, using the Harris \etal~and Couture \etal~colors). 
Observations of the GC systems of galaxies throughout the 1990s provided mounting evidence 
that bimodality was ubiquitous in massive galaxies. The primary catalyst of this research was 
the advent of the \emph{Hubble Space Telescope (HST)}. The Wide Field and Planetary Camera 2 (WFPC2) 
provided the spatial resolution and accurate photometry needed to reliably identify GC 
candidates in galaxies as distant as the Virgo Cluster at 17 Mpc (e.g., Whitmore \etal~1995). 
At this distance, GCs (with typical half-light radii of 2--3 pc $\sim 0.03-0.04$\arcsec) are 
resolvable with \emph{HST} and their sizes are measurable with careful modeling of the PSF. 
This drove down the contamination from background galaxies and foreground stars to low levels 
and was a substantial improvement over multi-band optical photometry from the ground.

Among the larger and more comprehensive photometric studies using \emph{HST}/WFPC2 were 
Gebhardt \& Kissler-Patig (1999), Larsen \etal~(2001a) and Kundu \& Whitmore (2001a). 
Using data from the \emph{HST} archive, Gebhardt \& Kissler-Patig showed that bimodality 
was a common phenomenon. However, since the imaging was shallow for many of the galaxies 
in their sample, they failed to find bimodality in $\sim 50$\% of their 50 galaxies. 
Taking advantage of deeper data, Larsen \etal~(2001a)  and Kundu \& Whitmore (2001a) found 
statistically significant bimodality in most of their sample galaxies, the majority of 
which were of early-type. Galaxies that were tentatively identified as unimodal in these 
studies were later, with improved photometric precision, shown to conform to the 
bimodality ``rule". Indeed, it is important to note that \emph{no} massive elliptical (E) 
galaxy has been convincingly shown to lack GC subpopulations. An absence of metal-poor GCs 
was suggested for both NGC 3311 (Secker \etal~1995) and IC 4051 (Woodworth \& Harris 
2000), and an absence of metal-rich GCs for NGC 4874 (Harris \etal~2000). However, 
\emph{HST}/WFPC2 imaging of NGC 3311 (Brodie, Larsen, \& Kissler-Patig 2000) revealed a 
healthy subpopulation of metal-poor GCs. It is now clear that the WFPC2 photometry of the 
Coma E IC 4051 was not deep enough to securely argue for a uni- or bimodal fit. Finally, 
the NGC 4874 result was due to a photometric zeropoint error (W.~Harris, private 
communication). The discovery of a massive E which indeed lacked a metal-poor (or 
metal-rich) subpopulation would be important, but so far no such instances have been 
confirmed.

The majority of these \emph{HST} studies were carried out in $V$- and $I$-equivalent bands. 
This choice was largely driven by efficiency considerations (shorter exposure times needed to 
reach a nominal S/N), despite the fact that other colors, such as $B-I$, offer much better metallicity 
sensitivity for old stellar populations.

It has been known for some time that the GC system of the Milky Way is also bimodal. The 
presence of GC subpopulations in the Milky Way was codified by Zinn (1985; see also 
Armandroff \& Zinn 1988) who identified two groups of GCs. ``Halo" GCs are metal-poor, 
non-rotating (as a system), and can be found at large galactocentric radii. ``Disk" GCs 
are metal-rich and form a flattened, rotating population. Later work on the spatial and 
kinematic properties of the metal-rich GCs by Minniti (1995) and C{\^ o}t{\' e} (1999) 
identified them with the Milky Way bulge rather than its disk (as we shall see below, this 
association seems to hold for other spirals as well, although see the discussion in \S 
3.3). In addition to their sample of early-type galaxies, Larsen \etal~(2001a) also 
discussed the GC systems of the Milky Way and NGC 4594 in some detail, pointing out that 
the 
locations of the GC color peaks in these spirals were indistinguishable from those of 
massive early-type galaxies.

The blue (metal-poor) and red (metal-rich) peaks in massive early type galaxies typically 
occur at $V-I$ = 0.95$\pm0.02$ and 1.18$\pm0.04$ (Larsen \etal~2001a). These colors 
correspond to [Fe/H] $\sim -1.5$ and $-0.5$ for old GCs (or a bit more metal-rich, 
depending on the metallicity scale and color-metallicity relation adopted).  Figure 1 
shows a histogram of the $V-I$ colors of GCs in the Virgo gE M87, which clearly shows 
bimodality (Larsen \etal~2001a). However, the peak locations are not exactly the same for 
all galaxies. Before GC bimodality was discovered, van den Bergh (1975) suggested and 
Brodie \& Huchra (1991) confirmed a correlation between the mean color/metallicity of GC 
systems and the luminosity of their parent galaxies. Brodie \& Huchra (1991) also showed 
that the slope of this relation was very similar to the relation connecting galaxy color 
and galaxy luminosity, but the GC relation was offset toward lower metallicities by about 
0.5 dex. They noted that the similarity in slope suggests a close connection between the 
physical processes responsible for the formation of both GCs and galaxies. Subsequently, a 
correlation between the color of just the metal-rich GCs and host galaxy luminosity was 
found by Forbes \etal~(1997), Larsen \etal~(2001a), and Forbes \& Forte (2001). The slope 
of this relation was again found to be similar to that of the color-magnitude relation for 
early-type galaxies ($V-I \propto -0.018 M_V$), suggesting that metal-rich GCs formed 
along with the bulk of the field stars in their parent galaxies.

With the exception of Larsen \etal~(2001a), little or no correlation between the color of 
the \emph{metal-poor} GCs and host galaxy luminosity was reported in these studies, 
although Burgarella \etal~(2001) and Lotz, Miller, \& Ferguson (2004) suggested such a 
relation might be present, but only for the dwarf galaxies. Larsen \etal~found a shallow 
relation for the metal-poor GCs in their sample of 17 massive early-type galaxies, albeit 
at moderate (3$\sigma$) statistical significance. Strader, Brodie, \& Forbes (2004a) 
compiled and reanalyzed high-quality data from the literature and found a significant ($> 
5\sigma$) correlation for metal-poor GCs, extending from massive Es to dwarfs over $\sim 
10$ magnitudes in galaxy luminosity. The relation is indeed relatively shallow ($V-I 
\propto -0.009 M_V$, or $Z \sim L^{0.15}$), making it difficult to detect, especially in 
heterogeneous data sets. This same slope was confirmed by Strader \etal~(2006) and Peng 
\etal~(2006) for early-type galaxies in Virgo. Figure 2 shows [Fe/H] vs.~$M_B$ for both 
subpopulations; the GC peaks are taken from Strader \etal~(2004a) and Strader \etal~(2006) 
and have been converted from $V-I$ and $g-z$ using the relations of Barmby \etal~(2000) 
and Peng \etal~(2006), respectively. These data, together with ancillary information
about the GC systems, are compiled in Table 1. The true scatter at fixed $M_B$ is unclear, since the 
observational errors vary among galaxies, and there may an additional component due to 
small differences between the $V-I$ and $g-z$ color--metallicity relations. The cutoff in 
the metal-rich relation at $M_B \sim -15.5$ primarily reflects the magnitude limit of the 
sample; it may continue to fainter magnitudes, although many such galaxies have only 
metal-poor GCs. The remarkable inference to be drawn from Figure 2 is that the peak 
metallicities of both subpopulations are determined primarily by galaxy luminosity (or 
mass) across the entire spectrum of galaxy types.

Using their new color-metallicity transformation between $g-z$ and [Fe/H], Peng 
\etal~found $Z \sim L^{0.25}$ for metal-rich GCs, which is also consistent with the 
previous estimates of the slope already noted. The color-metallicity relation appears to be 
quite nonlinear, as discussed below. Thus, even though the slopes of the metal-poor and 
metal-rich relations are significantly different in the GC color--galaxy luminosity plane, 
they are similar in the GC metallicity--galaxy luminosity plane (see Figure 2). In \S 11 
we discuss the constraints on galaxy formation implicit in these relations.

The Advanced Camera for Surveys (ACS) on \emph{HST} has significantly advanced our 
understanding of the color distributions of GC systems, offering a wider field of view and 
improved photometric accuracy compared to WFPC2. Three large studies of Es utilizing 
\emph{HST}/ACS have recently been published. As mentioned above, Peng \etal~(2006) and 
Strader \etal~(2006) studied the GC systems of early-type galaxies (ranging from dwarf to 
giant) using $g$ and $z$ data taken as part of the ACS Virgo Cluster Survey (C{\^ o}t{\' 
e} \etal~2004). Peng \etal~investigated all 100 (E and S0) galaxies, while Strader 
\etal~focused solely on the ellipticals (Es). Harris \etal~(2006) used $BI$ ACS photometry 
to analyze GCs in eight ``BCGs", galaxies which are among the brightest in their 
respective groups or clusters.

These studies resulted in several new discoveries. First, a correlation was found between 
color and luminosity for {\it individual} metal-poor GCs in some giant Es (the ``blue 
tilt"; see Figure 3). This is the first detection of a mass--metallicity relation for GCs. 
The blue tilt was found by Strader \etal~in the Virgo giant Es (gEs) M87 and NGC 4649 and 
by Harris \etal~in their sample, although the interpretations of the findings differ. The 
mass--metallicity relation for individual metal-poor GCs may argue for self-enrichment. 
Strader \etal~speculated that these metal-poor GCs were able to self-enrich because they 
once possessed dark matter halos that were subsequently stripped (see also discussion in 
\S 12). The M87 data are well-fit by a relation equivalent to $Z \propto M^{0.48}$ over 
the magnitude range $20 < z < 23.2$, where the turnover of the GC luminosity function 
(GCLF) is at $z \sim 23$. Harris \etal~found a similar relation ($Z \propto M^{0.55}$) but 
suggested that the trend was only present at bright luminosities ($M_I \la -9.5$ to $-10$, 
corresponding to $z \la 22$ in the Strader \etal~Virgo dataset). The CMDs in Strader 
\etal~for M87 and NGC 4649 appear consistent with a continuation of the correlation to 
magnitudes fainter than $z = 22$, but do not strongly distinguish between the two 
interpretations. We do know that the blue tilt phenomenon is not confined to galaxies in 
high density environments or even just to E galaxies. It has recently been reported for 
NGC 4594 (Spitler \etal~2006), a luminous Sa galaxy that lies in a loose group. Curiously, 
the Virgo gE NGC 4472 (also studied by Strader \etal) shows no evidence for the blue tilt. 
If this lack of a tilt is confirmed with better data, it will be a strong constraint on 
any potential ``universal" model for explaining the phenomenon in massive galaxies. The 
Milky Way itself does not show evidence for the tilt, but this could be due to the small 
number of metal-poor GCs ($\sim 100$) compared to massive galaxies or to the inhomogeneity 
of metallicities and integrated photometry in current catalogs.

Harris \etal~(2006) found that the metal-poor GC sequence lay slightly redward of, but 
parallel to, the luminosity--color sequence of dE nuclei from Lotz \etal~(2004). The dE 
data were converted from $V-I$ to $B-I$ for this comparison, and a small zero-point shift 
in the color conversion would line these up. Indeed, in Figure 3 the M87 $z$ vs.~$g-z$ 
color-magnitude diagram is shown superimposed with dE nuclei from Strader \etal~(2006). 
Here the sequence of dE nuclei falls right on top of the bright end of the metal-poor GC 
sequence, although the dE sequence is broader. This could be consistent with a scenario in 
which at least some of the metal-poor GCs in gEs are stripped nuclei of dEs. Since the 
size range of dE nuclei overlaps that of GCs, it may be impossible to ascertain the 
provenance of every luminous cluster.

Harris \etal~suggested that the tilted metal-poor GC relation caused the metal-poor and 
metal-rich peaks to merge at the brightest GC luminosities, turning a bimodal distribution 
into a nominally unimodal one. By contrast, Strader \etal~argued that at these high 
luminosities there is a separate population of objects with larger-than-average sizes and 
a range of colors, spanning the metal-poor to metal-rich subpopulations. Indeed, Harris 
\etal~find that $\sim 20-30$ of the brightest objects in the nearest galaxy in their 
sample, NGC 1407 (at $\sim 21$ Mpc), appear to be extended with respect to normal GCs. The 
size measurements suggest that there is something qualitatively different about (at least) 
a subset of the brightest GCs, which has also been recognized in NGC 1399 (Dirsch 
\etal~2003) and NGC 4636 (Dirsch, Schuberth, \& Richtler 2005). The sizes and luminosities 
of the bright intermediate-color objects in these galaxies suggest a relation to the 
``ultra-compact dwarf" galaxies (UCDs) discovered in both the Fornax and Virgo clusters 
(e.g., Phillips \etal~2001).

The second significant finding was that the color dispersion of the metal-rich GCs is nearly 
twice as large as that of the metal-poor GCs. Peng \etal, Strader \etal, and Harris 
\etal~reported essentially the same dispersions in the color distributions of both 
subpopulations. However, Peng \etal~and Harris \etal~adopted different color-metallicity 
relations, and these led to divergent conclusions about the metallicity distributions of these 
populations. Harris \etal~fit a linear relation between $B-I$ and metallicity using Galactic 
GCs. This has the advantage of being independent of stellar population models but the 
disadvantage of being yoked to the metallicity distribution of Galactic clusters. There are no 
low-reddening Galactic GCs with [Fe/H] $\ga -0.5$, so the empirical relation is unconstrained 
at these metallicities, and the data are poorly fit by a linear relation in the very 
metal-poor regime. Peng \etal~used a piecewise linear relation broken at $g-z = 1.05$ or 
[Fe/H] $\sim -0.8$. This utilized GCs in the Galaxy, M87, and NGC 4472 with both $g-z$ colors 
and spectroscopic metallicities (these are still ultimately tied to the Galactic GC [Fe/H] 
scale). The qualitative effect was to ``flatten" the relation at low metallicities compared to 
a linear fit---so small color changes correspond to large metallicity changes---and to 
``steepen" it at high metallicities.

Consequently, Peng \etal~found the metal-poor GCs to have a larger metallicity dispersion 
than the metal-rich GCs: 68\% half-width [Fe/H] intervals of $\sim 0.6$ dex and 0.3 dex 
for the metal-poor and metal-rich GCs in massive galaxies, respectively (these were 
derived from a nonparametric analysis and thus are not exactly equivalent to a $1\sigma$ 
dispersion for a normal distribution). By contrast, Harris \etal~deduced mean $1\sigma$ 
dispersions of $\sim 0.3$ dex and 0.4 dex, and noted that the metal-poor and metal-rich 
subpopulations in the Galaxy have $\sigma = 0.34$ dex and $\sigma = 0.16$ dex. The 
relative widths of the two Galaxy subpopulations are more consistent with the Peng 
\etal~results, though smaller in an absolute sense. However, even if the metallicity 
dispersion in the metal-poor GCs is larger, the \emph{absolute} metallicity spread is much 
smaller. The implications of these differences for the enrichment histories of the two 
subpopulations remain to be seen.

It seems clear from the new Galactic GC data in Peng \etal~that a single linear fit is not 
optimal, but the exact form of the relation at metallicity extremes is poorly constrained. 
Clearly, identifying the correct form of the $g-z$ to [Fe/H] conversion (and, indeed, 
conversions for other colors) is essential, since the metallicity spreads in individual 
subpopulations have important implications for the formation and assembly histories of GC 
systems. The different color dispersions for the subpopulations also indicate that it is 
necessary to use heteroscedastic (unequal variance) fits in mixture modeling. Homoscedastic 
fits will give systematic errors in the peak values, but may be the best option for systems 
with few GCs. Another important effect of nonlinear color-metallicity relations is that 
bimodal color distributions can be enhanced or even created from metallicity distributions 
that are not strongly bimodal (Richtler 2006; Yoon, Yi, \& Lee 2006).

The last principal finding in these new studies, reported in both Strader \etal~and Peng 
\etal, is that many dEs have metal-rich GC subpopulations. Their colors are consistent 
with an extrapolation of the GC color--galaxy luminosity relation for massive galaxies, 
though the fraction of metal-rich GCs tends to be smaller in dEs than in massive Es. Peng 
\etal~found that the median fraction of metal-rich GCs is $\sim 0.15-0.2$ in dEs and rises 
steeply towards more luminous galaxies. Some of this increase is due to the different 
radial distributions of the two subpopulations; the \emph{HST}/ACS data preferentially 
sample the more centrally-concentrated metal-rich GCs. \emph{Global} fractions of 
metal-rich GCs in massive Es are likely to be closer to 0.3--0.4 (Rhode \& Zepf 2004). The 
changing fraction of metal-rich GCs with galaxy luminosity, combined with the correlations 
of GC colors with galaxy luminosity for both subpopulations, fully explains the classic 
correlation between the mean color/metallicity of a GC system and parent galaxy luminosity 
(Brodie \& Huchra 1991). The overall slope actually measured depends 
on how the sample is defined. If \emph{HST} data are used, the metal-rich GCs will be 
overrepresented and the steep slope of the metal-rich GC relation will dominate the 
overall relation.

These new results suggest that considerable undiscovered detail may still be hidden in GC 
color distributions. Intensive use should be made of \emph{HST}/ACS while it is still operational. 
It is likely to be the best instrument available for the next $\sim 15$ years for studying 
the optical color distributions of GCs.

\subsection{Scenarios for Bimodality}

Once bimodality was observed to be a common phenomenon, several scenarios were presented 
to explain it. In this section, we briefly describe the leading scenarios (see also the 
review of West \etal~2004), but leave detailed discussion until \S 11. The major merger 
model of Ashman \& Zepf (1992) has the distinction of being the only model to predict 
bimodality {\it before} it was observed. This model evolved from early work suggesting 
that E galaxies formed in gas-rich major mergers of disk galaxies (Toomre \& Toomre 1972; 
Toomre 1977; Schweizer 1987). Burstein (1987) and Schweizer (1987) suggested that new GCs 
might be formed in large quantities during the merger process. Ashman \& Zepf (1992) and 
Zepf \& Ashman (1993) developed this idea into a predictive model in which the metal-poor 
GCs are donated by the progenitor spirals and the metal-rich GCs are formed in the 
gas-rich merger. This model gained enormous support when new \emph{HST} observations of 
merging galaxies found large numbers of young massive star clusters (YMCs). The most 
famous example of this is the Antennae (Whitmore \& Schweizer 1995), but several other 
cases were discovered in the early-to-mid 1990s (e.g., NGC 1275, Holtzman \etal~1992; NGC 
7252, Miller \etal~1997). The interpretation of these YMCs as ``proto-GCs" was widely 
adopted. Determining the extent to which these YMCs have properties consistent with 
``normal" old GCs is still an active area of research; see \S 8 below.

Several problems with the major merger model were pointed out in Forbes, Brodie \& 
Grillmair (1997), who showed that, when examined in detail, the number and color 
distributions of GCs in massive Es appeared to be inconsistent with the merger model 
predictions (see \S 11.1.1). They instead suggested that bimodality could arise as a 
consequence of a multi-phase dissipational collapse. In their scenario, the metal-poor 
globular clusters were formed in gaseous fragments during the earliest phases of galaxy 
formation. GC formation was then truncated at high redshift and resumed after a dormant 
period of a few Gyr. During this second phase the metal-rich GCs and the bulk of the 
galaxy field stars were formed. Forbes \etal~discussed this truncation process in terms of 
feedback, with gas being expelled from the early cluster-forming clumps. Such gas would 
later cool and recollapse into the more fully-formed galactic gravitational potential 
until the local conditions were conducive to renewed star formation. Subsequently, Santos 
(2003)  suggested cosmic reionization as the mechanism for truncating metal-poor GC 
formation. Although some details differ, similar ``in situ" models for GC formation were 
also presented by Harris \& Pudritz (1994) and Harris, Harris, \& Poole (1999).

In the accretion scenario of C{\^ o}t{\' e}, Marzke \& West (1998), the metal-rich GCs 
were formed in situ in a massive seed galaxy, while the metal-poor GCs were acquired in 
the dissipationless accretion of neighboring lower-mass galaxies (see also Hilker 1998; 
Hilker, Infante, \& Richtler 1999, and earlier work by Muzzio 1987 and references 
therein). This works in principle because of the long-known relation between the mean 
metallicity of the GC system and the mass of the host galaxy (van den Bergh 1975; Brodie 
\& Huchra 1991; Forbes \& Forte 2001). Stripping of GCs without the accompanying galaxy 
light is also a possibility in dense clusters. C{\^ o}t{\' e} \etal~(1998; 2000; 2002) 
explored the accretion model in detail using Monte Carlo simulations. They showed that the 
bimodality observed in massive galaxies can be reproduced provided that (i) each galaxy 
has an intrinsic ``zero-age" population of GCs, whose metallicity increases with the 
galaxy's mass, and (ii) the primordial galactic mass function for low-mass galaxies is a 
rather steep power law (with $\alpha \sim -2$). The slope is consistent with the halo mass 
functions predicted by standard $\Lambda$CDM models, but much steeper than that actually 
observed for present-day low-mass galaxies (but see \S 11.2).

This triad of scenarios: major merger, in situ/multiphase, and accretion, were the ones 
most frequently discussed to explain bimodality throughout the last decade. In a very real 
sense, the distinctions between these models could be blurred by placing the merger or 
accretion events at high redshift and allowing for significant gas in the components, so 
to some extent the debate was semantic. Indeed, Harris (2003) noted that these scenarios 
could be reclassified in terms of the amount of gas involved. Nevertheless, we have presented 
these scenarios here to give some context for the discussion of observations that follows.

\section{Global Properties}

\subsection{Specific Frequency}

Harris \& van den Bergh (1981) introduced specific frequency ($S_N = N_{GC} \times 10^{0.4\,(M_V
+ 15)}$) as a measure of the richness of a GC system normalized to host galaxy luminosity. This
statistic has been widely used in toy models to assess the feasibility of galaxy formation
mechanisms, e.g., the formation of gEs from disk-disk mergers. $N_{GC}$ was originally
calculated by doubling the number of GCs brighter than the turnover of the GCLF. This definition
makes both practical and physical sense. The faint end of the GCLF is usually poorly defined
(suffering both contamination and incompleteness), and $\sim 90$\% of the mass of the GC system
resides in the bright half. This approach is implicit when fitting a Gaussian (or $t_5$) functions
to observed GCLFs, since observations rarely sample the faintest clusters.

$S_N$ comparisons among galaxies are only valid if all galaxies have the same stellar 
mass-to-light (M/L) ratios. For this reason, Zepf \& Ashman (1993) introduced the quantity 
$T$---the number of GCs per $10^9 M_{\odot}$ of galaxy stellar mass. Since M/L is not 
generally known in detail for a particular galaxy, it is usually applied as a scaling 
factor that is different for each galaxy type, e.g., stellar $M/L_{V}$=10 for Es and 5 for 
Sbc galaxies like the Galaxy. In this section we quote observational results in terms 
of $S_N$, but convert to $T$ for comparisons among galaxies. An even better approach would 
be to directly estimate stellar masses for individual galaxies. Olsen \etal~(2004) discuss 
the use of $K$-band magnitudes for this purpose.

Soon after the merger model was proposed, van den Bergh (1982) argued that elliptical galaxies could 
not arise from the merger of spiral galaxies, since spirals have systematically lower $S_N$ values 
than ellipticals. Schweizer (1987) and Ashman \& Zepf (1992) suggested that this problem could be 
solved if new GCs were formed in the merging process. Observations of large numbers of YMCs in recent 
mergers seemed to support this solution. However, the $S_N$ will only increase through the merger if 
the GC formation efficiency is higher than it was when the existing GC system was formed. Studies of 
YMCs in nearby spirals (Larsen \& Richtler 2000) suggest that galaxies with larger star formation 
rates have more of their light in young clusters, but it seems unlikely that local mergers are more 
efficient at forming GCs than in the gas-rich, violent early universe (see Harris 2001).

In recent years there have been few wide-field imaging studies of the sort needed to accurately
estimate $S_N$. Nonetheless, a trend has emerged that suggests reconsideration of some earlier
results. Newer $S_N$ values tend to be lower than older ones. This revision
stems, for the most part, from improved photometry that is now deep enough to properly define
the GCLF turnover and reject contaminants. Imaging studies that cover a wide field of view are also
important, because they avoid uncertain extrapolations of spatial profiles from the inner regions
of galaxies. In some cases, like the Fornax gE NGC 1399, the luminosity of the galaxy was
underestimated. When corrected, the $S_N$ value for this galaxy was revised downward
by a factor of two, from $S_N \sim 12$ to $5-6$ (Ostrov \etal~1998; Dirsch \etal~2003).

It has been argued for some time that the evolutionary histories of central cluster 
galaxies are different than other Es of similar mass. Somehow this special status has 
resulted in high $S_N$ (or $T$). In addition, galaxies in high-density environments tend 
to have higher $S_N$ than those in groups.  
McLaughlin (1999) argued that high $S_N$ 
values in central galaxies arise because bound hot gas has been ignored for these 
galaxies, and that they do not reflect an increased GC formation efficiency with respect 
to other galaxies (see Blakeslee, Tonry, \& Metzger 1997 for additional arguments in favor 
of the hypothesis that high $S_N$ is due to large galactic mass-to-light ratios). The 
properties of the E galaxy NGC 4636 may also be consistent with this view. Despite its 
relatively low-density environment, it has $S_N \sim 6$, a value typical of central cluster 
galaxies (Dirsch \etal~2005). However, the galaxy has a dark matter halo (traced by a halo 
of hot X-ray gas) that is unusually massive for its luminosity.

The classic Harris (1991) diagram of $S_N$ vs.~luminosity implied a monotonic increase of $S_N$ with 
galaxy mass for high-mass galaxies; that diagram also showed an increase toward
low luminosities for dwarf galaxies. Both of these trends are less apparent in newer data. The 
situation for dwarf galaxies is discussed in more detail in \S 10, though it is worth noting here 
that it is difficult to make robust estimates of the number of GCs in low-mass galaxies
outside of the Local Group. In our view, the run of $S_N$ with galaxy mass and environment remains
uncertain. Additional discussion may be found in McLaughlin (1999) and the reviews by Elmegreen
(1999) and Harris (2001).

\subsubsection{Subpopulations}

Many of the problems with direct $S_N$ comparisons can be circumvented by considering the 
metal-poor and metal-rich subpopulations separately. In particular, recent mergers should 
only affect the metal-rich peak; independent of the details of new star formation, the 
$S_N$ of metal-poor GCs will not increase. Rhode, Zepf, \& Santos (2005) have exploited 
this fact by studying $T$ for metal-poor GCs in 13 massive nearby galaxies, nearly equally 
split between early- and late-type. In Figure 4 we show both $T_{blue}$ and $T_{red}$ 
vs.~stellar galaxy mass. The $T_{blue}$ values are taken from their paper, and the 
$T_{red}$ data were kindly provided by K.~Rhode. Rhode \etal~found an overall correlation 
between $T_{blue}$ and galaxy mass. The spirals are all consistent with $T_{blue} \sim 1$, 
while cluster Es lie higher at $T_{blue} \sim 2-2.5$. NGC 4594 also has $T_{blue} \sim 2$ 
(note in their classification NGC 4594 is an S0, not an Sa). Other field/group Es, 
including NGC 5128, NGC 1052, and NGC 3379, have values similar to those of the spirals.  
Since $M/L_{V}$ increases with galaxy mass (with a relation as steep as $\propto 
L^{0.10}$; e.g., Zepf \& Silk 1996), it is reasonable to expect at least a weak trend of 
$T$ with galaxy mass. However, Rhode \etal~(2005) argue that this can account for only 
$\sim 1/3$ of the observed trend. Because no \emph{global} GC color studies of M87 have 
been published, this galaxy was not included in the Rhode \etal~study. Its $S_N$ value is 
$\sim 3$ times larger than the Virgo gE NGC 4472 (Harris, Harris \& McLaughlin 1998). If 
both galaxies have similar total fractions of metal-poor GCs, the $T_{blue}$ value for M87 
would be $\sim 8$, though this value would be much lower if the mass of its hot gas halo 
were included along with the stellar mass (McLaughlin 1999).

Based solely on the metal-poor GCs, these comparisons seem to rule out the formation of 
cluster gEs (and some massive Es in lower-density environments) by major mergers of disk 
galaxies. However, the relative roles of galaxy mass and environment are still unclear. 
Spirals in clusters like Virgo and Fornax and more massive field/small group Es remain to 
be studied in detail. The metal-poor GC subpopulations of some lower-mass Es in 
low-density environments are still consistent with merger formation. The biggest caveat to 
these interpretations is the effect of biasing---that structure formation is not 
self-similar. Present-day spirals are mostly located in low-density environments like 
loose groups and the outskirts of clusters. High-$T_{blue}$ disk galaxies may have been 
common in the central regions of proto-galaxy clusters at high redshift but have merged 
themselves out of existence (or could, in some cases, have been converted to S0s) by the 
present day. Rhode \etal~(2005) argue that the observed trend of high $T_{blue}$ for 
cluster galaxies might be expected in hierarchical structure formation, since halos in 
high-density environments will collapse and form metal-poor GCs first (see also West 
1993). If GC formation is then truncated (by reionization, for example), such halos will 
have a larger number of metal-poor GCs than similar mass halos in lower density 
environments. See \S 11 for additional discussion of biasing.

The $T_{red}$ data show a similar correlation with galaxy mass, although with a smaller 
dynamic range. Note that the $T_{red}$ values for the spirals ($\sim 0.5$--1) are 
normalized to the \emph{total} galaxy mass in stars. Most have bulge-to-total ratios of $< 
0.3$, so if $T_{red}$ had been normalized just to the spheroidal component, it would be 
substantially higher than plotted. These data appear consistent with the hypothesis of 
near-constant formation efficiency for metal-rich GCs in both spirals and field Es with 
respect to spheroidal stellar mass (Forbes, Brodie, \& Larsen 2001; see also Kissler-Patig 
\etal~1997). The massive cluster Es have \emph{both} higher $T_{red}$ and $T_{blue}$. 
Again, however, because of the sample under study (with few field Es and no cluster 
spirals), it is unclear whether environment or mass is the predominating influence. This 
distinction is moot for the most massive Es since they are found almost exclusively in 
clusters, but is still relevant for typical Es.

Estimating the spread in $T_{red}$ at a given galaxy mass should help constrain the star-formation 
histories of early-type galaxies. The stellar mass of an E might have been built entirely through 
violent, gas-rich mergers (with metal-rich GC formation), or, alternatively, many of the stars could 
instead have formed quiescently in mature spiral disks (with little metal-rich GC formation). 
$T_{red}$ for the latter E should be significantly lower than for the former E.

\subsection{Radial and Azimuthal Distributions}

Most of the existing information on the global spatial distributions of GCs dates from 
older studies that could not separate GCs into subpopulations. The projected radial 
distributions are often fitted with power laws over a restricted range in radius, and it 
is clear that more luminous galaxies have shallower radial distributions (Harris 1986; see 
the compilation in Ashman \& Zepf 1998). Considering the GC system as a whole, typical 
projected power law indices range from $\sim -2 $ to $-2.5$ for some low-luminosity Es 
(this is also a good fit to the Galactic GC system; Harris 2001) to $\sim -1.5$ or a bit 
lower for the most massive gEs.  However, it should be kept in mind that power laws 
provide a poor fit over the entire radial range. Most GC radial distributions have cores, 
and gradually become steeper in their outer parts. King models capture some of this 
behavior. In nearly all cases, the GCs have a more extended spatial distribution than the 
galaxy field stars.

There have been a few wide field imaging programs which considered GC subpopulations 
separately. In their study of the Virgo gE NGC 4472, Geisler \etal~(1996) were the first to 
show clearly that color gradients in GC systems are driven solely by the different radial 
distributions of the subpopulations. The metal-poor and metal-rich GCs themselves show no 
radial color gradients. Rhode \& Zepf~(2001) found a color gradient for the total GC system in 
NGC 4472 interior to $\sim 8$\arcmin, but no gradient when the full radial extent of the GC 
system (22\arcmin $\sim 110$ kpc) was considered. Dirsch \etal~(2003) studied the GC system of 
the Fornax gE NGC 1399 out to $\sim 25$\arcmin ($\sim 135$ kpc); this work was extended to 
larger radii by Bassino \etal~(2006). The radial distributions of the two subpopulations are 
shown in Figure 5, along with the profile of the galaxy itself. The metal-poor and metal-rich 
subpopulations have power law slopes of $\sim -1.6$ and $\sim -1.9$, respectively. These 
differences persist to large radii, and lead to an \emph{overall} color gradient in the GC 
system. The radial distribution of the metal-rich GCs is a close match to that of the galaxy 
light, and suggests that they formed contemporaneously.

Another notable finding is the rather abrupt truncation of GC systems at large 
galactocentric radius. Rhode \& Zepf (2001) found that the surface density of GCs in 
NGC~4472 falls off faster than a de Vaucoleurs or power law fit at $\sim 20$\arcmin ($\sim 
100$ kpc). A similar drop-off occurs at 11\arcmin\ in NGC 4406 (Rhode \& Zepf 2004), and at 
9\arcmin\ in NGC 4636 (Dirsch \etal~2005). This may be evidence of truncation by the tidal 
field of the cluster.

Dissipationless mergers (whether disk-disk or E-E) tend to flatten the radial slopes of 
existing GC systems and create central cores (Bekki \& Forbes 2005). These cores are 
observed and have sizes of a few kpc (Forbes \etal~1996). These results apply to 
pre-existing GCs, which certainly includes metal-poor GCs, as well as metal-rich GCs that 
existed before the merger. Thus, the trends predicted by the Bekki \& Forbes simulation 
appear qualitatively consistent with observations. The observations are also consistent 
with more extensive merger histories for more massive galaxies; these gradually produce 
larger cores and flatter GC radial distributions. A desirable extension to this work would 
be to place these simulations in a cosmological framework in which merging occurs according 
to a full N-body merger tree. Also, as data become available, comparisons between 
observations and simulations could be restricted to metal-poor GCs. With this approach, any 
new metal-rich GCs that might have formed in the merger are irrelevant.

Ashman \& Zepf (1998) noted that little was known about the two-dimensional spatial 
distributions of GC systems. Scant progress has been made in the intervening years. 
Existing data are consistent with the hypothesis that both subpopulations have 
ellipticities and position angles similar to those of the spheroids of their parent 
galaxies (e.g., NGC 1427, Forte \etal~2001; NGC 1399, Dirsch \etal~2003; NGC 4374, 
G{\'o}mez \& Richtler 2004; NGC 4636, Dirsch \etal~2005). This also holds for the dEs in 
the Local Group (Minniti \etal~1996). It may be that in some galaxies (e.g., the E4 NGC 
1052; Forbes, Georgakakis, \& Brodie 2001) the metal-rich GCs follow the galaxy 
ellipticity more closely than the metal-poor GCs, but whether this is a common phenomenon 
is unknown. Naively, if the metal-rich GCs formed along with the bulk of the galaxy field 
stars, they should closely trace the galaxy light. The spatial distribution of the 
metal-poor GCs will depend in detail upon the assembly history of the galaxy.

\subsection{Variations with Galaxy Morphology}

\subsubsection{Spirals}

Our views on the GC systems of spiral galaxies are heavily shaped by the properties of the 
Milky Way (and, to a lesser degree, M31). This is discussed in detail in \S 5. A principal 
result from the Galaxy is that the metal-poor and metal-rich GCs are primarily associated 
with the halo and bulge, respectively. Forbes \etal~(2001) introduced the idea that the 
bulge GCs in spirals are analogous to the ``normal" metal-rich GCs in early-type galaxies, 
and that in both spirals and field Es the metal-rich GCs have $S_N \sim 1$ when normalized 
solely to the bulge luminosity. The constancy of bulge $S_N$ appeared consistent with 
observations of a rather small set of galaxies (Milky Way, M31, M33, NGC 4594) but needed 
further testing.

Goudfrooij \etal~(2003) provided such a test with an \emph{HST}/WFPC2 imaging study of 
seven edge-on spirals, ranging across the Hubble sequence (the previously-studied NGC 4594 
was part of the sample). Edge-on galaxies were chosen to minimize the effects of dust and 
background inhomogeneities on GC detection. Corrections for spatial coverage were carried 
out by comparison to the Galactic GC system; this does not appear to introduce systematic 
errors (see Goudfrooij \etal~for additional discussion). Kissler-Patig \etal~(1999) had 
previously found that one of these galaxies, NGC 4565, has a total number of GCs similar 
to the Galaxy. The small WFPC2 field of view and the low $S_N$ (or $T$) values of spirals 
as compared to Es resulted in the detection of only tens of GCs in some of the Goudfrooij 
\etal~galaxies. While bimodality was not obvious in all of the color distributions, each 
galaxy had GCs with a range of colors, consistent with multiple subpopulations. Using a 
color cut to divide the samples into metal-poor and metal-rich GCs, Goudfrooij \etal~found 
that all of the galaxies in their study had (i) a subpopulation of metal-poor GCs with 
$S_N \sim 0.5-0.6$ (normalized to total galaxy light), and (ii) constant bulge $S_N$, with 
the exception of the rather low-luminosity Sa galaxy NGC 7814, which appeared to have few 
GCs of any color. NGC 7814 was later the target of a ground-based study by Rhode \& Zepf 
(2003) with the WIYN telescope. They found a significant metal-rich GC subpopulation with 
a bulge $S_N$ squarely in the middle of the range found for the other galaxies, which 
showed that the single WFPC2 pointing on the sparse GC system of NGC 7814 gave an 
incomplete picture of the galaxy. In this case even small radial leverage was 
important: Rhode \& Zepf found that the surface density of GCs dropped to zero at just 12 
kpc (3\arcmin) projected.

Chandar, Whitmore, \& Lee (2004) studied GCs in five nearby spirals. Their galaxies (e.g., 
M51, M81) tended to be closer and generally better-studied than those in the Goudfrooij 
\etal~sample, but they were also at less favorable inclination angles. As a result, their 
GC candidate samples were more prone to contamination and more affected by (sometimes 
unknown amounts of) reddening. This was partially mitigated by their wide wavelength 
coverage, including (in some cases) $U$-band imaging, to help constrain the reddening of 
individual GCs. Chandar \etal~found evidence for bimodal GC color distributions in M81 and 
M101. Interestingly, M51, which had rather deep imaging, showed no evidence for metal-rich 
GCs. They found that NGC 6946 and M101 had subpopulations of clusters with sizes similar 
to GCs but extending to fainter magnitudes; the typical log-normal GCLF was not seen. In 
NGC 6946 the imaging was quite shallow and these faint objects were found to be blue, 
suggesting that they might be contaminants. However, in M101 the imaging was deeper and 
the faint objects had red colors, consistent with old stellar populations. This may be 
evidence that some spirals possess clusters unlike those typical of Es, and these clusters 
may be related to the faint red objects seen in some S0s (see next subsection). Chandar 
\etal~also compiled data from the literature on the GC systems of spirals, and argued that 
$S_N/T$ depends on Hubble type, but not on galaxy mass. This is consistent with the 
findings of Goudfrooij \etal~(2003).

In principle, the $S_N$ value should depend on whether a particular bulge formed 
``classically", with intense star formation, or through secular processes in quiescent gas 
disks (e.g., Kormendy \& Kennicutt 2004). In the former case we might expect a bulge $S_N$ 
similar to that found for Es, but in the latter case the star formation is likely to be 
sufficiently slow and extended that few or no GCs are formed along with the bulge stars. 
This may result in a rather low bulge $S_N$ compared to Es of similar mass. The 
implication is that all the bulges of galaxies in the Goudfrooij \etal~sample formed 
predominantly through the ``classical" route. The generally old ages of the metal-rich GCs 
in spirals (see \S 4) is evidence that the majority of bulge star formation, by whatever 
mechanism, happened at relatively early epochs.

Kormendy \& Kennicutt (2004) argue that a large fraction of spiral bulges are built by 
secular evolution, and that these ``pseudobulges" are especially common in late-type 
spirals. The diagnostics for pseudobulges are many, but include cold kinematics, surface 
brightness profiles with low Sersic indices, and, in some cases, young stellar 
populations. It is worth emphasizing that many of these pseudobulges could be composite 
bulges with young to intermediate-age stars superposed on an old classical bulge. In this 
case, the bulge $S_N$ could serve as a diagnostic of the degree to which a given bulge can 
have been built by classical or secular processes. The Milky Way itself could be an 
example of such a bulge. Its bulge is dominated by an old stellar population, but has a rather 
low velocity dispersion for its mass. The kinematics of the metal-rich GCs could be 
consistent with association with either a bulge or a bar (C{\^ o}t{\' e} 1999).

\subsubsection{S0s}

The leading theory for the formation of most S0s involves their transformation from spirals 
as groups and clusters virialize (e.g., Dressler \etal~1997). This can occur in a variety 
of ways, including ram pressure stripping and minor mergers that disrupt the disk 
sufficiently to halt star formation. In this context, it may be more appropriate to compare 
the GC systems of S0s to those of spirals, rather than make the traditional comparison with 
Es. Nonetheless, the GC systems of S0s appear to be quite similar to those of Es when 
compared at fixed mass. Kundu \& Whitmore (2001b) studied a variety of S0s with WFPC2 
snapshot imaging, and in many galaxies found broad color distributions consistent with 
multiple subpopulations. Peng \etal~(2006b) used deeper imaging from the ACS Virgo Cluster 
Survey and found color bimodality in nearly all of the massive S0s in their sample. These 
S0s fall right on the GC color--galaxy luminosity relations of the Es. If indeed S0s 
descend from spirals, this is yet another piece of evidence that massive galaxies of all 
types along the Hubble sequence have very similar GC color distributions, and hence are 
likely to have experienced similar violent formation processes at some point in their 
history.

One interesting finding so far confined to S0s was the serendipitous discovery of a new 
class of star cluster, now known informally as the Faint Fuzzies (FFs). These objects were 
first detected in the nearby (10 Mpc) S0 NGC 1023. Along with a normal, bimodal system of 
compact GCs, this galaxy hosts an additional population of faint ($M_V > -7$) extended 
($R_{eff} \sim 7-15$ pc) star clusters. In deep \emph{HST}/WFPC2 images, these objects are 
confined to an annular distribution closely corresponding to the galaxy's isophotes (Larsen 
\& Brodie 2000). Spectroscopic follow-up with Keck/LRIS showed that the FFs are metal-rich 
([Fe/H] $\sim -0.5$), old ($> 8$ Gyr), and rotating in the disk of the galaxy (Brodie \& 
Larsen 2002). With old ages, inferred masses of $\la 10^{5} M_{\odot}$, and sizes $\sim 5$ 
times larger than a typical globular or open cluster, these objects occupy a distinct 
region of age--size--mass parameter space for star clusters. As a population, they have no 
known analogs in the Milky Way or elsewhere in the Local Group. Similar objects have been 
found in the S0 galaxy NGC 3384 (Brodie \& Larsen 2002) and NGC 5195, a barred S0 
interacting with M51 (Lee, Chandar \& Whitmore 2005; Hwang \& Lee 2006). Peng \etal~(2006) 
found FFs in $\sim 25$\% of the S0s in the ACS Virgo Cluster Survey. Due to biases in 
sample selection, however, this fraction is probably not yet well-constrained. The FFs have 
relatively low surface brightness, and their properties may be consistent with $M \propto 
R^{2}$ (unlike GCs, which show no $M-R$ relation). In some cases, their colors are redder 
than those of the metal-rich GCs in the same galaxy, which may suggest higher 
metallicities.

Brodie, Burkert, \& Larsen (2004) and Burkert, Brodie \& Larsen (2005) showed that the 
properties of the FFs in NGC 1023 were consistent with having formed in a rotating 
ring-like structure and explored their origin. Numerical simulations suggest that objects 
with the sizes and masses of FFs can form inside giant molecular clouds, provided star 
formation occurs only when a density threshold is exceeded.  Such special star forming 
conditions may be present during specific galaxy-galaxy interactions, in which one galaxy 
passes close to the center of a disk galaxy, precipitating a ring of star formation. They 
speculated that the FFs might then be signposts for the transformation of spiral galaxies 
into lenticulars via such interactions. Alternatively, such conditions might also occur in 
the inner resonance rings associated with the bars at the centers of disk galaxies. In 
this case, the old ages of the FFs would suggest that barred disks must have been present 
at early times.

\section{Spectroscopy}

The detailed properties of individual extragalactic GCs---ages, metallicities, and chemical 
abundances---are important constraints on theories of GC and galaxy formation. Emphasis has 
been placed on accurate age-dating of GCs. To the extent that Es formed in recent gas-rich 
major mergers, this should be reflected in young age measurements for their metal-rich GCs, 
while both in situ and accretion scenarios predict old ages for both subpopulations.

In principle, integrated-light spectroscopy of individual GCs offers much stronger 
constraints on ages and elemental abundances than broadband photometry. In practice, even 
low-resolution spectroscopy is challenging. At the distance of Virgo ($\sim 17$ Mpc), GCs 
are already faint, with the turnover of the GCLF occurring at $V \sim 23.5$. Except within 
the Milky Way and M31, it has so far been possible to assemble only small samples of 
high-quality GC data, and these have required significant commitments of 8 to 10-m telescope 
time. Historically, spectroscopic samples have been not only small but also biased to the 
brightest and/or reddest objects in a given galaxy, forcing a reliance on photometric 
studies for global conclusions about metallicity distributions. As discussed below, this 
combination of photometry and small-sample spectroscopy has been used to establish that the 
ubiquitous color bimodality is due primarily to metallicity differences between the 
subpopulations, without the need to invoke age differences. There are also biases in spatial 
sampling: the fields of view of the spectrographs used in most studies are too small to 
sample the outermost GCs, and the GCs in the central regions are generally undersampled due 
to the minimum length of slitlets and the high central concentration of most GC systems.

There is a fundamental difference between metallicity and age studies in terms of the 
conclusions that can be derived from small spectroscopic samples. Age substructure can be 
identified in biased samples of metal-rich GCs, since younger GCs (with a standard GC mass 
function) will be brighter than their older counterparts. Thus magnitude-limited samples set 
an upper limit on the proportion of younger GCs in the system. By contrast, to properly 
sample the metallicity distribution, it is necessary to obtain spectra over a color range 
representative of the entire system. There is no \emph{technical} reason why large-sample 
spectroscopic studies with the new generation of highly-multiplexing spectrographs like 
Keck/DEIMOS and VLT/VIMOS are not feasible, and such work may proliferate in the near 
future. In Figure 6, we show representative spectra of three GCs in M31: an old metal-poor 
GC, an old metal-rich GC, and an intermediate-metallicity, $\sim 2$ Gyr GC (Beasley 
\etal~2005).

Most spectroscopic work on extragalactic GCs has utilized Lick/IDS indices (Burstein 
\etal~1984; Worthey \etal~1994; Trager \etal~1998). These were developed to measure 
absorption features from $\sim 4000-6400$ \AA\ in the spectra of early-type galaxies. Such 
galaxies typically have large velocity dispersions; thus the low resolution of the Lick 
system ($\sim 8-11$ \AA) and wide index bandpasses (tens of \AA) required by the IDS 
(Robinson \& Wampler 1972) were not severe impositions. Indices (essentially equivalent 
widths) are defined in terms of a central bandpass which contains the feature of interest 
and flanking bandpasses that set the local pseudocontinuum. Observations of distant 
systems are compared to simple stellar population (SSP) evolutionary models using a set of 
fitting functions (e.g., Worthey 1994) that predict Lick indices as a function of stellar 
parameters (e.g., $T_{\textrm{eff}}$, log $g$, [Fe/H]). Indices measured on modern spectra 
must be ``corrected" to the Lick/IDS system through observations of standard stars.

This system is not optimal for GCs, which have low velocity dispersions and metallicities 
compared to galaxies.  Although the typical S/N of extragalactic GC spectra has not justified 
the use of narrower indices until very recently, the opportunity now exists to improve the 
placement of index and pseudocontinuum bandpasses and to define new indices in the feature-rich 
wavelength region below 4000 \AA. A number of groups are currently defining new index systems 
on the basis of higher resolution ($\sim 1-3$ \AA) stellar libraries (e.g., the 2.3 \AA\ MILES 
library; Sanchez-Blazquez \etal~2006), as well as pursuing the direct fitting of models to 
observed spectra, avoiding indices entirely (e.g., Wolf \etal~2005).

In the following subsections we discuss observational results on metallicities, ages, and 
individual elemental abundances of extragalactic GCs.

\subsection{Metallicities and Ages}

Before proceeding, it is worth noting that the term metallicity is not precisely defined in 
the context of GCs.  Some estimates (e.g., composite PCA metallicities; Strader \& Brodie 
2004) are tied to Galactic GC metallicities on the Zinn \& West (1984) scale. This scale 
measures a nonlinear combination of metals and is unlikely to reflect either 
[Fe/H] or ``true" [$Z$/H]. The Carretta \& Gratton (1997) and the Kraft \& Ivans (2004) Fe 
scales have yet to be extended to the metal-rich regime and calibrations are limited to the 
metallicity distribution of the Milky Way GC system. This barely touches solar and is poorly 
matched to the metal-rich regime of GCs in gEs, which extends to solar (and perhaps 
beyond). The metal-rich GCs in the Galaxy are also few in number and generally highly 
reddened, as they are concentrated in the bulge. The alternative to direct calibration is to 
derive metallicities solely from models. A 
myriad of issues remain with this approach, including uncertain isochrones, meager stellar 
libraries, corrections to the Lick system, and adjustments for non-solar abundance ratios 
(see discussion in \S 4.2). Some authors have adopted the more agnostic [m/H] (with ``m'' for 
metallicity) to describe their values; others calculate a ``corrected" [Fe/H] from [m/H] or 
[$Z$/H] by subtracting an assumed or derived [$\alpha$/Fe] (e.g., Tantalo, Chiosi, \& 
Bressan 1998). Direct comparisons of metallicities derived from different methods can 
easily be uncertain at the $\sim0.2-0.3$ dex level.

One method to estimate metallicities for GCs is to simultaneously use multiple lines indices; a 
weighted combination of six indices was used to derive metallicities for individual GCs in the 
Milky Way and M87 (Brodie 1981; Brodie \& Hanes 1986). Brodie \& Huchra (1990, 1991) applied a 
refined version of this approach to GCs in a variety of galaxies. Two other studies derived 
metallicity calibrations using principle components analysis of indices of Galactic GCs: Gregg 
(1994) used a large set of narrow indices, and Strader \& Brodie (2004) used Galactic GC spectra 
from Schiavon \etal~(2004) to derive a metallicity estimator that is a linear combination of 11 
Lick indices. The PCA methods provide accurate metallicities on the Zinn \& West (1984) scale in 
the range $-1.7 \la$ [Fe/H] $\la 0.0$. The other principal method is to simultaneously derive metallicity
and age through comparisons to SSP models (see discussion below).

The bulk of spectroscopic studies of extragalactic GCs have been conducted with two 
instruments, Keck/LRIS and VLT/FORS. Ages, metallicities, and [$\alpha$/Fe] ratios have been 
estimated predominantly by measuring Lick indices.

In a Keck/LRIS study, Kissler-Patig \etal~(1998) inferred that the majority of GCs in the 
Fornax gE NGC 1399 were old, but that a small percentage might be as young as a few Gyr. 
However, the robustness of these conclusions was hampered by the low S/N of the spectra. In 
another Keck/LRIS program, Cohen, Blakeslee, \& Ryzhov (1998) studied a sample of $\sim 
150$ GCs in M87. Despite large errors on the absorption line indices measured in many of 
their spectra, the size of the sample allowed a good statistical comparison with spectra of 
Milky Way GCs. The M87 and Milky Way GCs were found to populate similar areas in Mg$b$ 
vs.~H$\beta$ and Mg$b$ vs.~Fe5270 diagrams, suggesting that the M87 GCs have ages and 
[Mg/Fe] ratios comparable to those of the Galactic GCs. These ages and [$\alpha$/Fe] ratios 
are known, from detailed analysis of individual cluster stars, to be $\ga 10$ Gyr and $\sim 
+0.2-0.3$, respectively (e.g., Carney 2001). Beasley \etal~(2000) found old ages for 
coadded (on the basis of $C-T_{1}$ color) WHT spectra of GCs in NGC 4472. Kuntschner 
\etal~(2002) presented a study of the S0 NGC 3115 showing that GCs in both the metal-poor 
and metal-rich subpopulations are old.

Subsequent studies undertaken with Keck/LRIS include: Ellipticals---NGC 1399 (Forbes 
\etal~2001); NGC 4472 (Cohen, Blakeslee, \& C{\^ o}t{\' e} 2003); NGC 4365 (Larsen 
\etal~2003; Brodie \etal~2005); NGC 3610 (Strader \etal~2003; Strader, Brodie, \& Forbes 
2004b); NGC 1052 (Pierce \etal~2005); NGC 1407 (Cenarro \etal~2006); S0s---NGC 1023 (Brodie 
\& Larsen 2002); NGC 524 (Beasley \etal~2004); Spirals---M81 (Schroder \etal~2002); NGC 4594 
(Larsen \etal~2002). In all cases it was concluded that at most a small fraction of the 
observed samples of GCs were young or of intermediate age ($\la 5-6$ Gyr). In the case 
studies of the intermediate-age merger remnant E NGC 3610, Strader \etal~(2003, 2004b) found 
that only two out of ten prime candidate intermediate-age GCs had ages consistent with 
formation in the merger. Figure 7 shows a typical index-index plot used to derive 
metallicities and ages for GCs, in this case, from NGC 1407 (Cenarro \etal~2006). Here the 
GCs studied appear to have ages consistent with Galactic GCs, but extend to higher 
metallicities.

In an effort to use more of the information available in a spectrum than is contained in a 
traditional index-index plot, Proctor, Forbes, \& Beasley (2004; see also Proctor \& 
Sansom 2002) developed a $\chi^2$ minimization routine that simultaneously fit all of the 
Lick indices to SSP models, exploiting the different sensitivities of each of the indices 
to age, metallicity, and [$\alpha$/Fe]. Metallicities derived in this manner are 
complementary to ones derived using the PCA methods described above, and, for GCs in the 
range over which the PCA metallicities are calibrated, the agreement is excellent (e.g., 
Cenarro \etal~2006). Multi-index approaches offer the advantage of minimizing random and 
systematic problems in any individual index, such as may arise in low S/N data. Moreover, 
where the GC background is an emission line region (from either sky or the host galaxy), 
or when individual element abundance anomalies or extreme horizontal branch morphologies 
may be present in the GC, such precautions are essential (see below).

The best of the Keck data were compiled in a meta-analysis by Strader \etal~(2005). This 
work synthesized more than a decade's worth of effort on the GCs systems of a 
heterogeneous sample of galaxies, spanning the range from dwarfs to gEs (all but NGC 4594 
were of early-type). A typical galaxy had only 10--20 spectra of sufficiently high S/N to 
be included in the analysis. Lick indices measured from these spectra were combined, and 
ages were derived through direct differential comparison to indices of Galactic GCs, in 
part to avoid SSP model uncertainties. This study showed that both the metal-poor and 
metal-rich GC subpopulations had mean ages as old as (or older than) Galactic GCs. In 
Figure 8, from Strader \etal~(2005), PCA metallicity is plotted against a composite Balmer 
line index of H$\beta$ and H$\delta_{A}$ (H$\gamma$ was excluded because the strong G band 
lies in its blue pseudocontinuum bandpass). The implication from their sample of galaxies 
is that GC formation, and by extrapolation, the bulk of star formation in spheroids, took 
place at early times ($z \ga 2$). As discussed in the Annual Review of Renzini (2006, this volume), fossil 
evidence from the local universe and in situ studies at high redshift lead to similar 
conclusions: most of the stars in massive Es formed at $z \sim 2-5$. Of course, studies of 
larger samples of GCs in a wide range of galaxies are needed to further test this result.

This conclusion is still consistent with the existence of young early-type galaxies in the 
local universe (Trager \etal~2000), because galaxy age estimates are luminosity-weighted. A 
small ``frosting" of recent star formation can easily result in the measurement of young 
spectroscopic ages for galaxies whose stars are predominately old, and there is an inverse 
relationship between the age and required burst strength. In addition, radial age gradients in 
early-type galaxies are often measured. Younger ages are derived from spectra extracted from 
small (1/8 of the effective radius; $r_e$) apertures than from larger ($r_{e}/2$) ones. In any 
case, a small amount of star formation is expected to produce a correspondingly small 
subpopulation of young GCs, which could easily have escaped detection in existing samples. 
High quality spectra of $> 100$ GCs per galaxy, now obtainable with highly multiplexing 
instruments such as Keck/DEIMOS and VLT/VIMOS, would be necessary to probe these low-level 
star-forming events.

Puzia \etal~(2005; see also Puzia 2003) presented a spectroscopic analysis of a more 
homogeneous sample of galaxies, comprising 5 Es and 2 S0s, all with $-19.2 > M_B > -21.2$, 
and mostly in small groups. The low end of this luminosity range is near $L^*$, so their 
sample, like that of Strader \etal~(2005), was dominated by early-type galaxies brighter 
than the knee of the galaxy LF. They included a detailed discussion of the advantages of the 
different Balmer lines in terms of dynamic range, age-metallicity degeneracy, and ability to 
correct to the Lick system. Using a combined Balmer line index and the $\alpha$-insensitive 
metallicity proxy [MgFe]\arcmin\ (Thomas, Maraston, \& Bender 2003), Puzia \etal~found 4 GCs 
(out of their total of 17 high-quality spectra) with index measurements consistent with ages 
less than 9 Gyr (their Figure 7). While this sample is not sufficient to address the 
proportion of young GCs that might be present in any individual galaxy, the presence of some 
younger GCs in these typically group galaxies would be consistent with ``downsizing", 
whereby lower-mass galaxies in lower-density environments tend to have younger mean ages 
(Cowie \etal~1996). There remains disagreement on the proportion of young GCs present in 
galaxies of moderate luminosity.
 
The evidence that most GCs in massive galaxies are quite old is relevant to the distinction 
between ``core" and ``power-law" E galaxies seen in the local universe. These designations stem 
from the surface brightness profiles in the very inner parts of the galaxy. The core galaxies 
tend to be bright ($M_V \la -21.5$), have boxy isophotes, and do not rotate; the power-law 
galaxies are fainter, have disky isophotes, and are at least partially rotationally supported 
(Kormendy \& Bender 1996; Faber \etal~1997). The working hypothesis for the creation of central cores is 
scouring by 
merging binary black holes. Kormendy \etal~(2006) have unified this scenario by demonstrating 
that power-law galaxies have excess light in their central parts, above an extrapolation of a 
best-fit Sersic profile in the outer parts. Core galaxies have no such excess light. They argue 
that the distinction between core and power-law galaxies represents Es whose last major merger 
was dry or wet, respectively (i.e., without or with star formation). In power-law galaxies, 
black hole scouring may still have occurred, but it is swamped by the light from the stars in the 
center that formed in the merger. The old GC ages inferred from observations tentatively indicate 
that, even in 
power-law galaxies, the wet merger that formed the E either occurred long ago, or produced only a 
relatively small amount ($\la 10$-20\%) of the galaxy's current stellar mass. However, much 
larger spectroscopic samples of GCs in power-law galaxies are needed to better constrain the 
fraction of mass produced in wet mergers.

\subsubsection{Uncertainties}

An important uncertainty in age determinations is the level of contribution from hot stars 
to the integrated spectra of GCs. These can ``artificially" enhance the Balmer line 
strengths beyond the values set by the main sequence turnoff and can lead to an 
underestimation of the GC age. Blue stragglers and blue horizontal branch stars are the 
primary offenders---hotter stars, including extreme HB, AGB-manque, and post-AGB stars 
have little flux in the optical region where ages are usually derived. At least in the 
Milky Way, blue stragglers have quite a small effect on a GC's optical integrated spectrum 
(e.g., Schiavon \etal~2002). It is worth noting that this may not necessarily be the case 
in extragalactic GCs that have high binary fractions or are very compact.

This leaves blue HBs stars as the main systematic source of uncertainty in GC age estimates. 
HB morphologies are primarily set by GC metallicity, with Balmer line strengths peaking at 
[Fe/H] $\sim -1.3$ and decreasing toward lower metallicities as the stars become hotter and H 
is ionized. The classic ``second-parameter problem"---that GCs of a given metallicity can have 
quite different HB morphologies---is currently unresolved, despite intensive research efforts 
over several decades. It has frequently been suggested that age is the second parameter (e.g., 
Chaboyer, Demarque, \& Sarajedini 1996). However, this does not appear to be true in the few 
cases where direct turnoff age comparisons can be made by normalizing the main sequence 
luminosity functions (e.g, Stetson, Vandenbergh, \& Bolte 1996). While age differences 
certainly will produce changes in HB morphology, it is dangerous to use the HB as an age 
indicator. As an example, several metal-poor GCs in M33 were studied with \emph{HST}/WPFC2 by 
Sarajedini \etal~(2000) and found to have unusually red HBs for their metallicities. 
Sarajedini \etal~suggested intermediate ages might be the cause. However, Larsen \etal~(2002) 
derived dynamical masses for these GCs and instead found that they had M/L ratios typical of 
old GCs. At present there is no known a priori predictor of GC HB morphology. Consequently, 
various HB diagnostics have been suggested. These include (i) the Ca H+K line strength 
inversion; enhanced H$\epsilon$ from hot stars can increase the Ca H line equivalent width 
compared to the Ca K line (Rose 1985), (ii) Near-ultraviolet ($NUV-V$) colors, and (iii) a 
tendency for progressively younger ages to be inferred from increasingly higher-order Balmer 
lines because of an increasing contribution from putative blue HB stars (Schiavon \etal~2004). 
The effect of BHB stars on integrated Balmer line strength can be substantial: Puzia 
\etal~(2005) estimated that $H\beta$ may be increased by up to 0.4 \AA (based on their 
analysis of Galactic GCs), while Maraston (2005) inferred an even greater effect from her SSP 
models at high metallicities (see also Thomas, Maraston, \& Korn 2004).

Aside from the classic second-parameter effect, there is some evidence that the HB 
morphologies of GCs in other galaxies can be quite different from those in the Milky Way. For 
example, in a \emph{HST}/WFPC2 study by Rich \etal~(2005), a sample of 12 M31 GCs appeared 
offset from the Milky Way HB--metallicity relation. One potential explanation is that these 
M31 GCs are $\sim 1-2$ Gyr younger than their Galactic counterparts. More startling is the 
finding by Sohn \etal~(2005), from an \emph{HST}/STIS FUV imaging study of bright M87 GCs, 
that all the GCs have much bluer $FUV-V$ colors than Galactic GCs of the same metallicity. 
Increasing the age of the M87 GCs relative to Galactic GCs is one potential solution, but it 
would require implausibly large age differences of 2--4 Gyr to bring the two samples into 
line. Fuel consumption arguments, and the fact that spectroscopic ages of these GCs are old 
(Cohen \etal~1998), suggest that the hot stars in M87 GCs are a redistribution of stars 
blueward of a normal BHB, and not simply an extension of it. A caveat to this finding is the 
UV-flux limited nature of their sample, which tends to select the most extreme GCs.

A night of 8--10-m telescope time can generate spectra of GCs at the distance of Virgo 
($\sim 17$ Mpc) from which H$\beta$ line index strengths can be measured to an accuracy of 
$\sim 0.1$\AA\ for GCs with $V \sim 21.5$ (for fainter GCs, the errors are correspondingly 
larger). As Figure 7 illustrates, this translates into an error of $\sim 1-2$ Gyr at old 
ages. Since the isochrones are more widely separated at younger ages, the ability to 
discriminate fine age differences improves with decreasing age. In these SSP model grids, 
BHB effects actually cause the isochrones to cross at low metallicities and old 
ages, setting a fundamental upper limit on our age estimates of $\sim 10$ Gyr at these 
metallicities. It is important to remember that the actual ages of the majority of GCs in 
the universe could well be older than this.

\subsection{[$\alpha$/Fe]}

It is widely recognized that measurements of the [$\alpha$/Fe] ratio of a stellar population 
can provide valuable insight into its star formation history, particularly by constraining the 
timing and duration of a starburst. Most $\alpha$-elements (e.g., Mg, Ti, Ca, Si) and 1/3 of 
the Fe (for the solar mixture) are generally thought to form in Type II SNe. Since these come 
from the explosions of massive stars, they closely trace the star formation rate, and begin to 
ignite within tens of Myr of the onset of a starburst. The remainder of the Fe is produced in 
Type Ia SNe, which typically occur over Gyr timescales. Enhanced [$\alpha$/Fe] ratios would 
thus indicate rapid star formation, occurring before there was time for significant enrichment 
of the interstellar medium by Type Ia SNe. This is generally the case in early-type galaxies 
(e.g., Worthey, Faber, \& Gonzalez 1992; Matteucci 1994; Trager \etal~2000; Thomas 
\etal~2005). A complication is that the timescale over which fresh ejecta become incorporated 
into interstellar gas is not necessarily known. Other nucleosynthetic sites, operating on a 
variety of time scales, also contribute to the cosmic mix (e.g., AGB stars produce a 
significant amount of s-process and light elements).

The naive expectation for metal-poor GCs in external galaxies is that they will have 
supersolar [$\alpha$/Fe], since they formed early in the universe, before there was 
substantial metal enrichment. The situation for metal-rich GCs is less clear, as it will 
depend upon the exact formation mechanism. If they formed along with most of the field 
stars in massive Es, they should share the [$\alpha$/Fe] $\sim +0.2-0.4$ of their parent 
galaxies. They could perhaps reach even higher values if they formed preferentially early 
in the starburst. Metal-rich GCs formed at lower redshift from $\sim$ solar metallicity 
gas in major disk-disk mergers might be expected to have [$\alpha$/Fe] closer to 0, since 
it is more difficult to enhance [$\alpha$/Fe] when starting from a high metallicity 
(although metallicity gradients in spiral disks must also be considered). The results from 
extragalactic GCs do \emph{not} generally agree with these expectations, though much of 
this may be due to the observational and technical difficulties in accurately estimating 
[$\alpha$/Fe], as discussed below.

In theory, it should be very easy to determine [$\alpha$/Fe] from low-resolution GC 
spectra. Mg is the element of choice because of the strong Mg$b$ triplet and MgH bandhead 
in the optical. Trager \etal~(2000) described a method for estimating [Mg/Fe] using models 
by Worthey (1994) and corrections from Trippico \& Bell (1995).  However, the widespread 
estimation of this quantity in extragalactic GCs did not occur until Thomas, Maraston \& 
Bender (2003) published the first models to explicitly include the effects of 
$\alpha$-enhancement.

In practice, the interpretation of the observations using these models has been 
complicated. The GCs in the metal-rich subpopulations in a variety of massive galaxies 
have been found to have [$\alpha$/Fe] ranging from 0 to $\sim +0.3$ (e.g., Kuntschner 
\etal~2002; Beasley \etal~2005) or higher (Puzia \etal~2005 find $\sim +0.45$ but with 
large scatter). By contrast, the metal-poor GCs frequently appear to have [$\alpha$/Fe] 
$\sim 0$ or even lower (e.g., Olsen \etal~2004; Pierce \etal~2005), although the 
convergence of SSP model lines at low metallicities increases the errors on these 
determinations of [$\alpha$/Fe]. If [$\alpha$/Fe] for metal-poor GCs were indeed this low, 
it would be quite surprising. In the simple view of nucleosynthesis outlined above, 
supersolar [$\alpha$/Fe] results from star formation over ``short" ($< 1$ Gyr) timescales 
while solar or subsolar [$\alpha$/Fe] indicates very extended star formation histories. 
Thus we expect the metal-poor GCs to have high [$\alpha$/Fe].

Galactic GCs from both subpopulations have [$\alpha$/Fe] $\sim +0.3$ (e.g., Carney 1996) or 
perhaps a little lower for some metal-rich bulge GCs. These values are quite closely 
reproduced by the popular Thomas \etal~(2003) SSP models for the Galactic GC data of Puzia 
\etal~(2002), although the results from applications to extragalactic GCs are mixed (as 
described above). Current stellar libraries and the Worthey (1994) fitting functions utilize 
few stars with metallicities in the range of metal-poor GCs, and SSP models may be more 
uncertain in this regime.

Schiavon (2006) gave a good summary of the issues involved in creating non-solar [$\alpha$/Fe] 
models. These include (i) the importance of employing the proper isochrones and luminosity 
functions, and (ii) the need to correct the index predictions of the fitting functions. Item 
(ii) is usually carried out differentially, using stellar models for stars in representative 
parts of the CMD. A preliminary attack on this problem by Trippico \& Bell (1995), using just 
three stars, produced broadly similar results to the more thorough treatment of Korn 
\etal~(2005). Korn \etal~also discussed in detail the elemental sensitivities of the 
individual Lick indices.

A potentially confusing difference among models is the treatment of 
[$\alpha$/Fe] vs.~[Fe/H]. Thomas \etal~(2003) calculated their models at 
fixed total metallicity ([$Z$/H]) and produced supersolar [$\alpha$/Fe] by 
\emph{lowering} [Fe/H], since the dominant component of $Z$ is the 
$\alpha$-element O. By contrast, Schiavon (2006) calculated models at 
fixed [Fe/H]. Schiavon's approach has the benefit of a direct link to the 
measurable quantity [Fe/H].

These SSP uncertainties, when combined with the wide Lick index 
bandpasses, which always admit contributions from other elements in 
addition to the targeted feature, make most current estimates of 
[$\alpha$/Fe] untrustworthy. We suggest intensive study of nearby GC 
systems (e.g., M31), where potentially more accurate results from 
high-resolution integrated spectroscopy of GCs can be directly compared to 
low-resolution spectra.

\subsection{Abundance Anomalies}

In addition to the $\alpha$-elements, GCs show a variety of abundance anomalies with respect 
to the solar mixture. This is most clear in the Galaxy, where detailed study of individual 
stars in GCs is possible (see the Annual Review of Gratton, Sneden, \& Carretta 2004). In 
extragalactic GCs, the most obvious anomaly is CN-enhancement. This was first reported in M31 
GCs by Burstein \etal~(1984), and, at least in this galaxy, appears to be due to an excess of 
N above even the levels in Galactic GCs (see \S 6.1), which themselves are enhanced in N over 
the solar mixture. Trager (2004) has argued that a similar CN anomaly (presumably due to N) is 
present in many GC systems, including NGC 3115 (Kuntschner \etal~2002), the old GCs in NGC 
3610 (Strader \etal~2003a), and the Fornax dSph (Strader \etal~2003b). 
It is also present in GCs in the gE NGC 1407 (Cenarro \etal~2006). We must consider very high N abundances to be a 
generic feature of the early chemical evolution of GC systems (and perhaps their host 
galaxies). Interestingly, of the two confirmed metal-rich intermediate-age GCs in NGC 3610, 
one shows the CN anomaly, and one does not.

\section{Near-IR Imaging}

A new tool for studying GCs emerged during the last few years with the advent of 
large-format infrared detectors. Kissler-Patig (2000) pointed out that it should be 
possible to use near-IR (NIR) photometry of GCs to break the age-metallicity degeneracy 
inherent in optical colors. The $V-K$ color of an old GC is a measure of the temperature 
of its red giant branch. This temperature is strongly dependent on metallicity but has 
little sensitivity to age. The prospect of large NIR imaging surveys is exciting because 
it offers the potential to accurately measure individual ages and mean metallicities for 
large numbers of GCs in galaxies out to the distance of the Virgo cluster and beyond.

Kissler-Patig and collaborators, primarily using VLT/ISAAC, have taken the lead in such NIR 
studies. Puzia \etal~(2002) reported a comparison in $VIK$ between the GC systems of the 
group S0 NGC 3115 and the Virgo E NGC 4365. The GC system of NGC 3115 appeared bimodal in 
color-color space, while the color distribution of GCs in NGC 4365 looked largely unimodal, 
with a formal peak at supersolar metallicity and intermediate age ($\sim 2-5$ Gyr). 
However, the presence of a large subpopulation of intermediate-age GCs was confusing, as 
the central light of the galaxy itself is uniformly old (Davies \etal~2001). How could a 
large number of GCs form without an accompanying field star component? This puzzling result 
precipitated a flurry of activity on the GC system of NGC 4365. Spectroscopy of a subset of 
the candidate intermediate-age GCs seemed to confirm their young ages, while demonstrating 
that other young candidates (selected from their optical/NIR colors) were in fact old 
(Larsen \etal~2003).

However, follow-up spectroscopy with somewhat higher S/N found no evidence for any 
intermediate-age clusters. Moreover, when considered in combination with new 
\emph{HST}/ACS imaging, these data seemed to point instead to three subpopulations of 
\emph{old} GCs, with a very centrally-concentrated, intermediate-metallicity subpopulation 
filling in the gap between the normal metal-poor and metal-rich GCs (Brodie \etal~2005). A 
similar intermediate-metallicity subpopulation has been discovered by Beasley \etal~(2006) 
in the gE NGC 5128. This interpretation is consistent with wider-field $K$-band photometry 
obtained by Larsen, Brodie, \& Strader (2005). These authors discussed all of the 
available data and the details of the error estimates, and showed evidence for systematic 
errors in optical-NIR SSP models. Nonetheless, based on a comparison of \emph{HST}/NICMOS 
$H$-band photometry of 70 GCs in NGC 4365 and 11 GCs NGC 1399 with SSP models, Kundu 
\etal~(2005) concluded that a large number of intermediate-age (2--8 Gyr) GCs with 
metallicities up to [Fe/H] = +0.4 are present in NGC 4365.  Young metal-poor GCs would 
also be inferred from the distribution of their data on SSP model grids. Since metal-poor 
GCs are expected to be old under nearly all formation scenarios, this further suggests 
that at least some of the offset of metal-rich GCs to young ages may be due to systematic 
errors in the SSP models (Larsen \etal~2005). Clearly, the last word has not yet been 
spoken on this intriguing galaxy.

Hempel \etal~(2003) studied the early-type galaxies NGC 5846 and NGC 7192 with NIR 
photometry, reporting that the former galaxy hosts a large population of intermediate-age 
GCs. Followup $U$-band photometry in NGC 5846 and NGC 4365 (Hempel \& Kissler-Patig 2004) 
was argued to support the earlier results, although the number of GCs detected was low. 
Follow-up spectroscopy does not appear to support the presence of a large subpopulation of 
intermediate-age GCs in NGC 5846 (Puzia \etal~2005). The fraction of GCs with ages 
formally $< 8$ Gyr in the observed sample of luminous GCs is 15-20\% (T.~Puzia, private 
communication).

Goudfrooij \etal~(2001a) presented the best evidence to date of a significant 
population of intermediate-age GCs in any post-merger galaxy. They used a combination 
of \emph{HST} and ground-based optical and NIR photometry of GCs in NGC 1316, a 
merger remnant E in the Fornax cluster. They found metal-rich GCs that were 
substantially brighter than normal GCs at that distance and deduced photometric ages 
of $\sim 3$ Gyr for these bright red objects. These photometric ages were confirmed 
by spectroscopy of a small number of the brightest clusters (Goudfrooij \etal~2001b). 
However, even in this galaxy, the strongest pieces of evidence for an 
intermediate-age subpopulation are (i) the presence of unusually luminous GCs and 
(ii) the power-law (rather than log-normal) GC luminosity function of the metal-rich GCs 
(see also Goudfrooij \etal~2004).

Improvements in NIR SSP models will brighten the prospects for future NIR studies and help 
this field to reach its full potential. In addition, the launch of the \emph{James Webb 
Space Telescope (JWST)} in the next decade will allow the collection of high-quality 
$K$-band photometry for many GC systems.

\section{Globular Cluster--Field Star Connections}

Other than the Milky Way, the GC systems of only two massive galaxies have been 
observed in detail, in the sense that a significant number of their GCs have been 
observed with high S/N spectroscopy and we can study their field stars directly. M31, 
our sister spiral galaxy in the Local Group, and NGC 5128, one of the nearest massive 
E galaxies, are both sufficiently well-studied that they can 
elucidate the role of GCs in tracing star formation in their parent galaxies.

\subsection{M31}

Though the total mass of M31 is similar to that of the Galaxy (Evans \& Wilkinson 
2000), it has a GC system $\sim 3$ times as large, with $\sim 400-450$ GCs (van den 
Bergh 1999; Barmby \& Huchra 2001). A comprehensive catalog of photometric and 
spectroscopic data for M31 GCs was assembled by Barmby \etal~(2000) and used to show 
that, broadly speaking, the M31 GC system is quite similar to that of the Galaxy; M31 
has an extended halo subpopulation and a more centrally-concentrated rotating 
bulge/disk subpopulation.

The kinematics of the M31 GC system may be complex. Perrett \etal~(2002) found that the 
metal-rich GCs have a velocity dispersion similar to the bulge ($\sim 150$ km/s) and a 
rotational velocity of 160 km/s. Surprisingly, they also found a large rotational velocity 
for the \emph{metal-poor} GCs, $\sim 130$ km/s. By contrast, the metal-poor GCs in the 
Galaxy show little evidence for rotation. Visually, the spatial distribution of the Perrett 
\etal~metal-poor GC sample appears to be the superposition of a centrally-concentrated 
spherical population and a relatively thin disk component. Indeed, Morrison \etal~(2004) 
used the Perrett \etal~velocities to argue that M31 possesses a rapidly rotating ``thin 
disk" of old metal-poor GCs. Beasley \etal~(2004) presented high-S/N spectra of a small 
number of these putative old metal-poor GCs and showed that they have spectra more similar 
to young ($< 1$ Gyr), approximately solar metallicity objects. Their luminosities suggest 
stellar masses comparable to massive Galactic open clusters or low-mass GCs. The presence of 
young GCs has also been claimed by Burstein \etal~(2004) and Puzia, Perrett, \& Bridges 
(2005) (both based in part on data from Beasley \etal~2004) and by Fusi-Pecci \etal~(2005). 
Cohen \etal~(2005) used Keck NIR/AO imaging to show that four of these very young clusters 
are actually ``asterisms"---either chance groupings of bright stars or a few brighter stars 
superimposed upon sparse open clusters. A more detailed examination of archival \emph{HST} 
imaging indicates that perhaps half of the candidate young clusters are real; previous 
\emph{HST}/WFPC2 had confirmed the existence of four such clusters (Williams \& Hodge 2001). 
A high-resolution imaging survey of the disk will be needed to assess the true number of 
young GCs. In any case, if these interloping objects (asterisms and/or YMCs) are removed 
from the metal-poor GC candidate list, much of the rotational signature observed by Perrett 
\etal~would be erased. The apparent lack of an old metal-poor thin disk of GCs is important, 
since the existence of this disk would rule out any significant merger having taken place in 
the last $\sim 10$ Gyr. As discussed below, such a merger is the working theory for the 
existence of intermediate-age stars in M31's halo.

Due to selection effects, including reddening in the inner disk and the paucity of GC 
candidates at large radii, the overall fraction of metal-poor to metal-rich GCs is 
still poorly constrained in M31. From their kinematic and metallicity study, Perrett 
\etal~(2002) found that only $\sim 25$\% of a sample of $\sim 300$ GCs belong to the 
metal-rich (bulge) peak, compared to one third in the Galaxy. If, as discussed above, 
many of the candidate metal-poor GCs in M31 are interlopers, then the ratio of 
metal-poor to metal-rich GCs in M31 could be quite consistent with that found for the 
Galaxy (2:1). There are several ongoing studies of the M31 GC system with MMT/Hectospec 
that should provide a clearer picture of this situation.

Burstein \etal~(1984) found evidence for Balmer line anomalies in M31 GCs compared to GCs 
in the Milky Way. Their preferred explanation was that the M31 GCs might be younger than 
their Milky Way counterparts. Brodie \& Huchra (1991) and Beasley \etal~(2004) compared 
integrated-light spectra of Milky Way and M31 GCs at the same metallicity and found no 
evidence for enhanced Balmer lines in M31 GCs. Beasley \etal~(2005) extended these results 
by fitting the Lick indices measured from their sample GCs to stellar population models. 
They employed the multi-index $\chi^2$ minimization method developed by Proctor, Forbes, 
\& Beasley (2004) to derive ages and metallicities for these clusters. In addition to the 
normal old metal-poor and metal-rich GC subpopulations, they found a group of six GCs with 
intermediate metallicities and ages of 3--6 Gyr. A larger number of GCs with similar ages 
were also reported by Puzia \etal~(2005). This group of GCs forms a coherent chemical and 
kinematic group of objects in M31, consistent with the accretion of an SMC-type galaxy 
within the last several Gyr (in the Galaxy, there are several several comparable GCs 
associated with the Sgr and CMaj dSphs; \S 10). Ashman \& Bird (1993) had previously 
analyzed M31 GC kinematics and found evidence for $\sim 7$ distinct groups of GCs. The 
clustering analysis of Perrett \etal~(2003) found $\sim 8$ such groups, but Monte Carlo 
simulations indicated that statistically half or more of these should be chance groupings. 
In an addition to the rapidly growing zoo of stellar clusters, Huxor \etal~(2005) have 
identified 3 metal-poor objects in the halo of M31 that have luminosities typical of GCs, 
but sizes from 25--35 pc. They suggest that these could be stripped nuclei of dwarfs, 
analogous to $\omega$ Cen in the Galaxy (Majewski \etal~2000).

Brown \etal~(2004) obtained extremely deep \emph{HST}/ACS photometry of the M31 GC 379-312 
([Fe/H $\sim -0.6$) and through isochrone fitting derived a formal age of 
$10^{+2.5}_{-1}$ Gyr, perhaps $\sim 1$ Gyr younger than Galactic metal-rich GCs. Rich 
\etal~(2005) carried out photometry deeper than the horizontal branch for a sample of 
12 
M31 GCs. They found the classic second parameter effect, but with an offset that 
might be explained if the GCs were $\sim 1-2$ Gyr younger than their Galactic 
counterparts. These results hint at an interesting relative age difference between 
the two GC systems, but hard evidence will be difficult to come by. It would require 
a UV/optical space telescope with an aperture larger than \emph{HST} to derive turnoff ages 
for significant numbers of M31 GCs.

Burstein \etal~(1984) also discovered CN enhancements in M31 GCs (with respect to 
Galactic GCs). This result was later confirmed by Brodie \& Huchra (1991), who
pointed out that CH at 4300 \AA\ did not appear to be anomalous, and argued that the 
observed CN enhancement might be due to an overabundance of N. Beasley 
\etal~(2004) did not see the enhancement clearly in the Lick CN$_{2}$ index, which 
measures the weaker 4215 \AA\ bandhead, but it is obvious in a different index 
sensitive to the 3883 \AA\ transition. That N is the culprit in CN variations was 
confirmed directly by near-UV spectra of Galactic and M31 GCs around the NH band at 
3360 \AA\ (Burstein \etal~2004). Observations in this region of the spectrum have been limited by the low fluxes of GCs 
and the low UV efficiencies of 
telescopes, instruments, and detectors. As a result, the SSP 
modeling effort has proceeded slowly, and quantitative [N/Fe] values do not yet 
exist. Thomas, Maraston, \& Bender (2003) found that [N/Fe] $\sim +0.8$ was required 
to match CN in some Galactic GCs; in M31 GCs it must be even higher. Since NH (in the 
UV) is dominated by light from stars around the main sequence turnoff, the 
N enhancements are probably primordial in origin and cannot be attributed to non-canonical 
mixing processes on the red giant branch.

M31's stellar halo, studied in fields from $\sim 10-30$ projected kpc, appears 
predominately metal-rich ([m/H] $\sim -0.5$), a full 1 dex higher than the Milky Way 
(e.g., Mould \& Kristian 1986; Durrell \etal~1994). In one M31 halo field, Brown 
\etal~(2003) found that $\sim 30$\% of the stars were of intermediate-age, again unlike 
the Milky Way. Given the large disk and bulge of M31, it has been suggested that these 
results may be explained by the contamination of putative halo samples by disk/bulge stars 
(Worthey \etal~2005). However, this interpretation does not appear to be consistent with 
recent kinematic studies of a subset of the intermediate-age stars (Rich \etal~2006). In 
any case, M31 also appears to have a metal-poor projected $r^{-2}$ stellar halo, like the 
Galaxy (Guhathakurta \etal~2005; Kalirai \etal~2006). Numerical simulations of M31 
indicate that a relatively minor accretion event (involving an LMC-sized galaxy) would 
suffice to redistribute the observed number of metal-rich, intermediate-aged stars from 
the disk into the halo (Font \etal~2006).

Before discussing how field star properties relate to GCs, it is helpful to define 
the quantity $T^{n}$, which is the Zepf \& Ashman (1993) $T$ parameter normalized to the 
mass of a particular stellar population (vs.~the original definition, which 
normalizes a single GC subpopulation to the total stellar mass in the galaxy). For 
spheroids with few (or no) metal-poor stars, $T^{n}_{red}$ is the same as $T_{red}$, 
modulo uncertainties in the adopted $M/L$.

The Galaxy has $\sim 100$ metal-poor GCs and a halo mass of $\sim 10^9 M_{\odot}$ (Carney, Latham, \& 
Laird 1990), resulting in a metal-poor $T^{n}_{blue} \sim 100$. By contrast, the corresponding 
metal-rich value is $T^{n}_{red} \sim 5$ (there are only $\sim 50$ metal-rich GCs and bulge mass 
$\sim 10^{10} M_{\odot}$; Kent 1992). Thus the metal-poor GCs formed with an efficiency twenty times 
higher than the metal-rich GCs with respect to field stars.

In the Galaxy, the association of the metal-poor GCs with halo stars is clear; they 
have similar peak metallicities, spatial distributions, and kinematics
(though the metallicity \emph{dispersion} of the field stars is much larger than that
of the GCs). The situation 
for the metal-rich GCs is somewhat less well-defined but, in general, their spatial 
distribution and kinematics are more similar to the bulge than the thick 
disk (see \S 7.2). However, the metallicity distribution of the metal-rich GCs does 
\emph{not} 
match that of the bulge. Fulbright \etal~(2006) used Keck/HIRES spectra to 
recalibrate the metallicity distribution of red clump stars in Baade's Window (Sadler 
\etal~1996). They found a peak at [Fe/H] $\sim -0.15$, roughly $0.3-0.4$ dex higher than 
the peak of the metal-rich GCs. Baade's Window is located beyond one bulge effective 
radius, so if the bulge has a negative metallicity gradient the mean offset could be 
even larger.

The metallicity of the M31 bulge is similar to the bulge of the Galaxy (Sarajedini \& 
Jablonka 2005). The GC color distributions in the Milky Way and M31 show no 
significant differences, suggesting a similar metallicity offset in M31 between the 
metal-rich GCs and the bulge (as borne out by spectroscopy of individual GCs in 
M31). The metallicity of the metal-poor component of the stellar halo in M31 is not 
well-constrained, so no direct comparison can yet be made with the GCs.

\subsection{NGC 5128}

As one of the nearest (albeit disturbed) giant ellipticals ($\sim 4$ Mpc), NGC 5128 holds 
special significance for stellar population studies. W.~and G.~Harris and coworkers 
have taken the lead in studying the relationship between GCs and field stars in this galaxy.

Their \emph{HST}/WFPC2 photometry of red giants in fields from $\sim 8$ to 31 kpc (projected radius)
give metallicity distributions that peak at [m/H] $\sim -0.2$ to $\sim -0.4$, with a 
slight negative metallicity gradient with galactocentric radius (Durrell, Harris \& 
Pritchet 
2001; Harris \& Harris 2002). Such metallicities are consistent with those of the 
metal-rich GCs, modulo uncertainties in the relative zeropoints of the metallicity 
scales. However, the presence of a radial metallicity gradient in the field stars 
complicates any comparison with the flat metallicity distribution of the GCs. At what 
radius should the two be compared? The field star metallicity distribution is 
reasonably well-fit with an accreting box model like the one needed to solve the 
Galactic G-dwarf problem (Lynden-Bell 1975). Rejkuba \etal~(2005) used an \emph{HST}/ACS 
outer halo (38 kpc) 
pointing to derive a metallicity distribution for NGC 5128 field stars very similar 
to that of Harris \etal, 
except that they found a larger number of metal-rich stars ($\ga$ solar). These stars
were missed by Harris \etal~because of large bolometric corrections in $V-I$ at high metallicities.
An age analysis of the 
Rejkuba \etal~field, based primarily on the locations of the ABG bump and the red 
clump, yielded a mean age of $\sim 8$ Gyr. However, the data are also consistent with 
a two-component model, comprising an older base population plus a small percentage of 
5-7 Gyr stars. Based on the number of AGB stars (Soria \etal~1996) and Mira variables 
(Rejkuba \etal~2003), a figure of $\sim 10$\% has been suggested for the fraction of the
total stellar mass in intermediate-age stars. This is consistent with the 7\% fraction of 
intermediate-age GCs in the galaxy (see below).

NGC 5128 possesses only a relatively minor tail of metal-poor ``halo" stars, even in the 
outermost pointing, but (as in M31) it is unclear whether this offers a constraint on the 
presence or absence of a Milky Way-like metal-poor projected $\sim r^{-2}$ halo.  Harris \& 
Harris (2002) inferred from the paucity of metal-poor stars that $T^{n}_{blue}$ must be much 
larger than $T^{n}_{red}$. They argued for a scenario in which the metal-poor GCs formed at the 
beginning of starbursts in small potential wells. These could be efficiently evacuated by 
supernova feedback before many accompanying metal-poor stars had a chance to form. In principle,
other truncation mechanisms (including reionization) could produce a similar result (see 
\S 11 and 12).

There are good theoretical reasons to believe that GCs might form in the initial stages of 
major star forming events. This may be when the gas pressure is highest, favoring the creation 
of massive compact clusters (e.g., Elmegreen \& Efremov 1997; Ashman \& Zepf 2001). By 
contrast, the metal-rich GCs are expected to form during a disspational starburst in a more 
fully-assembled potential well. Here feedback is less effective, and many field stars form. 
The scenario in which GCs form in the initial stages of a starburst could also qualitatively 
reproduce a metallicity offset between metal-rich GCs and field stars. Such an offset is 
observed between the zeropoints of the metallicity--galaxy luminosity relations for metal-rich 
GCs and field stars (Forbes \etal~1997). However, a difficulty with any such comparison is the 
existence of radial metallicity gradients in galaxies, since metal-rich GCs show no such 
gradient. If both metal-poor GC and field star formation are simultaneously truncated at high 
redshift, their metallicity distributions may show no offset, as is the case in the Galaxy.

GCs in NGC 5128 appear to fall on the structural fundamental plane defined by 
Galactic and M31 GCs (Harris \etal~2002), although the mean cluster ellipticity in 
the current (small) sample is higher than that of Galactic GCs and more similar to
the distribution in the old GCs of the LMC. Harris \etal~also identified 
six massive GCs that had evidence of extratidal light. This could suggest that these 
objects are stripped nuclei of dEs, or simply that they are normal GCs, but that King 
models provide a suboptimal fit to their surface brightness profiles. Wide-field 
photometry in the Washington system by Harris \etal~(2004), although complicated by the 
proximity of NGC 5128 and its low galactic latitude, revealed a total population of 
$\sim 1000$ GCs ($S_N = 1.4$), evenly divided between metal-poor and metal-rich GCs. Peng 
\etal~(2004) presented a large photometric and spectroscopic survey of more than 200 
GCs in NGC 5128. 20--25 of the GCs had high enough S/N to determine spectroscopic 
ages and metallicities, and they found a wide range of ages for the metal-rich GCs, 
with a mean age of $\sim 5$ Gyr. Through direct comparison with integrated spectra of 
Galactic GCs from Cohen \etal (1998), Peng \etal~found some evidence for [Mg/Fe] $> 
0$, although perhaps not as high as in the Galaxy.

Beasley \etal~(2005) estimated ages and metallicities for $\sim 200$ GCs in NGC 5128 from deep 
2dF spectroscopy (with some overlap with the Peng \etal~sample). Using the multi-line fitting 
method of Proctor \etal~(2004), they deduced that the bulk of the GCs in this galaxy can be 
assigned to one of three \emph{old} subpopulations: the usual metal-poor and metal-rich 
subpopulations, plus an intermediate-metallicity subpopulation reminiscent of one discovered 
in the Virgo gE NGC 4365 (Brodie \etal~2005). Evidence for trimodality is also presented in 
Woodley, Harris, \& Harris (2005). In Figure 9, histograms of the GC $B-I$ and metallicity 
distributions show the trimodality in the metallicity distribution. As is readily apparent, 
the nonlinear color--metallicity relationship combines the metal-poor and 
intermediate-metallicity peaks into a single peak in the $B-I$ histogram. The presence of old 
intermediate-metallicity GCs is unusual among massive Es. If these GCs have an accompanying 
subpopulation of field stars, the field star metallicity distribution could be a combination 
(albeit not necessarily well-mixed) of intermediate-metallicity and metal-rich stars. However, 
this would be difficult to detect in present data. In addition to the old subpopulations, 
about 7\% of the GCs in the Beasley \etal~sample appear to be metal-rich with intermediate 
ages (2--8 Gyr), contrary to previous results. Candidate intermediate-age GCs were specially 
targeted, so this fraction is probably an upper limit.

The kinematics of the GCs are similar to those observed in the Galaxy: significant rotation is seen 
in the metal-rich GCs (as defined in Peng \etal~2004), but only weak rotation is apparent in the 
metal-poor GCs. This result is contrary to that expected to result from a disk-disk merger. In a 
merger remnant, the inner parts of the galaxy should display very little rotation because angular 
momentum is transferred to the outer regions. Of course, the remnant properties will depend on the 
details of the merger, and a statistical sample of Es that might plausibly have formed in recent 
major mergers does not yet exist. See \S 7 for addition discussion of GC kinematics.

\subsection{Other Galaxies}

An alternative approach to studying the relationship between GCs and field stars has 
been taken by Forte, Faifer, \& Geisler (2005), who assume that the field star 
subpopulations follow the radial and color distributions of the GC subpopulations
and use galaxy surface photometry to decompose the integrated light into metal-poor 
and metal-rich constituents. For NGC 1399, they find $T^{n}_{blue} \sim 25$ and 
$T^{n}_{red} \sim 4$. This metal-rich value is virtually identical to that of the 
Galaxy and, as before, the metal-poor value is much larger than the metal-rich one.  
While this approach is potentially a powerful tool for studying field stars in 
galaxies with unresolved stellar populations, substantial future work will be 
required to establish the veracity of the key underlying assumption: that the radial 
and metallicity distributions of GCs and their associated field stars are the same.

\section{Kinematics}

\subsection{Ellipticals}

Since GCs are more extended than the stellar light of galaxies, they are useful kinematic 
tracers of the dark matter halo at large galactocentric radii. GC kinematics also encode 
detailed signatures of the assembly histories of galaxies; the implications of such 
observations are just beginning to be understood. Large kinematic samples of GCs ($> 50$) 
have been observed in only four gEs: NGC 1399, M87, NGC 4472, and NGC 5128 (discussed 
above). Small samples have been studied in several other galaxies, but the numbers are too 
few to allow the subpopulations to be effectively separated. Aside from noting that there 
appear to be signatures of strong rotation in the S0s NGC 3115 (Kuntschner et al. 2002) and 
NGC 524 (Beasley et al. 2004), we will confine our detailed discussion to these four 
well-studied galaxies.

The largest sample of GC velocities at present exists for NGC 1399, the central gE in 
the Fornax cluster. Richtler \etal~(2004) have analyzed kinematics for 468 GCs at 
projected radii of $2-9$\arcmin~($\sim 11-49$ kpc). The velocity dispersions for the 
metal-poor and metal-rich GCs are $\sim 290$ km/s and $\sim 255$ km/s, respectively, 
and do not appear to vary with galactocentric radius. They report a slight tangential 
bias for the metal-poor GCs, but both subpopulations are generally consistent with 
isotropic orbits. The larger dispersion for the metal-poor GCs is consistent with 
their extended spatial distribution. Neither subpopulation shows significant 
rotation, although a weak signature is observed in metal-poor GCs beyond 6\arcmin. 

It has been suggested from simulations (Bekki \etal~2003) that NGC 1399 might have stripped 
the outermost GCs from the nearby E NGC 1404, leaving it with an anomalously low $S_N$ (NGC 
1399 has a high $S_N$ of $\sim 5-6$, typical of cluster Es). Bassino \etal~(2006) have found 
that three other Fornax Es---NGC 1374, NGC 1379, and NGC 1387---all have $S_N$ lower than 
typical of cluster Es, and also suggest that they may have suffered GC loss to NGC 1399. New 
GC velocities for NGC 1399, out to $\sim 80$ kpc, show an asymmetric velocity distribution of 
the metal-poor GCs at these large radii (Schuberth \etal~2006). This could be interpreted as a 
signature of GCs that have been stripped from nearby Es but have not yet reached equilibrium 
in the NGC 1399 potential.

NGC 4472 has been studied by Zepf \etal~(2000) and C{\^ o}t{\' e} \etal~(2003), with 
$\sim 140$ and $\sim 260$ velocities, respectively, measured over projected radii of 
$\sim 1.5-8$\arcmin\ ($\sim 7-40$ kpc). The velocity dispersions of the metal-poor 
and metal-rich GCs are $\sim 340$ km/s and $\sim 265$ km/s, respectively. The 
metal-poor GCs rotate around the galaxy's minor axis with a velocity of $\sim 90-100$ 
km/s; the metal-rich GCs show weak evidence for counter-rotation around the same 
axis. There is a hint that the outermost (beyond $\sim 25$ kpc) metal-poor GCs may 
rotate about the major axis, but this conclusion relies on few GCs. The orbits of 
both subpopulations are consistent with isotropy to the radial limit of the data.

The kinematics of GCs in M87 are rather different from those in either NGC 4472 or NGC 1399. From a 
sample of $\sim 280$ velocities, C{\^ o}t{\' e} \etal~(2001) find that \emph{both} the metal-poor and 
metal-rich GCs rotate around the minor axis at $\sim 160-170$ km/s, with respective velocity 
dispersions of $\sim 365$ and $\sim 395$ km/s, when averaged over the whole system. The metal-poor GCs 
actually appear to rotate around the \emph{major} axis within $\sim 16$ kpc but switch to minor axis 
rotation only at larger radii. C{\^ o}t{\' e} \etal~suggest that this change is due to the increasing 
gravitational dominance of the cluster potential beyond $\sim 18$ kpc. The GC system as a whole 
appears quite isotropic, but the data are consistent with a tangential bias for the metal-poor GCs and 
an opposing radial bias for the metal-rich GCs; these biases are poorly constrained in the present 
sample.

The results for four well-studied gEs (including NGC 5128, discussed in the previous 
section) present a heterogeneous picture. Although no two galaxies seem to be kinematically 
similar thus far, we will briefly discuss the extent to which the accumulated information 
can constrain galaxy formation. K.~Bekki and collaborators have used numerical simulations 
to study the properties of GCs in merging galaxies. Bekki \etal~(2002) found that, in 
disk-disk major mergers, newly formed metal-rich GCs are centrally concentrated and extended 
along the major axis of the remnant. Angular momentum transferred to the metal-poor GCs 
causes them to rotate and extend their spatial distribution. These results are strongly 
dependent on the numerical details of the simulations, and could change if, for example, a 
different prescription for star formation was used. Simulations of dissipationless major 
mergers for galaxies with both a disk and bulge extended these results (Bekki \etal~2005). 
However, the initial conditions (velocity dispersion and anisotropy) of the GC system 
strongly affect the outcome of such simulations. Pre-existing metal-poor and metal-rich GCs 
acquire significant amounts of rotation beyond $\sim 10$ kpc, regardless of the orbital 
properties of the merging galaxies. Mergers with larger mass ratios leave relatively 
spherical GC distributions with less rotation. With such a generic prediction of rotation, 
it is unclear how galaxies like NGC 1399, which shows no rotation in either subpopulation, 
could have been created. It may be that the merging histories of gEs leave a variety of 
complex rotational signatures, and that much of the observed differences are due to natural 
variations in remnant properties and projection effects.

It is instructive to consider the results of cosmological simulations of the assembly of 
dark matter and stellar halos of massive galaxies.  Even though the GCs are more extended 
than the stellar light of the galaxy, they are still more centrally concentrated than the 
dark matter, so are expected to have a lower velocity dispersion than the dark matter at 
fixed radii (if the anisotropy of the dark matter and GCs are similar). While providing some 
constraints, present data sets are still too small to fully determine the anisotropy ($\beta 
= 1-v^2_{\theta}/v^2_{r}$) of the GC system (e.g., Wu \& Tremaine 2006). A wide 
range of numerical simulations 
suggest that a general relation holds between $\beta$ and galactocentric radius for both 
halo tracers and the dark matter:  $\beta \sim 0$ (isotropy) in the inner parts, with radial 
anisotropy increasing outward to $\beta \sim 0.5$ at the half-mass radius of the tracer 
population (Hansen \& Moore 2004; Dekel \etal~2005; Diemand, Madau, \& Moore 2005; Abadi 
\etal~2005). Very large samples of GCs ($\sim 500-1000$) will be needed to test these 
predictions.

Studies of lower-mass Es are also important, and several groups are pursuing hybrid 
approaches, using both GCs and planetary nebulae (PNe) to constrain the potential. Romanowsky 
\etal~(2003) suggested, on the basis of PNe kinematics in three ``normal" Es (NGC 821, NGC 
3379, and NGC 4494), that these galaxies lacked dark matter. A more conventional 
interpretation is that the observed PNe originate in stars ejected during mergers and are on 
highly radial orbits (Dekel \etal~2005). GCs (especially those in the metal-poor 
subpopulation) are expected to have lower orbital anisotropy than PNe, due both to their 
extended spatial distribution and the increased probability of destruction for GCs on radial 
orbits. Two recent studies of GC kinematics in NGC 3379 suggest that the galaxy possesses a 
``normal" $\Lambda$CDM dark halo and are more consistent with the Dekel \etal~interpretation 
(Pierce \etal~2006; Bergond \etal~2006). Thus, GCs may be the best tracers of the mass 
distribution of Es at large radii.

\subsection{Disk Galaxies}

The best recent review of the kinematics of Galactic GCs is found in Harris (2001). The 
metal-rich GCs rotate at $\sim 90-150$ km/s, and the rotation rises out to a radius of 
$\sim 8$ kpc, beyond which there are few metal-rich GCs.  This increasing rotational 
velocity may represent a transition from bulge to thick disk, though the maximum 
velocity of $\sim 150$ km/s is still less than expected for a typical thick disk in a 
massive spiral. There is little net rotation over the metal-poor subpopulation as a whole. 
Moreover, no individual radial bin of metal-poor GCs rotates, but somewhat 
surprisingly, a strong signature of prograde rotation (140 km/s) is seen in the most 
metal-poor GCs ([Fe/H] $< -1.85$). This effect is dominated by very metal-poor GCs in 
the inner halo. The degree of rotational support is similar in this very metal-poor 
group ($v/\sigma \sim 1.2$) to the total subpopulation of metal-rich GCs ($\sim 1.3$). 

M31 has been discussed above: the metal-rich GCs rotate with $v/\sigma \sim 1.1$, and the 
kinematic state of the metal-poor GCs remains unclear. If the identification of the disk 
objects as young clusters (no matter what the mass) is secure, then the old metal-poor 
subpopulation is probably pressure-supported.

The situation in M33 is also uncertain, since the ages of many \emph{HST}-confirmed star 
clusters are unknown. Chandar \etal~(2002) found that the old metal-poor GCs have a 
velocity dispersion of $\sim 80$ km/s and are not rotating. They suggested that there is 
a small population of old inner GCs with disk-like kinematics, but an adequate test of 
this must await a significant increase in the kinematic sample.

Olsen \etal~(2004) have suggested that GCs in several Sculptor Group spirals have 
kinematics consistent with the rotating HI gas disks in these galaxies (in 
NGC 253, the kinematics are consistent with asymmetric drift from an initial cold 
rotating disk). Candidates were selected from outside the bright optical disk, so 
contamination from open clusters in the disk is unlikely to have occurred. Few GCs 
were observed in each galaxy, and the low systemic velocities of the galaxies 
inevitably biased the GC selection; those with low velocities cannot be efficiently 
distinguished from foreground stars. Nonetheless, even with this bias, the 
line-of-sight velocity dispersions of the GCs are low: from $\sim 35-75$ km/s out to 
$\sim 10$ kpc or so. Given the mixed evidence for old disk GCs presented thus far, 
this preliminary result is worth following up.

The disk galaxies present a somewhat cleaner kinematic picture than the Es, though the 
number of galaxies studied is still small. If indeed the metal-poor GCs in M31 turn out 
to be pressure-supported, then this appears to be a common feature of disk galaxies. In 
the two galaxies with bulges (M31 and the Galaxy) the metal-rich GCs have substantial 
rotational support. At the very least, these results can serve as valuable starting 
conditions for simulations of major mergers.

\section{Luminosity Functions}

Excellent reviews of the GC luminosity function (GCLF), including its use as a distance 
indicator, are given in Harris (2001) and Richtler (2003). The NED D database (2006 release)
gives a comprehensive compilation of GCLF distances.
Here we summarize some basic facts and discuss the \emph{evolution} of the GCLF.

In many massive galaxies studied to date, the GCLF can be well-fit by a Gaussian or $t_5$ 
distribution; typical parameters for the normal distribution in Es are peak $M_V \sim -7.4$, 
with $\sigma \sim 1.4$. In spirals, the peak is similar, but $\sigma \sim 1.2$ may be a more 
accurate value for the dispersion. The peak luminosity is a convolution of the peaks for the 
two GC subpopulations; line-blanketing effects (especially in bluer bands) cause the 
metal-poor GC peak to be slightly brighter than the metal-rich peak, so the exact location of 
the peak depends on the color distribution of the GC system (Ashman, Conti, \& Zepf 1995). The 
standard peak or turnover of this distribution corresponds to a mass of $\sim 2 \times 10^5 
M_{\odot}$. In absolute luminosity or mass space, this distribution is a broken power law with 
a bright-end slope of $\sim -1.8$ and a relatively flat faint-end slope of $\sim -0.2$ 
(McLaughlin 1994). In some massive galaxies there appears to be a departure below a power law 
for the most massive GCs (e.g., Burkert \& Smith 2000).  The expectation of significant 
dynamical evolution at the faint end of the GCLF has led to suggestions that only the bright 
half of the GC mass function (GCMF) represents the initial mass spectrum of the GC system. The 
faint half would then represent the end product of a Hubble time's worth of dynamical 
destruction. Recent observational and theoretical work on this problem offers some supporting 
evidence, but also raises some interesting questions.

In the Galaxy, the total sample of GCs is relatively small, especially when divided into 
subpopulations. This limits our ability to draw general conclusions from this best-studied 
galaxy. Harris (2001) provided a good overview of the current situation. Within the 8 kpc 
radius that contains both metal-poor and metal-rich GCs, the subpopulations have quite similar 
LFs. Considering only the metal-poor GCs, the turnover of the GCLF becomes brighter by nearly 
$\sim 0.5$ mag out to 8--9 kpc, then decreases to its initial value. The inner radial trend is 
the \emph{opposite} of naive expectations for dynamical evolution, which should preferentially 
disrupt low-mass GCs as shocks accelerate mass loss through two-body relaxation. There are too 
few metal-rich GCs to study the radial behavior of the GCLF in detail. The GCs in the far 
outer halo (beyond $\sim 50$ kpc or so) are all quite faint ($M_V < -6$), except for the 
anomalous GC NGC 2419. This cluster has $M_V = -9.6$ and may be the stripped nucleus of a 
dwarf galaxy (van den Bergh \& Mackey 2004). Of course, this level of detail is not expected 
to be visible in external galaxies, whose GC systems are generally seen in projection about an 
unknown axis. This is an important limitation in interpreting the results of the theoretical 
simulations discussed below.

Fall \& Zhang (2001) presented a semi-analytic study of GC system evolution in a Milky 
Way-like galaxy. The mass-loss rate by two-body relation is taken to depend only on the 
details of the cluster's orbit, and not on the mass or concentration of the GC. This appears 
to be consistent with results from more detailed N-body simulations (e.g., Baumgardt \& 
Makino 2003). Fall \& Zhang found that a wide range of initial GCMFs, including single and 
broken power laws, eventually evolved to turnover masses similar to those observed. They 
found that the evolution of the turnover mass was rapid in the first Gyr or two, but that 
subsequent evolution was slow. The turnover is expected to change substantially with 
galactocentric radius, due to the preferential destruction of low-mass GCs towards the 
galaxy center. This can only be avoided if the outer GCs display strong radial anisotropy. 
However, the assumptions made by Fall \& Zhang need to be considered when applying their 
results to real data. For instance, they used a static spherical potential, while a live 
and/or nonradial potential could increase phase mixing and erase some radial signatures of 
evolution.

Vesperini (2000; 2001) modeled the evolution of GCLFs of two initial forms, log-normal and 
power-law functions. The evolution of individual GCs was determined using analytic 
formulae derived from the simulations of Vesperini \& Heggie (1997). To summarize their 
results: GCLFs that are initially log-normal provide a much better fit to the observed 
data. Dynamical evolution can indeed carve away the low-mass end of a power-law GCLF to 
produce a log-normal function, but the resulting turnover masses are generally small ($\la 
10^5 M_{\odot}$), and vary significantly with galactocentric radius and from galaxy to 
galaxy. All three predictions for the evolution of an initial power-law function are 
inconsistent with observations. However, a GCLF that is initially log-normal suffers 
little evolution in shape or turnover, and that only in the first few Gyr. If constant 
initial $S_N$ is assumed, GC destruction leads to a $S_N$--galaxy luminosity relation of 
the form $S_N \propto L^{0.67}$. Vesperini \etal~(2003) explored M87 in detail, and 
(reminiscent of Fall \& Zhang 2001) found that the observed log-normal GCLF, which has a
constant turnover with radius (see also Harris \etal~1998),
could only evolve from a power-law initial GCLF if there were 
(unobserved) strong radial anisotropy in the GC kinematics. Of course, the details of the 
simulations are quite important. For example, Vesperini \& Zepf (2003) showed that a 
power-law GCLF and a concentration--mass relation for individual GCs can result in a 
log-normal GCLF with little radial variation in turnover mass. This is because 
low-concentration GCs are preferentially destroyed at all radii. This is consistent with 
the study of Smith \& Burkert (2002), who found that the slope of the low-mass part of the 
GCLF in the Galaxy depends upon GC concentration. Another relevant piece of evidence is 
the similarity in the GCLF turnover among spirals, Es, and even dEs (see \S 10.1). This 
suggests either initial GCLFs close to log-normal (such that little dynamical evolution 
occurs), or that the dominant destruction processes are not specific to particular galaxy 
types--for example, disk shocking in spirals.

The sole galaxy with a significant subpopulation of intermediate-age GCs and evidence 
for dynamical evolution is the merger remnant NGC 1316. \emph{HST}/ACS observations
by Goudfrooij \etal~(2004) have clearly demonstrated dynamical evolution of its GC
system. When they divided the red ($1.03 \le V-I \le 1.40$) GCs into two equal radial
bins, they found that the outer (beyond $\sim 9.4$ kpc) GCs have the power law LF
seen for many systems of YMCs in merging and starbursting galaxies. By contrast, the 
inner GCs show a LF turnover characteristic of \emph{old} GC systems. As
Goudfrooij \etal~argued, these observations would appear to provide the conclusive  
link between YMCs in mergers and old GCs in present-day Es.

However, the difficulty with this interpretation is that the observed location of the 
metal-rich peak of the inner GCs in NGC 1316 is at $M_V \sim -6$. If the metal-rich GCs were 
formed in the merger, are $\sim 3$ Gyr old, and have solar metallicity (consistent with 
spectroscopic results; Goudfrooij \etal~2001), Maraston (2005) models with either a Salpeter 
or Kroupa IMF predict that age-fading to 12 Gyr will result in a peak at $M_V \sim -4.6$. 
Old metal-rich GCs are observed to have a GCLF turnover at $M_V \sim -7.2$ (e.g., Larsen 
\etal~2001). Thus $\sim 2.6$ mag of additional evolution of the GCLF turnover ($\sim 1.2$ 
mag of dynamical evolution and $\sim 1.4$ mag of age fading) would be required to turn the 
new metal-rich GC subpopulation of NGC 1316 into that of a normal E. The 1.2 mag of 
dynamical evolution needed is far beyond that predicted by even the most ``optimistic" 
models for GC destruction. For example, Fall \& Zhang (2001) models for GC system evolution, 
assuming a power-law initial GCMF, predict $< 0.2$ mag of evolution in the GCLF peak from 3 
to 12 Gyr using the same Maraston (2005) models. This is consistent with the finding of 
Whitmore \etal~(2002) that little evolution in the GCLF peak is expected after the first 
1.5--2 Gyr, as age-fading balances GC destruction. The implication is that in a Hubble time, 
the metal-rich GC system of NGC 1316 may not look like that of a normal E galaxy. Fall \& 
Zhang (2001) models, designed to study GC destruction in a Milky Way-like galaxy, may even 
have limited applicability to NGC 1316. Other models (e.g., those of Vesperini discussed 
above) predict less evolution, so the expected difference between the suitably evolved young 
GCs in NGC 1316 and the old metal-rich GCs in local Es could be even greater. A caveat here 
is that the observed GCLF peak could be the convolution of a brighter, more evolved GC 
population in the innermost regions with a less evolved population in the outer parts of the 
bin. Such a convolution would tend to lessen the difference between the observed and 
expected turnover luminosity.

Thus, while there is some evidence for the paradigm of the evolution of a power-law LF to a 
log-normal LF, there are still unresolved issues. The installation of WFC3 on a potential 
future \emph{HST} servicing mission would allow much more efficient and accurate age-dating 
of YMCs through its wide-field $U$-band imaging capability. This would also make it possible 
to investigate the mass function as a function of age in more detail, and hence directly 
address questions of dynamical evolution.

\section{Sizes}

The present-day appearance of a GC is a complicated convolution of the initial conditions of its 
formation with subsequent internal and external dynamical effects. The surface brightness
distributions of GCs are reasonably well-fit by single-mass King (1966) models. These form
a dimensionless one-parameter family as a function of the concentration $c=r_t/r_0$ (where $r_t$ is the tidal radius
and $r_0$ is the scale radius). In dimensionless form, the half-light radius ($r_h$--``size") is
a monotonic function of $c$, and is the only one of these three radii that is relatively unaffected by dynamical evolution
(Spitzer 1987; Meylan \& Heggie 1997) and can serve as a probe of GC formation conditions.

A correlation between $r_h$ and galactocentric radius ($R$) for Galactic GCs was discovered 
by van den Bergh \etal~(1991). This could not be explained by dynamical evolution of the GC 
system, since diffuse inner GCs might be expected to be destroyed, but compact distant GCs 
should (if they existed) have remained intact. Thus, this result represents strong evidence 
for some degree of in situ formation of Galactic GCs.

\emph{HST}/WFPC2 imaging of NGC 3115 and M87 revealed that metal-poor GCs are $\sim 20$\% 
larger than metal-rich GCs (Kundu \& Whitmore 1998; Kundu \etal~1999). This was 
confirmed statistically in many early-type galaxies by Larsen \etal~(2001) and Kundu 
\& Whitmore (2001a). Explanations offered for this result have included:

(i) It represents an intrinsic formation difference, e.g., the metal-rich GCs 
formed in a higher-pressure environment.

(ii) It is a results of projection effects. Since the metal-rich GC spatial distribution is
 more centrally concentrated than that of the metal-poor GCs, within some given projected 
radius the metal-rich GCs will tend to lie at smaller $R$. If there is a strong correlation 
between size and $R$ (as in the Galaxy), the metal-rich GCs will appear smaller on average than 
the metal-poor GCs (Larsen \& Brodie 2003). This model predicts that size differences will be 
largest in the inner parts of galaxies and disappear in the outer regions.

(iii) It is a natural outcome of assuming metal-poor and metal-rich GCs have the same 
half-mass radii. Since the brightest stars in metal-rich GCs are more massive
than in metal-poor GCs, mass segregation leads to a more compact distribution and a smaller
half-light radius (Jord{\'a}n 2004). In this model there should be little change in the relative sizes with galactocentric 
distance.

Option (i) is not testable at present, so should be left as a fallback only if the other 
possibilities can be eliminated. Regarding (ii), Larsen \& Brodie (2003) showed that the 
$r_h$--$R$ correlation in the Galaxy could explain all of the observed size differences 
between the metal-poor and metal-rich GCs. However, in order to explain the $\sim 20$\% 
size difference in external galaxies, steep $r_h$--$R$ relations would be required, and 
the radial distribution of the GCs would need to have a central core like a King profile 
(this appears to be consistent with observations). Model (iii) has a number of critical 
assumptions upon which its conclusions depend, including identical GC ages and initial 
mass functions. Small changes in either of these parameters (e.g., an age difference of 
$\sim 2$ Gyr between the metal-poor and metal-rich GCs) could erase most of the expected 
size difference. Jord{\'a}n (2004) also used equilibrium King-Michie models to represent 
the GCs; full N-body modeling is a desirable next step. In Figure 10, taken from 
Jord{\'a}n (2004), we show the sizes of GCs in M87 together with a best-fit model of type 
(iii).

Several recent observational results have provided important new constraints. Jord{\'a}n 
\etal~(2005) studied GC sizes in 67 early-type galaxies with a wide range of luminosity 
from the ACS Virgo Cluster Survey (C{\^ o}t{\' e} \etal~2004). For bright metal-poor GCs 
they found a significant but rather shallow relationship between $r_h$ and projected $R$ 
(normalized to the effective radius of the galaxy). In log space the value of the slope is 
$0.07$, compared to $\sim 0.30$ for a similar sample of Galactic GCs. They did not list 
the fits for individual galaxies, and there are clearly variations, but the bulk of the 
galaxies do not appear to have $r_h$--$R$ relations as steep as observed in the Galaxy, so 
projection effects on the GC subpopulation sizes should be small, and option (ii) is not 
favored. They also found that GCs in bluer/fainter host galaxies tend to be slightly 
larger. Despite these variations, $r_h$ is still relatively constant among galaxies. Thus, 
with their observed correlations between $r_h$ and galaxy properties, they were able to 
calibrate $r_h$ as a distance indicator, as suggested by Kundu \& Whitmore (2001a).

Wide-field \emph{HST}/ACS data for NGC 4594 (covering $\sim 6$\arcmin $\times 10$\arcmin) do 
not appear consistent with option (iii), however. Spitler \etal~(2006) found that the ratio 
of metal-poor to metal-rich GC sizes declines steeply and steadily from $\sim 1.25$ in the 
center to $\sim 1$ at the edge of the complete observations. Thus it appears that, at least 
in NGC 4594, projection effects account for most of the observed size differences.

It seems clear from these results that both projection and segregation mechanisms can play a 
role in determining the sizes observed for extragalactic GCs. Since each is sensitive to the 
physical conditions of the GC system, galaxies will need to be studied on an individual basis 
to determine which effects are important. Particularly valuable would be high-resolution, 
wide-field imaging of GC systems, like the \emph{HST}/ACS mosaic of NGC 4594 discussed above.

\section{Dwarf Galaxies}

It has been known for a long time that many dwarf galaxies have GC systems (e.g., Fornax dSph; 
Shapley 1939). In the Galaxy formation model of Searle \& Zinn (1978) and in many subsequent 
studies of GC formation, it was envisioned that metal-poor GCs form in protogalactic 
dwarf-sized clumps (e.g., Harris \& Pudritz 1994; Forbes \etal~1997; C{\^ o}t{\' e} 
\etal~1998; Beasley \etal~2002). The dwarf satellites around massive galaxies like the Milky 
Way can then be interpreted as the remnants of a large initial population of such objects, 
most of which merged into the forming protogalaxy. If this process happened at high redshift, 
most of the fragments could still have been be gaseous and thus have formed stars and/or 
contributed gas as they merged. At lower redshift the process could be primarily 
dissipationless, as envisioned by C{\^ o}t{\' e} \etal~and seen in action through the 
present-day accretion of the Sgr dSph. What can GCs tell us about this process?

The faint end of the galaxy luminosity function is uniquely accessible in the Local Group. 
Forbes \etal~(2000) provided a good census of GCs in Local Group dwarfs, which has changed 
only marginally since that time. The candidate GC in the dIrr Aquarius is apparently a yellow 
supergiant (D.~Forbes, private communication), so the lowest-luminosity Local Group galaxies 
with confirmed GCs are the Fornax ($M_V = -13.1$) and Sgr ($M_V = -13.9$) dSphs, each of which 
has at least five GCs. The recently discovered CMaj dSph (Martin \etal~2004) appears to have 
at least four GCs whose properties are distinct from the bulk of the Galactic GC system 
(Forbes, Strader, \& Brodie 2004). At least two of the GCs in each of Sgr and CMaj are of 
intermediate age and metallicity. The LMC has a subpopulation of old metal-poor GCs and a 
famous ``age gap" between the old GCs and a subpopulation of intermediate-age GCs ($\sim 3$ 
Gyr). It also hosts a number of younger clusters, some of which, like GCs, might be massive 
enough to survive a Hubble time (e.g., Searle, Wilkinson, \& Bagnuolo 1980; van den 
Bergh 1994). The SMC has only one old GC but a more continuous distribution of massive 
clusters to younger ages (Mighell \etal~1998). Together these results suggest star formation 
histories that were at least moderately bursty (e.g., Layden \& Sarajedini 2000). Lower-mass 
Galactic dwarfs (e.g., Leo I; $M_V \sim -12$) do not have GCs. Less is known about the GC 
systems of similar-mass M31 dwarfs. Grebel \etal~(2000) suggest a candidate GC in And I ($M_V 
\sim -12$), but this GC could be a contaminant from M31. Outside the Local Group, Sharina 
\etal~(2003) spectroscopically confirmed a GC in the M81 dSph DDO 78, which has a mass 
intermediate between Fornax and And I. Karachentsev \etal~(2000) identified candidate GCs in a 
number of other M81 dwarfs, but these have not yet been confirmed.

As discussed in Strader \etal~(2005), these observations put important constraints on the 
minimum mass of halos within which metal-poor GCs could form. Fornax and Sgr have total 
masses of $\ga 10^8 M_{\odot}$ (Walcher \etal~2003; Law \etal~2005) and the total mass 
estimate for And I, assuming it has a similar M/L ratio, is a few $\times 10^7 M_{\odot}$. 
This suggests that, at least in a relatively low-density group environment, GCs formed in 
halos with minimum masses of $\sim 10^7-10^8 M_{\odot}$. Whether GCs typically form 
in groups of GCs or alone is unknown. Fornax and Sgr each have several GCs, and the dIrr NGC 
6822 ($M_V \sim -15.2$) has up to three old GCs, though two of these are located far from 
the main body of the galaxy (Cohen \& Blakeslee 1998; Hwang \etal~2005). The dIrr WLM ($M_V 
\sim -14.5$) has only one old GC, which is metal-poor (Hodge \etal~1999). A caveat to these 
arguments is that in some models (e.g., Kravtsov, Gnedin \& Klypin 2004), present-day dwarf 
satellites have undergone significant stripping of dark matter, and may have been much more 
massive initially ($\ga 10^9-10^{10} M_{\odot}$). However, detailed comparisons between 
observed velocity dispersion profiles and numerical simulations suggest little mass loss due 
to tidal stripping for well-studied Galactic dSphs (Read \etal~2006). It is important to 
realize that differences in baryonic mass loss (e.g., due to stellar feedback; Dekel \& Silk 
1986) may modify the amount of stellar mass in galaxies of similar halo mass. GC kinematics 
(see \S 10.3) offer one of the best routes to directly determine the total masses of dwarfs 
outside the Local Group.

\subsection{Specific Frequencies and Luminosity Functions}

GC systems were discovered around 11 Virgo dwarf ellipticals (dEs\footnote{In deference to 
common usage, here we utilize the term ``dwarf elliptical" for early-type galaxies with 
low luminosities ($M_B \la -18$). The structural parameters of many of these galaxies 
differ from those of ``classical" Es (e.g., Kormendy 1985; Kormendy 1987) and may well 
suggest a different formation history. They are sometimes called spheroidal (Sph) 
galaxies. A small number of galaxies in this luminosity range have structural parameters
consistent with power-law Es (see \S 4).}) using ground-based (CFHT) imaging (Durrell 
\etal~1996). The $S_N$ of these galaxies is relatively high, $\sim 3-8$, and the 
GC systems are very centrally concentrated: most GCs are within $< 30$\arcsec\ (2 kpc) at 
the distance of Virgo, while a typical dE has a half-light radius $\la 1$ kpc. Strader 
\etal~(2006) obtained radial distributions that were consistent with these earlier 
results, except for a few bright dEs ($M_V \la -18$) where the outermost GCs were found at 
7--9 kpc. This is near the limit of the radial coverage of \emph{HST}/ACS at the distance 
of Virgo, so it is possible that GCs may be found at even larger radii.

Miller \etal~(1998) used \emph{HST}/WFPC2 snapshot imaging to explore the specific 
frequencies of a large sample of dEs in the Virgo and Fornax clusters, including galaxies 
with luminosities as faint as $M_B \sim -13$. They found a dichotomy between nucleated 
(dE,N) and non-nucleated (dE,noN) galaxies. dE,noN dwarfs appeared to have low 
$S_N$ values ($\sim 3$), independent of galaxy luminosity. dE,N galaxies had higher $S_N$ 
and showed an inverse correlation between $S_N$ and luminosity. It has been suggested that 
dE galaxies may have originated as dIrrs or low-mass disk galaxies (e.g., Moore, Lake, \& 
Katz 1998; see discussion below). Miller \etal~argued that few dE galaxies could have formed 
in this manner, as the $S_N$ values even for dE,noN galaxies are larger than expected from 
age-fading such hosts.

Strader \etal~(2006) revisited these findings in an \emph{HST}/ACS study of Virgo Es which 
included 37 dEs. 32 of these have structural parameters consistent with ``true" dEs; the other 
five appear to be faint power-law Es (Kormendy \etal~2006 and private communication). The ACS 
images offered superior areal coverage and depth compared to those available to Miller \etal, but 
the Strader \etal~sample spanned a smaller luminosity range: $-15 \la M_B \la -18$. It was not 
possible to investigate the differences between dE,N and dE,noN galaxies, since many of the 
galaxies previously classified as dE,noNs either have faint nuclei or are power-law Es. There may 
be few true dE,noNs with $M_B \la -15$ in Virgo. Strader \etal~found no strong correlation between 
galaxy luminosity and $S_N$ for either dEs or faint power-law Es. The faintest galaxies might have 
larger $S_N$, but the effect is not strong. The lack of a $S_N-L$ trend could be due to the more 
restricted luminosity range of galaxies studied by Strader \etal compared to Miller \etal.

Interestingly, a bimodal distribution of $S_N$ values was discernible in their sample. As 
shown in Figure 11, more than half of the galaxies were found to have $S_N \sim 1$, while 
the $S_N$ values of the remainder ranged from 3 to 10, with a median at $\sim 5$. This 
difference spans the observed luminosity range and does not correlate with either the 
presence of a nucleus or the color distribution of the GCs. A natural interpretation of 
the $S_N$ differences is that they reflect multiple formation channels for dEs in Virgo. 
Mechanisms for forming dEs include ``harassment", the cumulative effect of many high-speed 
galaxy encounters (Moore \etal~1996), stripping or age-fading of low-mass disks (Kormendy 
1985), or processes similar to those responsible for the formation of more massive Es 
(this many be most applicable to the faint power-law Es). It is possible that the high 
$S_N$ galaxies (the Fornax dSph with $S_N$ $\sim 29$ is an extreme example) simply 
represent those in which stellar feedback during the first major starburst was very 
effective (e.g., Dekel \& Silk 1986). However, if this is the case, a signature should be 
apparent in the GC color distributions. In particular, the presence of metal-rich GCs (see 
below) and a continuous (rather than bimodal) distribution of $S_N$ might be expected. 
Certainly, feedback will be increasingly important with decreasing galaxy mass, and would 
provide a simple explanation for a relation between $S_N$ and luminosity, should one be 
confirmed. Photometric, structural, and kinematic studies of these same dEs will be needed 
to discriminate among the many possible explanations.

These same studies have also afforded the ability to study the GCLF in dEs. The power law 
slope ($\sim -1.8$) measured by Durrell \etal~(1996) for the massive end of the GCLF in 
their Virgo dEs is the same as that found in normal Es. However, they measured a GCLF 
turnover ($M_V \sim -7.0$) that is fainter by $\sim 0.4-0.5$ mag than that typical for 
massive galaxies. This result may have been influenced by the difficulty of rejecting 
contaminants in ground-based data. Strader \etal~(2006) constructed a composite of the 37 
dEs in their \emph{HST} sample, using the outer parts of the images to correct for 
background contamination. In contrast to Durrell \etal, they found that the dE GCLF peak 
occurs at the same value as in the massive gEs in their sample (M87, NGC 4472, NGC 4649). 
This comparison was made in $z$, where there is little dependence of cluster $M/L$ on 
metallicity over the relevant range, so the differences in GC color distributions between 
gEs and dEs should not affect the GCLF comparisons.

That the GCLF peaks for the dwarfs and the giants match so well in the Strader \etal~study 
is perhaps puzzling. The theoretical expectation is that in low-mass galaxies dynamical 
friction will act to deplete the GC system, preferentially destroying massive GCs. Such 
GCs will spiral into the center in less than a Hubble time, forming or contributing to a 
nucleus. Lotz \etal~(2001) performed a semi-analytic study of this phenomenon, and found 
that dynamical friction is expected to produce more luminous nuclei than observed. Strader 
\etal~(2006) found that the luminosities of a subset of dE nuclei are consistent with 
formation through dynamical friction, but that the majority appear to be formed by a 
separate mechanism. This point, together with the similarity of the GCLF turnovers, 
implies that dynamical friction has not had a substantial effect on the GC systems of dEs. 
Lotz \etal~(2001) suggested several explanations for the lack of observable consequences of 
dynamical friction, including extended dark matter halos around dEs, or tidal torquing of 
GCs (this latter explanation was also proposed for the Fornax dSph by Oh \etal~2000). 
Goerdt \etal~(2006) have used numerical simulations to show that the dynamical friction 
timescale in Fornax is longer than a Hubble time if its dark matter halo has a core 
(instead of the cusp generically predicted in galaxy formation in $\Lambda$CDM). Kinematic 
studies of GCs in dEs, such as those of Beasley \etal~(2005), are also beginning to build 
a better understanding of the halo potentials of dwarf galaxies.

So far we have included little discussion of dIrrs. Their GC systems are
quite difficult to study. Indeed, the contrast in our understanding of dEs and
dIrrs is analogous to the information gap between Es and spirals. The
relatively small GC systems of dIrrs, their ongoing star formation, and the
resulting inhomogeneity of the background are serious observational challenges.
A \emph{HST}/WFPC2 study of 11 Virgo and Fornax dIrrs by Seth \etal~(2004) found
typical $S_N$ values of $\sim 2$, but uncovered two galaxies with much higher  
$S_N$. Stellar M/L values are low in typical dIrrs, which suggests that the
$S_N$ values will become much higher after nominal age-fading of the dIrr, as
might be expected in transformation to a dE. However, many of the detected
objects are unlikely to be old GCs. In the Local Group, the Magellanic Clouds 
and NGC 6822 each have a population of massive intermediate-age GCs (in addition
to a small number of old GCs) that
reflect the extended star formation histories of these galaxies. The same may
be true of cluster dIrrs. Spectroscopy will probably be needed to determine the
present fraction of intermediate-age GCs, which is crucial for isolating the  
differences in the formation histories of the various classes of dwarf
galaxies. The combination of optical and NIR photometry, once the SSP models
have been sufficiently calibrated for old GCs, offers a promising future tool  
for identifying younger clusters.

\subsection{Color Distributions}

The classical view of dwarf galaxy GCs is that they are uniformly metal-poor.
This is supported by a ``combined" metallicity distribution of old GCs in Local
Group dwarfs (Minniti \etal~1996), which peaks at [Fe/H] $\sim -1.8$, 0.3 dex  
more metal-poor than that of the Galactic metal-poor GC subpopulation.

We now know that reality is more complicated. As noted above,   
the Sgr dSph (and perhaps the CMa dSph) have two GCs of intermediate
metallicity and age. An \emph{HST}/WFPC2 study of the color distributions of dEs in  
Virgo and Fornax (Lotz \etal~2004) revealed a rather wide spread in color,
consistent with the presence of metal-rich GCs. However, it was not possible to
distinguish subpopulations in their small GC samples.

Sharina \etal~(2005) have published a study of GCs in a large sample of dSphs and dIrrs 
using a heterogeneous set of \emph{HST}/WFPC2 images. The metal-poor GC peak appears to be 
$\sim 0.1$ redder in $V-I$ than the peak found by Lotz, Miller, \& Ferguson (2004) for their 
sample of dEs. There is no ready explanation for this finding, which is inconsistent with 
previous work. It may be that a photometric zero-point offset is to blame, as suggested by 
Sharina \etal~themselves. The GC color distributions of both galaxy types have a tail to 
redder values. This may reflect a small subpopulation of metal-rich GCs, or could 
be due to contamination by background objects. There may also be a few metal-rich GCs in the 
dIrrs studied by Seth \etal~(2004).

The existence of metal-rich GCs in dEs has been shown conclusively by
Peng \etal~(2006) and Strader \etal~(2006). A large fraction of Virgo dEs, down to
quite faint magnitudes ($M_B \sim -15$), were found to have bimodal color
distributions, analogous to those observed in massive Es. The slope of the
metal-rich GC color--galaxy luminosity relation is not well-constrained at
these low luminosities due to the small number of GCs associated with each
galaxy. The new data points are consistent with either: (i) a linear
extrapolation to lower magnitudes and bluer colors from the region of massive
galaxies, or (ii) a slight flattening at the low mass end of the relation. It
is reasonable then to ask whether these metal-rich GCs (or at least a subset)
could have intermediate ages, like the two younger GCs discovered in the Sgr
dSph. Beasley \etal~(2005) obtained high-quality spectra for three
metal-rich GCs in the Virgo dE VCC 1087. Their old ages are consistent with those
of the metal-poor GCs within the errors. Although spectra of similar quality
for a large sample of dEs clearly would be desirable, the results to date suggest
that there is no obvious dichotomy in the color distributions of dEs and
massive Es.

Using data from Peng \etal~(2006), Forbes (2005) has pointed out a possible link between GC 
bimodality in dEs and the galaxy color bimodality observed in large surveys (e.g., Bell \etal~2004) 
below a critical mass of $\sim 3 \times 10^{10} M_{\odot}$. Above this mass, nearly all galaxies 
in the Virgo Cluster Survey have bimodal GC systems; below this mass, an increasing fraction of 
galaxies have unimodal color distributions. One interpretation of this phenomenon is that the 
critical mass represents a transition from ``cold", smooth accretion of gas into halos
below the critical mass to ``hot" 
accretion of gas that shocks at the virial radius and is unable to form stars (e.g., Dekel \& 
Birnboim 2006).

\subsection{Kinematics}

Kinematic studies of GCs in dwarfs are challenging, principally because of the small GC 
systems and lack of luminous GCs. Puzia \etal~(2000) found that the velocity dispersion of 
seven GCs in the luminous dE NGC 3115 DW1 suggested a relatively high $M/L_{V} \sim 
22\pm13$. This could suggest the presence of dark matter, or that its parent S0 NGC 3115 is 
stripping the outermost GCs.

In the Virgo dE VCC 1087, the GCs rotate at $\sim 100$ km/s around the minor axis 
(Beasley\etal~2005). The sample of twelve GCs is dominated by metal-poor GCs, although it 
includes three GCs whose colors and spectroscopic metallicities are consistent with a 
metal-rich subpopulation (such subpopulations appear to be common in Virgo dEs; see \S 2 and 
10.2). Its GC system has the largest rotational support of any galaxy studied to date, with 
$v/\sigma \sim 3.6$, typical of a disk. This makes VCC 1087 a prime candidate for a dE that 
evolved from a disky dIrr. We note in passing that, although the LMC is often considered to 
have a rotating disk population of old metal-poor GCs (e.g., Schommer \etal~1992), van den 
Bergh (2004) has argued that the current data do not strongly discriminate between disk and 
halo kinematics for these GCs.

\section{Globular Cluster Formation}

\subsection{Classical Scenarios}

In \S 2.1 we described the three principal scenarios that have been suggested as explanations for 
GC bimodality: major disk-disk mergers, in situ formation through multiphase dissipational 
collapse, and dissipationless accretion. How do these models account for the other observed 
properties of GC systems? Here we discuss the arguments made in the literature for and against 
these scenarios, as well as additional constraints from newer data described in this article.

\subsubsection{Major Mergers}

As noted in \S 2.1, the observation of young massive star clusters in many merger remnants 
throughout the 1990s gave a significant boost to the major disk-disk merger model for GC 
bimodality (Ashman \& Zepf 1992). While some of these objects definitely have masses and sizes 
that should allow them to evolve into old GCs (e.g., Maraston \etal~2004; Larsen, Brodie, \& 
Hunter 2004), others may have abnormal IMFs that preclude their long-term survival (e.g., 
McCrady, Gilbert, \& Graham 2003; Smith \& Gallagher 2001; Brodie \etal~1998), though 
important uncertainties in dynamical mass estimates due to mass segregation remain (McCrady, 
Graham, \& Vacca 2005). In a broader context, despite the fact that YMCs and GCs are 
remarkably similar in many respects, it remains unclear whether (after a Hubble time of 
evolution) young GC \emph{systems} will have properties consistent with those of old GC 
systems in local galaxies. The issue here is that observations of intermediate-age GCs may be 
at odds with the expected signatures of a simple dynamical evolution scenario (see \S 8).

Even before color bimodality had been observed, several authors used GCs to constrain
the feasibility of the disk-disk merger picture for forming Es. Harris (1981) and van
den Bergh (1984) noted that typical Es had more populous GC systems than spirals.
Massive disk galaxies have $S_N \sim 1$; Es have $S_N \sim 2-5$ depending on
environment, with even higher values for brightest cluster galaxies (BCGs) like M87.
This is often termed the ``$S_N$ problem". Schweizer (1987) explicitly addressed this
concern by suggesting that many new GCs might be formed in the merger.
As has been pointed out by multiple authors, this will only raise the
$S_N$ if GCs form with a higher efficiency relative to field stars than they did in  
the protogalactic era. Since GC formation efficiency appears to increase with
star formation rate, $S_N$ may only increase if the star formation rate
in a present-day merger is higher than it was when the GCs in spirals were
originally formed.

Several other problems with the major merger model were pointed out in Forbes \etal~(1997). 
For example, they showed that there is a correlation between $S_N$ and the fraction of 
metal-poor GCs, such that the highest $S_N$ galaxies also have the highest proportion of 
metal-poor GCs. However, the major merger scenario predicts the opposite behavior: the 
mechanism to increase the $S_N$ of spirals is the formation of new metal-rich GCs in the 
merger; this should result in larger metal-rich GC subpopulations in more massive Es. Ashman 
\& Zepf (1998) gave a candid analysis of the then current situation on the merger front and 
suggested that the gEs that dominated the high $S_N$ end of the Forbes \etal~relation could be 
expected to have augmented their metal-poor GC population by the accretion of lower-mass 
galaxies (see below) during their complex formation histories.  Moreover, they pointed out 
that the $S_N$ values in the literature were likely to be very uncertain because so few 
galaxies had been scrutinized with high-quality wide-field imaging. As discussed in \S 3, more 
recent work confirms that $S_N$ values for Es tend to come down with improved observations. In 
general, however, there are still fewer metal-rich than metal-poor GCs in present-day Es, and 
this remains in conflict with the major merger prediction. Rhode \etal~(2005; see \S 3.1) 
considered the $S_N$ (or $T$) values of the individual subpopulations and concluded that 
massive cluster Es cannot have been formed from mergers of local spirals, although some 
lower-mass field Es could still have formed in this manner. Harris (2001) reached essentially 
the same conclusion from an analysis of the required gas content and GC formation 
efficiencies.

Another constraint on the major merger model can be found in the metal-poor GC 
metallicity--galaxy luminosity correlation (Strader \etal~2004a). In the mean, the 
metal-poor GC subpopulations of spirals have lower metallicities than those of massive Es. 
This seems to be a strong argument against the major merger scenario. However, this 
conclusion does not take into account the expected effects of biasing---see \S 11.2. It is 
notable that even some low-mass dEs have bimodal GC color distributions that follow the 
same peak relations as massive galaxies (\S 10.2), even though these galaxies have 
presumably not suffered a major merger. So, even if major disk-disk mergers were a viable 
route to producing bimodality in some cases, they could not be the sole process in 
operation. In addition, the ages of metal-rich GCs in Es (see \S 4) imply a formation epoch $z \ga 2$.
This restricts most putative major mergers to higher redshifts.

As discussed in \S 2.1, the Forbes \etal~(1997) multi-phase collapse scenario arose as a 
response to issues with the merger model. There has been little observational evidence to 
date against the Forbes \etal~scenario, but, to a considerable extent, this is because it 
made few specific predictions of observable quantities. Its usefulness was as a framework 
within which to consider alternative explanations for GC color bimodality, and it 
identified aspects of the picture still under consideration, e.g., the need to truncate GC 
metal-poor formation at high redshift. More recent scenarios described below in \S 11.2 
are generally consistent with this broad framework.

We emphasize that the arguments presented here against the major merger scenario apply 
principally to the formation of massive Es from \emph{present-day} spirals with relatively small 
bulges. Current GC observations are consistent with dissipational formation of Es at relatively 
high redshift ($z \ga 2$)---including major mergers, as long as the disk progenitors have higher 
$S_N$ than spirals in the local universe. Subsequent dissipationless merging could then form the 
most massive gEs, under the constraint of ``biased" merging discussed in \S 11.2.

\subsubsection{Dissipationless Accretion}

The accretion scenario of C{\^ o}t{\' e} \etal~(1998) was explicitly designed to be 
consistent with hierarchical structure formation. It assumes a protogalactic GC 
metallicity--galaxy mass relation produced through a dissipational process at high 
redshift. The GC systems of present-day galaxies are envisaged to have formed through 
subsequent dissipationless merging. Since in this scenario the intrinsic GC metallicities 
of massive protogalaxies are quite high, such galaxies must accrete large numbers of 
metal-poor GCs from dwarf galaxies to produce bimodality. C{\^ o}t{\' e} \etal~(2002) used 
Monte Carlo simulations to show that the acquisition of the necessary numbers of 
low-metallicity GCs required the low-mass end of the the protogalactic mass function to 
have a very steep slope ($\sim -2$). However, even with such a steep mass function, 
Ashman, Walker, \& Zepf (2006) found that, when they ran simulations similar to those of 
C{\^ o}t{\' e} \etal, color distributions like those observed in massive galaxies occurred 
in only a small fraction ($\sim 5$\%) of their simulations. Another potential problem is 
that the accreted dwarfs would be expected to contribute many metal-poor field stars that 
are not observed, unless the dwarfs are primarily gaseous (e.g., Hilker 1998).

The fact that metal-poor GCs in dwarfs have much lower metallicities than those in   
massive Es (by 0.5--0.6 dex) would, at first sight, seem to be direct evidence against the
accretion scenario. This argument was made in Strader \etal~(2004a). However, this
line of reasoning does not account for the effects of biased structure formation,    
which may be the key to properly understanding the implications of the metal-poor GC 
metallicity--galaxy mass relation.

As already emphasized, in the light of our current understanding of hierarchical
galaxy assembly, all galaxy formation scenarios must be accretion/merger scenarios at
some level.  The major merger and the accretion models (as published) both provided
an important focus for theoretical discussion and observational effort by making
fairly explicit predictions against which the observations could be compared. The 
preponderance of new evidence now suggests that, while elements of each remain
viable, the details are pointing us in new directions (see below). 

\subsection{Hierarchical Merging and Biasing: Recent Scenarios}

Beasley \etal~(2002) explored GC bimodality in a cosmological context using the semi-analytic 
galaxy formation model of Cole \etal~(2000), and this work contained elements of all three 
classic scenarios. While largely phenomenological, it makes the most specific predictions of 
any model proposed, and because the scenario is in the context of a full model of cosmological 
structure formation, it implicitly accounts for many of the issues discussed below (e.g., 
biasing). Metal-poor GCs were assumed to form in the early universe in gas disks in low-mass 
dark matter halos. As in Forbes \etal~(1997), Beasley \etal~found it necessary to invoke the 
truncation of metal-poor GC formation at high redshift (in this case, $z > 5$) in order to 
produce bimodality. The metal-rich GCs were generally formed during gas-rich mergers. Their 
predictions for the metal-rich subpopulation included: a correlation between GC metallicity 
and galaxy luminosity, significant age and metallicity substructure, and decreasing mean ages 
and metallicities in lower-density environments.

Following suggestions from Santos (2003), Strader \etal~(2005) and Rhode \etal~(2005) proposed 
hierarchical scenarios for GC formation intended to account for the metal-poor GC metallicity--galaxy 
mass relation and the correlation of metal-poor $S_N$ (or $T_{blue}$) with galaxy mass. Metal-poor 
GCs are proposed to form in low-mass dark matter halos at very high redshift, typically $z \sim 
10-15$. Halos in high-density environments collapse first. As discussed in Strader \etal, this 
scenario can reproduce the observed correlations with galaxy mass, given reasonable assumptions 
(including the truncation of metal-poor GC formation at high $z$, plausibly by reionization). It can 
also explain other observations, such as the radial distribution of metal-poor GCs (Moore \etal~2006) 
and possibly the mass--metallicity relation for individual metal-poor GCs (Strader \etal~2006; Harris 
\etal~2006). Metal-rich GCs form in the subsequent dissipational merging that forms the host galaxy. 
When the bulk of this ``action" took place is not well-constrained but, for galaxies at $\sim L^{*}$ 
and above, most GCs appear to have formed at $z \ga 2$ (Strader \etal~2005; Puzia \etal~2005). Some 
additional dissipationless merging for massive Es appears to be required, based on the evolution of 
the ``red sequence" luminosity function of early-type galaxies from $z \sim 1$ to the present (Faber 
\etal~2005; Bell \etal~2004) and the dichotomy of core parameters (\S 4.1). The ages of metal-rich 
GCs do not constrain such dissipationless merging, but in the future the radial distributions and 
kinematics may offer interesting insights. We call this picture of GC formation the synthesis 
scenario.

It has been mentioned several times in the preceding sections that biasing is a key factor
in understanding structure formation. Could the metal-poor GC subpopulations
of massive Es have been built from major mergers of present-day lower-mass disk
galaxies or by the accretion of many dwarf galaxies, since both dwarfs and spirals
have metal-poor GCs of lower metallicity than those in massive Es?

To simultaneously accommodate: (i) the metal-poor GC metallicity--galaxy mass relation and 
(ii) the theoretical and observational evidence that most massive galaxies have undergone some 
degree of merging/accretion since $z \sim 2$, we must argue that the metal-poor GC relation 
was \emph{different} at higher redshift. A present day $L^{*}$ galaxy cannot have been 
assembled from present-day sub-$L^{*}$ galaxies. Instead, the merging must have been biased, 
in the sense that galaxies with metal-poor GC systems that would lie above the relation 
connecting GC metallicity and host galaxy mass at $z=0$ would tend to have merged into more 
massive galaxies by the present (see Figure 12 for a schematic diagram of this process). This 
can be understood as a direct result of hierarchical structure formation: high-$\sigma$ peaks 
in the most overdense regions (destined to become, e.g., galaxy clusters) collapse and form 
metal-poor GCs first. These metal-poor GCs will be more highly enriched than those forming in 
halos that collapse later, either because they have more time to self-enrich, or because the 
density of nearby star-forming halos is larger and they could capture more outflowing enriched 
gas. These first-forming metal-poor GCs will tend to be concentrated toward the center of the 
overdensity and will quickly agglomerate into larger structures. Similar mass fluctuations in 
the less-overdense outer regions will tend to be accreted into larger structures more slowly. 
Some may survive to form more stars and become dwarf satellites of the central galaxy. This 
picture, at least as it relates to dark matter halos, is well-understood and accepted. But the 
important point for GC formation scenarios is that these surviving dwarfs \emph{are not} 
representative of the halos that merged to form the central galaxy. The latter collapsed first 
and may have very different star (and GC) formation histories from those that collapsed later.

This process will operate on a variety of scales. For example, the dwarf satellites of the 
Galaxy have metal-poor GCs with lower metallicities than those of halo GCs in the Galaxy. 
Moreover, the disk or E galaxies that merged to form gEs like M87 and NGC 4472 must have 
had metal-poor GCs with metallicities higher than those typical for Es and spirals in the 
Virgo cluster today. The metal-poor GCs, although they formed at very high redshift, 
already ``knew" to which galaxy they would ultimately belong. The metal-poor relation 
rules out merger and accretion models, but only in the local universe for structure 
forming at the present day. Nonetheless, it is consistent with hierarchical galaxy 
formation, and is a strong end constraint for any galaxy formation model. To illustrate 
biasing, Figure 13 shows two snapshots ($z=12$ and 0) of a high-resolution dark matter 
simulation of the formation of a $10^{12} M_{\odot}$ galaxy (Diemand \etal~2005; Moore 
\etal~2006). Low-mass, high-$\sigma$ peaks collapse first in a filamentary structure and 
end up centrally concentrated in the final galaxy.

\section{Cosmological Formation of Metal-poor Globular Clusters}

The preceding sections suggest that the bulk of GCs formed at high redshift. It then follows that 
they have enormous promise in a cosmological context. In this section we discuss current ideas on 
the ``cosmological" formation of GCs and how this has shaped our overall view of galaxy formation. 
In what follows, the term cosmological refers to models in which GCs form in low-mass dark matter 
halos, before the bulk of their parent galaxy has been assembled. 

Soon after CDM cosmology was proposed, Peebles (1984) argued that $10^8 M_{\odot}$ halos, each 
hosting a few $\times \, 10^6 M_{\odot}$ of gas, would be the first to collapse and form stars 
in the early universe. He suggested GCs as their progeny, and noted that the halos of these 
GCs might be stripped without disrupting the cluster itself. This basic idea has been 
sustained to the present day, though the mechanism is probably limited to metal-poor GCs. 
Rosenblatt, Faber, \& Blumenthal (1988) refined this scenario by suggesting that metal-poor 
GCs form in 2.8$\sigma$ halos. This gives a reasonable match to the observed radial 
distribution of metal-poor Galactic GCs and the mass fraction of metal-poor GCs (with respect to total 
stellar mass) in a variety of galaxies. The principal problem with this picture---indeed, of GC formation in individual 
dark matter halos in general---is that the minimum required baryonic collapse factor ($\sim 
10$ or more) would produce GCs with more rotation than that observed, unless they form preferentially 
in very low-spin halos, or some other mechanism acts to remove angular momentum. A speculative 
solution might be to form the GC in the core of a larger gas cloud, if the ``extraneous" 
material can be stripped later.

Moore (1996) wrote a short influential paper that described the use of N-body simulations to 
show that the faint tidal tails observed around some GCs (Grillmair \etal~1995) were 
inconsistent with the presence of extended dark matter halos, but consistent with the low M/L 
ratios observed in the central regions (e.g., 
Illingworth 1976; Pryor \etal~1989). The observed tidal tails around, e.g., Pal 5 
(Odenkirchen \etal~2002) demonstrate that there are at least some present-day GCs that lack dark 
matter. However, this does not prove that all (or even most) GCs are free of extended dark 
matter halos, nor does it rule out metal-poor GC formation inside halos that are later stripped 
away.

This latter idea has been developed in a number of papers by Mashchenko \& Sills (2005a,b),
who studied the formation and evolution of GCs with individual dark matter halos. 
With high-resolution N-body simulations, they found that (depending on the details of the 
actual collapse) many properties of simulated GCs with halos are similar to those of observed 
GCs. For example, the central mass-to-light ratios are expected to be quite low. Structure in 
the halo (e.g., triaxiality, or breaks in the outer parts of the density profile) could easily 
be manifested as tidal cutoffs, extratidal stars, or eccentric outer contours; such features 
are not incompatible with the presence of dark matter. When a GC with a dark matter halo 
evolves in a tidal field, their simulations indicated that it loses either most (for an NFW 
halo) or nearly all (for a Burkert halo) of its dark matter. This finding was confirmed in a 
very high resolution dark matter and gas simulation by Saitoh \etal~(2005).

Bromm \& Clarke (2002) used a simulation with both dark matter and gas to study GC 
formation at high redshift. As noted by Peebles (1984), $\sim 10^8 M_{\odot}$ minihalos 
are expected to collapse out of $3\sigma$ fluctuations at $z \sim 15$. At a high fixed 
gas density threshold Bromm \& Clarke created sink particles as ``GCs". These GCs
initially form inside of halos, but the simultaneous collapse of mass scales results in 
violent relaxation that erases most of the substructure. The resulting GC mass spectrum 
is set by that of the dark matter, and is a power law with index $\sim -1.8$. The main 
problem here is that it is not possible to tell, with the current level of sophistication 
of 
their simulations, whether the violent relaxation is real or merely an artifact of 
insufficient resolution.

Other authors have explored in more detail the triggering mechanism for putative GC formation in
dark matter halos. In the model of Cen (2001), ionization fronts from cosmic 
reionization shock gas in low-mass halos. The gas is compressed by a factor $\sim 100$ and 
collapses to form metal-poor GCs. To produce the observed numbers of GCs, a large population of 
low spin ($\lambda < 0.01$) halos is required. This condition appears to be satisfied when the 
halo number density is modeled with extended Press-Schecter theory, but whether it would hold 
in high-resolution cosmological simulations is unknown. This model predicts a power law GCMF 
with a slope of $\sim -2$, similar to that observed at the high mass end, and has the rather 
attractive property of predicting no GC mass--radius relation, consistent with the 
observations. C{\^ o}t{\' e} (2002) pointed out that this picture could predict a large number 
of (unobserved) intergalactic GCs; this objection might be addressed if the ionization fronts 
are effective above a threshold only met quite close to protogalaxies.

Scannapieco \etal~(2004) proposed a somewhat similar mechanism, in which gas in minihalos is 
shock-compressed by galaxy outflows. The momentum of the shock strips the gas from the halo, 
nicely solving the dark matter problem. However, this model predicts a mass--radius relation for 
individual GCs, and the observed lack of such a relation may deal this picture a fatal blow.

Ricotti (2002) suggested that GCs themselves could have reionized the universe. The predicted 
number of ionizing photons appears to be sufficient, assuming that the escape fraction of such 
photons is near unity. Such a high escape fraction is qualitatively feasible, given the 
extended spatial distribution of the GCs with respect to their parent galaxies, but the 
scenario requires detailed modeling (including radiative transfer) in the proper cosmological 
setup.

The high-resolution simulation of Kravtsov \& Gnedin (2005) offers a glimpse of what should be 
possible in the future. They performed a gas and dark matter simulation of the formation of a 
Milky Way analogue to $z \sim 3$. They were not able to resolve GC formation directly, but 
assumed GCs formed in the cores of GMCs when the dynamical time exceeded the cooling time. 
These GMCs were located in the flattened gas disks of protogalaxies. The resulting mass 
function appears to be consistent with that of massive GCs, but the metallicity distribution 
does not: at the end of the simulation, large numbers of (unobserved) [Fe/H] $\sim -1$ GCs were 
being formed, even though their simulation included feedback. A desirable future extension of 
such simulations is to test whether reionization might effectively end metal-poor GC formation 
where ``traditional" stellar feedback cannot.

Moore \etal~(2006) have shown that the observed radial distribution of metal-poor GCs in 
the Galaxy can be reproduced if the GCs are assumed to form in $\ga 2.5\sigma$ peaks of 
$\ga 2 \times 10^8 M_{\odot}$ that collapse and form stars before $z \sim 12$ (see Figure 
14). If the key assumption of truncation by reionization is supported by other evidence, 
then the surface density distributions of metal-poor GCs could be used to probe 
reionization in a variety of galaxies, setting limits on the epoch and homogeneity of 
reionization.

In summary, the cosmological formation of metal-poor GCs is supported by several lines of 
argument. (i) The ages of metal-poor GCs. The absolute ages of GCs are still poorly 
known, and must be continually revised in light of advances in a variety of subfields 
(for example, the recent revision of the $^{14}$N($p,\gamma$)$^{15}$O reaction rate 
increases GC ages by 0.7--1 Gyr; Imbriani \etal~2004). If the GC ages are sufficiently 
close to the age of the universe, cosmological formation becomes a necessity, since 
low-mass halos as described above are the only existing sites for star formation. (ii) 
The recently discovered correlation between GC metallicity and mass for bright metal-poor 
GCs in several massive galaxies (Strader \etal~2006; Harris \etal~2006). Self-enrichment 
is a potential explanation for this correlation, and it is possible that metals could 
only be retained in the potential well of a dark matter halo. (iii) The radial 
distribution of metal-poor GCs. Moore \etal~(2006) show that the metal-poor GCs in the 
Galaxy have a radial distribution consistent with formation in $2.5\sigma$ peaks in the 
dark matter distribution at $z > 12$. (iv) Observations of Local Group dwarfs show that 
the lowest-mass galaxies with GCs 
have total masses of $\la 10^8 M_{\odot}$ (see \S 10), as expected under cosmological 
formation.

Does this mean that two separate mechanisms are needed to explain metal-poor and 
metal-rich GC formation? The strongest similarity between the two subpopulations is
their mass function; for the less evolved high-mass part of the mass function, these are
approximately power laws with indices $\sim -1.8$ to $-2$. Since power law distributions 
are a consequence of a variety of physical processes, this similarity does not mandate an 
identical formation history for both subpopulations. For example, the power law slope 
observed in the GC mass function is the same as both that of GMCs in the Galaxy and of 
low-mass dark matter halos collapsing at $z \sim 15-20$.

\section{Future Directions}

Those of us actively trying to understand extragalactic globular clusters and their 
connection to galaxy formation are currently operating in a data-dominated rather than a 
theory-dominated field. We are accumulating a wealth of observational information that we 
cannot fully interpret.  An urgent need exists for improvements in numerical and 
semi-analytic simulations to help identify GC formation sites and track their spatial, 
kinematic, chemical, and structural evolution. Models that can resolve the masses and 
sizes of a typical GC are tantalizingly close to implementation, and their advent will 
signal a leap in progress toward placing the formation of GCs in its proper cosmological 
context. With the new generation of ``big telescope" wide-field multiplexing 
spectrographs, such as Keck/DEIMOS, Magellan/IMACS, MMT/Hectospec, and VLT/VIMOS, it is 
possible to study large samples of GCs in a wide variety of galaxies to carry out detailed 
tests of these future models, establishing ages, metallicities, and kinematics.

Other important developments will come from the community of SSP modelers. As already 
mentioned, there remain significant disagreements within this community on the treatment 
of $\alpha$-enhancement, horizontal branch morphology (the second parameter problem), and 
underlying stellar synthesis techniques. A new way forward to study extragalactic GCs in 
detail is the SSP modeling of high-resolution spectra (e.g., Bernstein \& McWilliam 2005). 
This offers the possibility of estimating abundances of light, $\alpha$-, Fe-peak, and 
even some strong r- and s-process elements. With an 8-10m class telescope, this technique 
can be applied to significant samples of GCs in the Local Group and nearby galaxies, and 
even to the brightest few GCs in Virgo. This program is in its infancy, but could 
represent a significant leap in our understanding of the detailed formation histories of 
galaxies. GCs may offer the only route to measuring the abundances of interesting elements 
that are unobservable in massive galaxies themselves due to their large velocity 
dispersions. A possibility enabled by multiplexing high-resolution spectroscopy is 
detailed dynamical modeling of individual Galactic GCs, to determine whether any contain 
halo dark matter. The discovery of dark matter in GCs would be a ``smoking gun" of 
cosmological GC formation.

With the development of new large-format detectors and CCD mosaics, wide field optical imaging 
is poised to address numerous outstanding issues. Obtaining global radial and color 
distributions for individual GC subpopulations is essential to test scenarios for GC and 
galaxy formation, especially as models emerge that predict these quantities in detail. Such 
imaging can also be used to probe the evolution of the GCLF, and provide a definitive test of 
the scenario in which GCs are formed with a power-law LF that subsequently evolves through 
various destruction processes to the log-normal LF observed for old GC systems. Since such 
processes are expected to operate more efficiently at small galactocentric radii, changes in 
the GCLF of GCs with galactocentric radius would be revealing. Such imaging has the potential 
to constrain the epoch of reionization from the spatial distribution of metal-poor GCs (as 
discussed in \S 12).

These ideas cover only a small fraction of the important advances likely to occur in the field 
over the next decade. The eventual availability of 30-m class telescopes and \emph{JWST} may 
allow us to reach the inspiring goal of Renzini (2002): ``To directly map the evolution of GCs 
in galaxies all the way to see them in formation, and eventually stick on the wall a poster 
with a million-pixel picture of a $z=5$ galaxy, with all her young GCs around.''\\

We thank many colleagues for reading drafts of this manuscript and for useful discussions, 
including Keith Ashman, Michael Beasley, Javier Cenarro, Laura Chomiuk, Juerg Diemand, 
Sandra Faber, Duncan Forbes, Genevieve Graves, Soeren Larsen, John Kormendy, Piero Madau, 
Joel Primack, Katherine Rhode, Tom Richtler, Brad Whitmore, and Steve Zepf. We also thank 
Michael Beasley, Andres Jord{\'a}n, and Tom Richtler for permission to use their figures. 
Michael Beasley, Soeren Larsen, and Katherine Rhode provided data used to create other 
figures. Finally, we thank Takayuki Saitoh and Marsha Wolf for providing advance copies of 
their articles. Support was provided by NSF Grants AST-0206139 and AST-0507729, an NSF 
Graduate Research Fellowship, and STScI grant GO-9766.\\

\clearpage
\begin{table}
\tiny
\begin{center}
\begin{tabular}{lccccccccccc}
Name&	Galaxy Type$^{a}$&	Environment$^{b}$& $M_B$$^{c}$	& MP color$^{d}$&	
MR color $^{d}$&	MP [Fe/H] $^{e}$	& MR	[Fe/H]$^{e}$&	
Color$^{f}$	 &	$S_N$$^{g}$	&	$S_N$ Refs$^{h}$	 \\ 
    &   &   &  (mag) & (mag) & (mag) & (dex) & (dex) & & & \\
\hline
\hline
NGC 4472&	core E	&	Virgo C	&	$-$21.9	&	0.951	&	1.411	&	$-$1.32	&	$-$0.17	&	$g-z$	&	$3.6\pm0.6$&	1	\\
NGC 1399&	core E	&	Fornax C&	$-$21.8	&	0.952	&	1.185	&	$-$1.37	&	$-$0.39	&	$V-I$	&	$5.1\pm1.2$&	2	\\
NGC 3309&	core(?) E&	Hydra C	&	$-$21.6	&	0.947	&	1.134	&	 $-$1.39	&	$-$0.60	&	$V-I$	&	\nodata	&	\nodata	\\
NGC 4486&	core E	&	Virgo C	&	$-$21.5	&	0.953	&	1.390	&	$-$1.31	&	$-$0.21	&	$g-z$	&	$14.1\pm1.5$&	3	\\
NGC 3311&	core(?) E&	Hydra C	&	$-$21.5	&	0.929	&	\nodata	&	$-$1.47	&	\nodata	&	$V-I$	&	\nodata	&	\nodata	\\
NGC 4406&	core E	&	Virgo C	&	$-$21.5	&	0.986	&	1.145	&	$-$1.23	&	$-$0.56	&	$V-I$	&	$3.5\pm0.5$&	1\\
NGC 4649&	core E	&	Virgo C	&	$-$21.4	&	0.964	&	1.424	&	$-$1.26	&	$-$0.14	&	$g-z$	&	$4.1\pm1.0$&	4\\
NGC 524	&	S0	&	N524 G	&	$-$21.4	&	0.980	&	1.189	&	$-$1.25	&	$-$0.37	&	$V-I$	&	\nodata	&	\nodata	\\
NGC 4374&	core E	&	Virgo C	&	$-$21.2	&	0.927	&	1.322	&	$-$1.45	&	$-$0.33	&	$g-z$	&	$1.6\pm0.3$&	5\\
NGC 5322&	core(?) E&	N5322 G	&	$-$21.2	&	0.942	&	\nodata	&	$-$1.41	&	\nodata	&	$V-I$	&	\nodata	&	\nodata	\\
NGC 4594&	S0/Sa	&	N4594 G	&	$-$21.2	&	0.939	&	1.184	&	$-$1.43	&	$-$0.39	&	$V-I$	&	$2.1\pm0.3$&	1\\
NGC 4365&	core E	&	Virgo C	&	$-$21.1	&	0.891	&	1.232	&	$-$1.63	&	$-$0.50	&	$g-z$	&	\nodata	&	\nodata\\
NGC 7619&	core(?) E&	Pegasus C&	$-$21.1	&	0.973	&	\nodata	&	$-$1.28	&	\nodata	&	$V-I$	&	\nodata	&	\nodata\\
NGC 7562&	core(?)	E&	Pegasus C&	$-$20.9	&	0.920	&	\nodata	&	$-$1.51	&	\nodata	&	$V-I$	&	\nodata	&	\nodata\\
NGC 2768&	S0	&	N2768 G	&	$-$20.8	&	0.919	&	\nodata	&	$-$1.51	&	\nodata	&	$V-I$	&	\nodata	&	\nodata	\\
NGC 4621&	transition E&	Virgo C	&	$-$20.7	&	0.927	&	1.305	&	$-$1.45	&	$-$0.36	&	$g-z$	&	\nodata	&	\nodata\\
NGC 5813&	core E	&	N5846 G	&	$-$20.6	&	0.935	&	\nodata	&	$-$1.44	&	\nodata	&	$V-I$	&	$5.7\pm1.8$&	6\\
IC 1459	&	core E	&	I1459 G	&	$-$20.6	&	0.955	&	\nodata	&	$-$1.36	&	\nodata	&	$V-I$	&	\nodata	&	\nodata	\\
NGC 3115&	S0	&	N3115 G	&	$-$20.4	&	0.922	&	1.153	&	$-$1.50	&	$-$0.52	&	$V-I$	&	\nodata	&	\nodata	\\
NGC 4494&	power E	&	N4565 G	&	$-$20.4	&	0.901	&	1.128	&	$-$1.59	&	$-$0.63	&	$V-I$	&	\nodata	&	\nodata	\\
NGC 4552&	core E	&	Virgo C	&	$-$20.3	&	0.951	&	1.334	&	$-$1.32	&	$-$0.31	&	$g-z$	&	\nodata	&	\nodata	\\
NGC 253	&	Sc	&	Sculptor G&	$-$20.3	&	0.912	&	\nodata	&	$-$1.54	&	\nodata	&	$V-I$	&	\nodata	&	\nodata	\\
NGC 1404&	core(?)	E&	Fornax C&	$-$20.3	&	0.938	&	1.170	&	$-$1.43	&	$-$0.45	&	$V-I$	&	$2\pm0.5$&	7\\
M31	&	Sb	&	Local G	&	$-$20.2	&	0.912	&	\nodata	&	$-$1.54	&	\nodata	&	$V-I$	&	1.3	&	8	\\
NGC 3379&	core E	&	Leo I G	&	$-$20.1	&	0.964	&	1.167	&	$-$1.32	&	$-$0.47	&	$V-I$	&	$1.2\pm0.3$&	1\\
NGC 4278&	core E	&	N4631 G	&	$-$20.1	&	0.908	&	\nodata	&	$-$1.56	&	\nodata	&	$V-I$	&	\nodata	&	\nodata	\\
NGC 4473&	power E	&	Virgo C	&	$-$20.0	&	0.942	&	1.310	&	$-$1.37	&	$-$0.35	&	$g-z$	&	\nodata	&	\nodata	\\
NGC 3608&	core E	&	N3607 G	&	$-$20.0	&	0.923	&	\nodata	&	$-$1.49	&	\nodata	&	$V-I$	&	\nodata	&	\nodata	\\
NGC 1400&	S0	&	Eridanus G&	$-$19.9	&	0.951	&	1.164	&	$-$1.38	&	$-$0.48	&	$V-I$	&	\nodata	&	\nodata	\\
Milky Way&	Sbc	&	Local G	&	$-$19.9	&	0.898	&	\nodata	&	$-$1.60	&	\nodata	&	$V-I$	&	0.7	&	8	\\
NGC 1023&	S0	&	N1023 G	&	$-$19.7	&	0.912	&	1.164	&	$-$1.54	&	$-$0.48	&	$V-I$	&	\nodata	&	\nodata	\\
NGC 4291&	core E	&	N4291 G	&	$-$19.6	&	0.940	&	\nodata	&	$-$1.42	&	\nodata	&	$V-I$	&	\nodata	&	\nodata	\\
NGC 3384&	S0	&	Leo I G	&	$-$19.5	&	0.942	&	1.208	&	$-$1.41	&	$-$0.29	&	$V-I$	&	\nodata	&	\nodata	\\
NGC 3607&	S0	&	N3607 G	&	$-$19.5	&	0.939	&	1.099	&	$-$1.43	&	$-$0.75	&	$V-I$	&	\nodata	&	\nodata	\\
NGC 1427&	power E	&	Fornax C&	$-$19.4	&	0.940	&	1.153	&	$-$1.42	&	$-$0.52	&	$V-I$	&	$3.4\pm0.6$&	9\\
NGC 4478&	power E	&	Virgo C	&	$-$19.2	&	0.882	&	1.195	&	$-$1.68	&	$-$0.56	&	$g-z$	&	\nodata	&	\nodata	\\
NGC 4434&	power E	&	Virgo C	&	$-$19.2	&	0.911	&	1.179	&	$-$1.53	&	$-$0.59	&	$g-z$	&	\nodata	&	\nodata	\\
NGC 3377&	power E	&	Leo I G	&	$-$19.2	&	0.936	&	1.103	&	$-$1.44	&	$-$0.74	&	$V-I$	&	\nodata	&	\nodata	\\
NGC 4564&	S0	&	Virgo C	&	$-$19.0	&	0.935	&	1.263	&	$-$1.40	&	$-$0.44	&	$g-z$	&	\nodata	&	\nodata	\\
NGC 4387&	power E	&	Virgo C	&	$-$19.0	&	0.859	&	1.112	&	$-$1.79	&	$-$0.72	&	$g-z$	&	\nodata	&	\nodata	\\
NGC 4660&	S0	&	Virgo C	&	$-$18.7	&	0.923	&	1.320	&	$-$1.47	&	$-$0.33	&	$g-z$	&	\nodata	&	\nodata	\\
NGC 247	&	Sd	&	Sculptor G&	$-$18.7	&	0.908	&	\nodata	&	$-$1.56	&	\nodata	&	$V-I$	&	\nodata	&	\nodata	\\
NGC 4733&	power E	&	Virgo C	&	$-$18.6	&	0.918	&	1.131	&	$-$1.52	&	$-$0.62	&	$V-I$	&	\nodata	&	\nodata	\\
NGC 4550&	S0	&	Virgo C	&	$-$18.6	&	0.883	&	1.145	&	$-$1.66	&	$-$0.56	&	$V-I$	&	\nodata	&	\nodata	\\
NGC 4489&	S0	&	Virgo C	&	$-$18.5	&	0.900	&	1.260	&	$-$1.58	&	$-$0.44	&	$g-z$	&	\nodata	&	\nodata	\\
NGC 4551&	power E	&	Virgo C	&	$-$18.5	&	0.890	&	1.219	&	$-$1.64	&	$-$0.52	&	$g-z$	&	\nodata	&	\nodata	\\
M33	&	Scd	&	Local G	&	$-$18.4	&	0.900	&	\nodata	&	$-$1.59	&	\nodata	&	$V-I$	&	\nodata	&	\nodata	\\
NGC 4458&	power E	&	Virgo C	&	$-$18.4	&	0.892	&	1.223	&	$-$1.63	&	$-$0.51	&	$g-z$	&	\nodata	&	\nodata	\\
NGC 55	&	Sm	&	Sculptor G&	$-$18.3	&	0.892	&	\nodata	&	$-$1.63	&	\nodata	&	$V-I$	&	\nodata	&	\nodata	\\
IC 3468	&	dE 	&	Virgo C	&	$-$18.1	&	0.925	&	1.130	&	$-$1.46	&	$-$0.68	&	$g-z$	&	1.1	&	10	\\
NGC 300	&	Sd	&	Sculptor G&	$-$18.1	&	0.892	&	\nodata	&	$-$1.63	&	\nodata	&	$V-I$	&	\nodata	&	\nodata	\\
NGC 4482&	dE	&	Virgo C	&	$-$18.1	&	0.884	&	1.065	&	$-$1.67	&	$-$0.80	&	$g-z$	&	1.6	&	10	\\
LMC	&	dIrr	&	Local G	&	$-$18.0	&	0.890	&	\nodata	&	$-$1.63	&	\nodata	&	$V-I$	&	0.8	&	8	\\
NGC 3599&	S0	&	Leo I G	&	$-$17.6	&	0.872	&	1.112	&	$-$1.71	&	$-$0.70	&	$V-I$	&	\nodata	&	\nodata	\\
IC 3019	&	dE	&	Virgo C	&	$-$17.2	&	0.864	&	1.093	&	$-$1.77	&	$-$0.75	&	$g-z$	&	1.8	&	10	\\
IC 3381	&	dE	&	Virgo C	&	$-$17.1	&	0.897	&	1.168	&	$-$1.60	&	$-$0.61	&	$g-z$	&	5.2	&	10	\\
IC 3328	&	dE	&	Virgo C	&	$-$17.1	&	0.905	&	1.114	&	$-$1.56	&	$-$0.71	&	$g-z$	&	0.9	&	10	\\
NGC 4318&	dE	&	Virgo C	&	$-$17.1	&	0.884	&	1.182	&	$-$1.67	&	$-$0.59	&	$g-z$	&	0.7	&	10	\\
IC 809	&	dE	&	Virgo C	&	$-$17.1	&	0.864	&	1.129	&	$-$1.77	&	$-$0.68	&	$g-z$	&	3.3	&	10	\\
IC 3653	&	power E	&	Virgo C	&	$-$17.1	&	0.851	&	N	&	$-$1.84	&	N	&	$g-z$	&	0.6	&	10	\\
IC 3652	&	dE	&	Virgo C	&	$-$16.9	&	0.879	&	1.173	&	$-$1.69	&	$-$0.60	&	$g-z$	&	3.7	&	10	\\
VCC 543	&	dE	&	Virgo C	&	$-$16.9	&	0.794	&	N	&	$-$2.13	&	N	&	$g-z$	&	0.4	&	10	\\
IC 3470	&	dE	&	Virgo C	&	$-$16.9	&	0.938	&	1.138	&	$-$1.39	&	$-$0.67	&	$g-z$	&	5.0	&	10	\\
NGC 4486b&	power E	&	Virgo C	&	$-$16.8	&	0.920	&	1.126	&	$-$1.51	&	$-$0.64	&	$V-I$	&	\nodata	&	\nodata	\\
IC 3501	&	dE	&	Virgo C	&	$-$16.8	&	0.872	&	1.177	&	$-$1.73	&	$-$0.60	&	$g-z$	&	5.3	&	10	\\
IC 3442	&	dE	&	Virgo C	&	$-$16.8	&	0.837	&	1.089	&	$-$1.91	&	$-$0.76	&	$g-z$	&	0.8	&	10	\\
VCC 437	&	dE	&	Virgo C	&	$-$16.7	&	0.894	&	1.042	&	$-$1.61	&	$-$0.85	&	$g-z$	&	1.1	&	10	\\
IC 3735	&	dE	&	Virgo C	&	$-$16.7	&	0.848	&	1.198	&	$-$1.85	&	$-$0.56	&	$g-z$	&	1.6	&	10	\\
SMC	&	dIrr	&	Local G	&	$-$16.7	&	0.919	&	\nodata	&	$-$1.51	&	\nodata	&	$V-I$	&	1.2	&	8	\\
IC 3032	&	dE	&	Virgo C	&	$-$16.6	&	0.861	&	1.169	&	$-$1.78	&	$-$0.61	&	$g-z$	&	0.9	&	10	\\
VCC 200	&	dE	&	Virgo C	&	$-$16.6	&	0.931	&	1.101	&	$-$1.42	&	$-$0.74	&	$g-z$	&	0.5	&	10	\\
IC 3487	&	dE	&	Virgo C	&	$-$16.5	&	0.824	&	1.086	&	$-$1.97	&	$-$0.76	&	$g-z$	&	1.2	&	10	\\
IC 3509	&	power E	&	Virgo C	&	$-$16.4	&	0.869	&	1.071	&	$-$1.74	&	$-$0.79	&	$g-z$	&	7.3	&	10	\\
VCC 1895&	dE	&	Virgo C	&	$-$16.3	&	0.818	&	1.071	&	$-$2.01	&	$-$0.79	&	$g-z$	&	1.6	&	10	\\
IC 3647	&	dE	&	Virgo C	&	$-$16.2	&	0.845	&	N	&	$-$1.87	&	N	&	$g-z$	&	1.0	&	10	\\
IC 3383	&	dE	&	Virgo C	&	$-$16.2	&	0.834	&	1.056	&	$-$1.92	&	$-$0.82	&	$g-z$	&	4.6	&	10	\\
VCC 1627&	power E	&	Virgo C	&	$-$16.2	&	0.755	&	N	&	$-$2.33	&	N	&	$g-z$	&	0.5	&	10	\\
IC 3693 &       power E &       Virgo C &       $-$16.2 &       0.779   &       1.177   &       $-$2.21 &       $-$0.60 &       $g-z$   &       1.0     &       10      \\	
IC 3101	&	dE	&	Virgo C	&	$-$16.1	&	0.842	&	1.134	&	$-$1.88	&	$-$0.67	&	$g-z$	&	4.9	&	10	\\
IC 798 &	power E	&	Virgo C	&	$-$16.1	&	0.854	&	N	&	$-$1.82	&	N	&	$g-z$	&	4.6	&	10	\\
IC 3779	&	dE	&	Virgo C	&	$-$16.1	&	0.880	&	1.060	&	$-$1.69	&	$-$0.81	&	$g-z$	&	2.0	&	10	\\
IC 3635	&	dE	&	Virgo C	&	$-$16.0	&	0.898	&	N	&	$-$1.59	&	N	&	$g-z$	&	5.0	&	10	\\
VCC 1993&	dE	&	Virgo C	&	$-$16.0	&	0.851	&	N	&	$-$1.84	&	N	&	$g-z$	&	0.3	&	10	\\
IC 3461&	dE	&	Virgo C	&	$-$15.8	&	0.919	&	1.142	&	$-$1.49	&	$-$0.66	&	$g-z$	&	12.1	&	10	\\
VCC 1886&	dE	&	Virgo C	&	$-$15.8	&	0.877	&	N	&	$-$1.70	&	N	&	$g-z$	&	1.5	&	10	\\
IC 3602	&	dE	&	Virgo C	&	$-$15.7	&	0.855	&	N	&	$-$1.82	&	N	&	$g-z$	&	1.1	&	10	\\
NGC 205	&	dE	&	Local G	&	$-$15.6	&	0.922	&	N	&	$-$1.50	&	N	&	$V-I$	&	3	&	8	\\
VCC 1539&	dE	&	Virgo C	&	$-$15.6	&	0.898	&	1.039	&	$-$1.59	&	$-$0.87	&	$g-z$	&	9.5	&	10	\\
VCC 1185&	dE	&	Virgo C	&	$-$15.6	&	0.890	&	1.077	&	$-$1.64	&	$-$0.78	&	$g-z$	&	6.3	&	10	\\
IC 3633	&	dE	&	Virgo C	&	$-$15.5	&	0.884	&	N	&	$-$1.67	&	N	&	$g-z$	&	3.3	&	10	\\
IC 3490	&	dE	&	Virgo C	&	$-$15.4	&	0.858	&	1.136	&	$-$1.80	&	$-$0.67	&	$g-z$	&	7.3	&	10	\\
VCC 1661&	dE	&	Virgo C	&	$-$15.3	&	0.838	&	1.130	&	$-$1.90	&	$-$0.68	&	$g-z$	&	2.3	&	10	\\
NGC 185	&	dE	&	Local G	&	$-$14.8	&	0.882	&	N	&	$-$1.67	&	N	&	$V-I$	&	4.6	&	8	\\
NGC 6822&	dIrr	&	Local G	&	$-$14.7	&	0.850	&	\nodata	&	$-$1.80	&	\nodata	&	$V-I$	&	1.2	&	11	\\
NGC 147	&	dE	&	Local G	&	$-$14.3	&	0.807	&	N	&	$-$1.98	&	N	&	$V-I$	&	3.6	&	8	\\
WLM	&	dIrr	&	Local G	&	$-$13.9	&	0.910	&	N	&	$-$1.55	&	N	&	$V-I$	&	1.7	&	8	\\
Sagittarius&	dSph	&	Local G	&	$-$12.8	&	0.871	&	\nodata	&	$-$1.71	&	\nodata	&	$V-I$	&	18.1	&	11	\\
Fornax	&	dSph	&	Local G	&	$-$12.6	&	0.858	&	N	&	$-$1.77	&	N	&	$V-I$	&	28.8	&	8	\\

\end{tabular}
\end{center}
\caption{Properties of GC color distributions.}
\end{table}
\clearpage

$^{a}${Power/Core E: Ellipticals with 
power-law or cored center surface brightness distributions. NGC 4621 is a transition 
between the two groups. The galaxies with (?) are not formally classified---the division 
has been made between core/power-law Es at $M_B = -20$ (Faber \etal~1997; Kormendy 
\etal~2006). Classifications are from Kormendy \etal~(2006, and private communication) for 
Virgo galaxies, Faber \etal~(1997) for many early-type galaxies, and NED for the remainder. 
The term dE (dwarf elliptical) is often used for all non-star forming galaxies with $M_B 
\ga -18$; here we use it only for galaxies with faint central surface brightness and 
Sersic n $\sim 1$ (exponential) profiles.}$^{b}${Local 
galaxy environment. C: cluster, G: group.}$^{c}${References for $M_B$ can be found in 
Strader \etal~(2004a) and Strader \etal~(2006).}$^{d}${Peak metal-poor (MP) and metal-rich 
(MR) colors derived from mixture modeling of the GC color distributions. ``N" indicates
that study of the galay has found no metal-rich GCs. Galaxies with ``..."
either have metal-rich GCs without a determined color peak (typically massive galaxies), or ambiguous evidence
for such a subpopulation.}$^{e}${[Fe/H] 
values converted from the listed colors using the relations of Barmby \etal~(2000) and Peng \etal~(2006)
for $V-I$ and $g-z$, respectively.}$^{f}${The color listed in columns 5 and 6. 
The $V-I$ colors are from Strader \etal~(2004a); the $g-z$ colors are from Strader 
\etal~(2006)}$^{g}${$V$-band specific frequency ($S_N$). Values are only listed if from a 
study with sufficient spatial coverage and photometric depth for an accurate estimate of 
the total GC population. The values from Strader \etal~(2006) have been converted from the 
$B$-band $S_N$ by dividing by 2.1; this assumes $B-V=0.8$.}$^{h}${Literature sources for 
$S_N$ estimates. 1: Rhode \& Zepf (2004) 2: Dirsch \etal~(2003). 3: Harris \etal~(1998).
4: Forbes \etal~(2004). 5: Gomez \& Richtler (2004). 6: Hopp \etal~(1995).
7: Forbes \etal~(1998). 8: Forbes \etal~(2000). 10: Strader \etal~(2006). 11: This work; the $S_N$ estimates
have been increased from Forbes \etal~(2000) to include new GCs in NGC 6822 and Sgr (see text).}


\begin{figure}
\epsscale{1}
\plotone{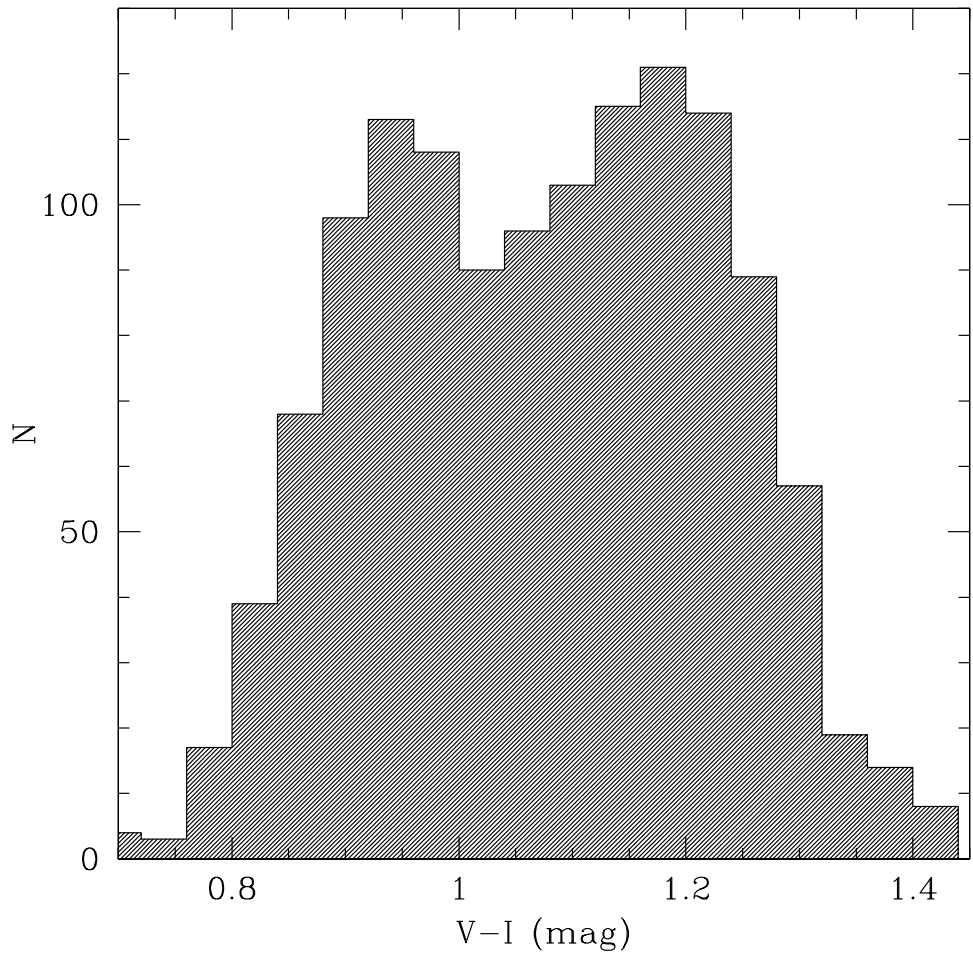}
\caption{$V-I$ color histogram of GCs in the Virgo gE M87, showing clear bimodality (Larsen
\etal~2001; figure from data courtesy of S.~Larsen).}
\label{fig1}
\end{figure}

\begin{figure}
\epsscale{1}
\plotone{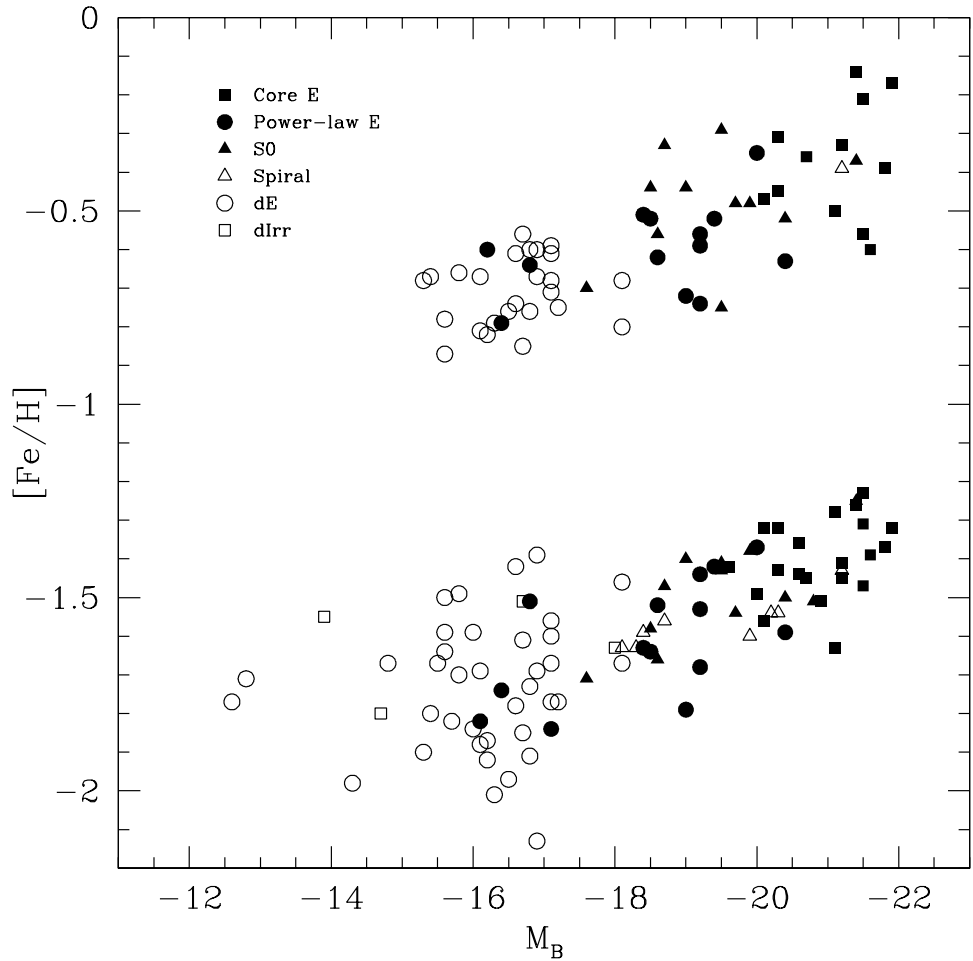}
\caption{Peak GC metallicity vs.~galaxy luminosity ($M_B$) for metal-poor and metal-rich 
GCs in a range of galaxies. The points are from Strader \etal~(2004a) and Strader 
\etal~(2006) and have been converted from $V-I$ and $g-z$ to [Fe/H] using the relations of 
Barmby \etal~(2000) and Peng \etal~(2006), respectively. Galaxy types are indicated in the 
figure key; classifications are in Table 1. Linear relations exist 
for both subpopulations down to the limit of available data.}
\label{fig2}
\end{figure}

\begin{figure}
\epsscale{1}
\plotone{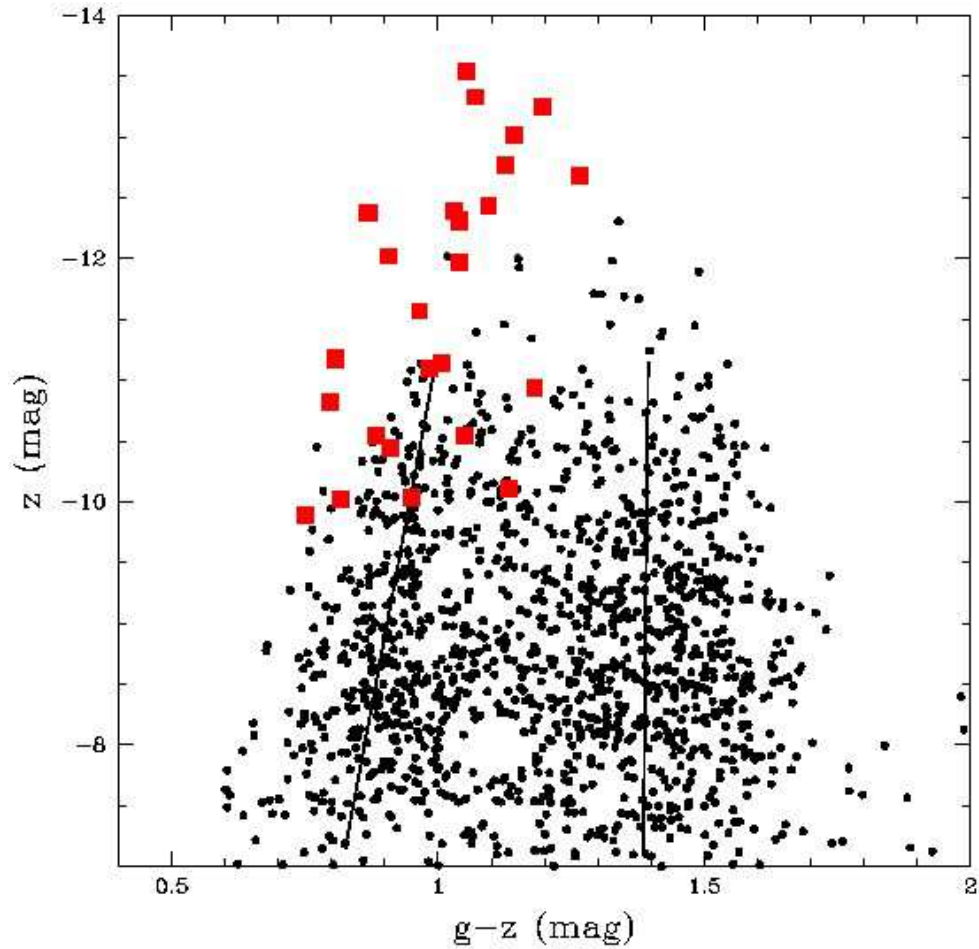}
\caption{$z$ vs.~$g-z$ color-magnitude diagram for M87 GCs (black circles) and Virgo dE nuclei (red 
squares) from Strader \etal~(2006). A correlation between color and luminosity for the bright 
metal-poor GCs is apparent (the ``blue tilt"). The solid lines are fitted linear relations.
The dE nuclei are generally consistent with 
the sequence of metal-poor GCs, but extend to higher luminosities and have a larger spread in color.}
\label{fig3}
\end{figure}

\begin{figure}
\epsscale{1}
\plotone{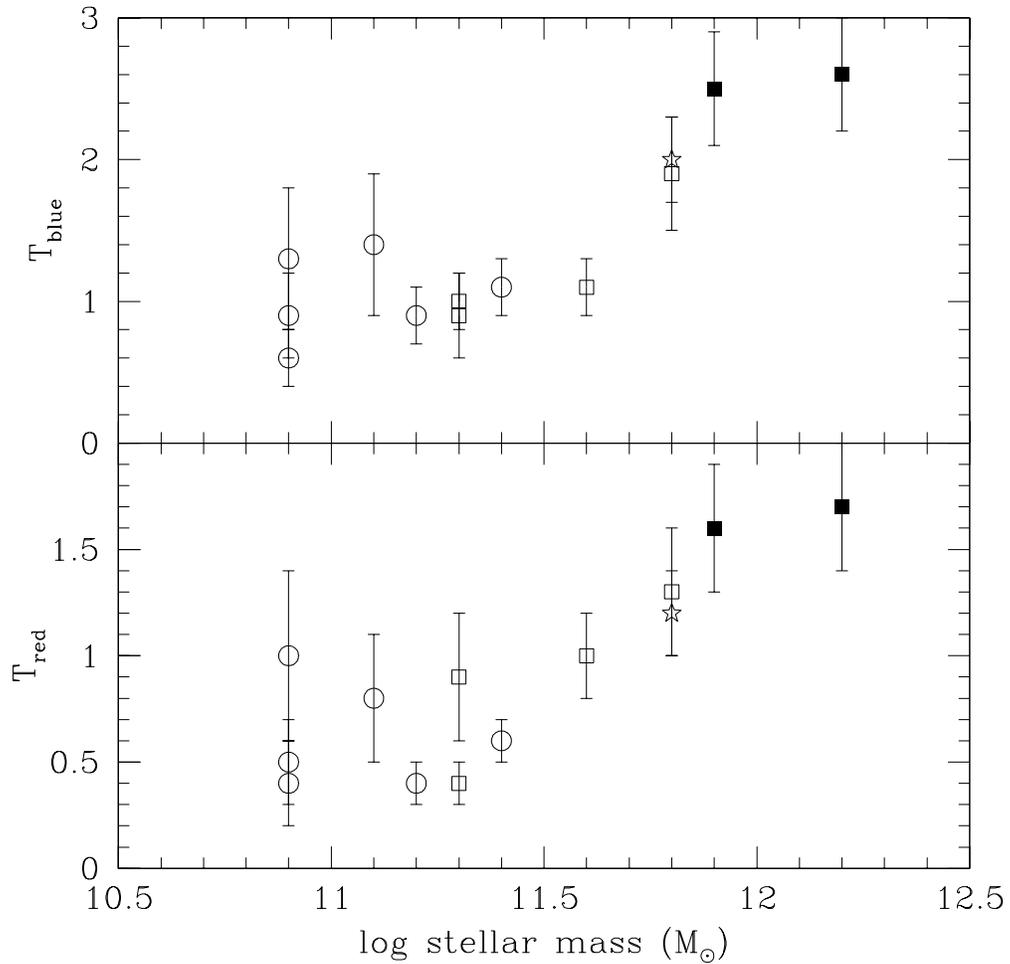}
\caption{$T_{blue}$ and $T_{red}$ (top and bottom panels, respectively) vs.~galaxy mass for a range 
of spirals and Es (Rhode \etal~2005). Filled squares are cluster Es, open squares are field/group 
Es and S0s, open 
circles are field/group spirals, and the open star is the Sa/S0 galaxy NGC 4594. There is a general 
trend of increasing $T_{blue}$ and $T_{red}$ with galaxy mass (Data courtesy K.~Rhode).}
\label{fig4}
\end{figure}

\begin{figure}
\epsscale{1}
\plotone{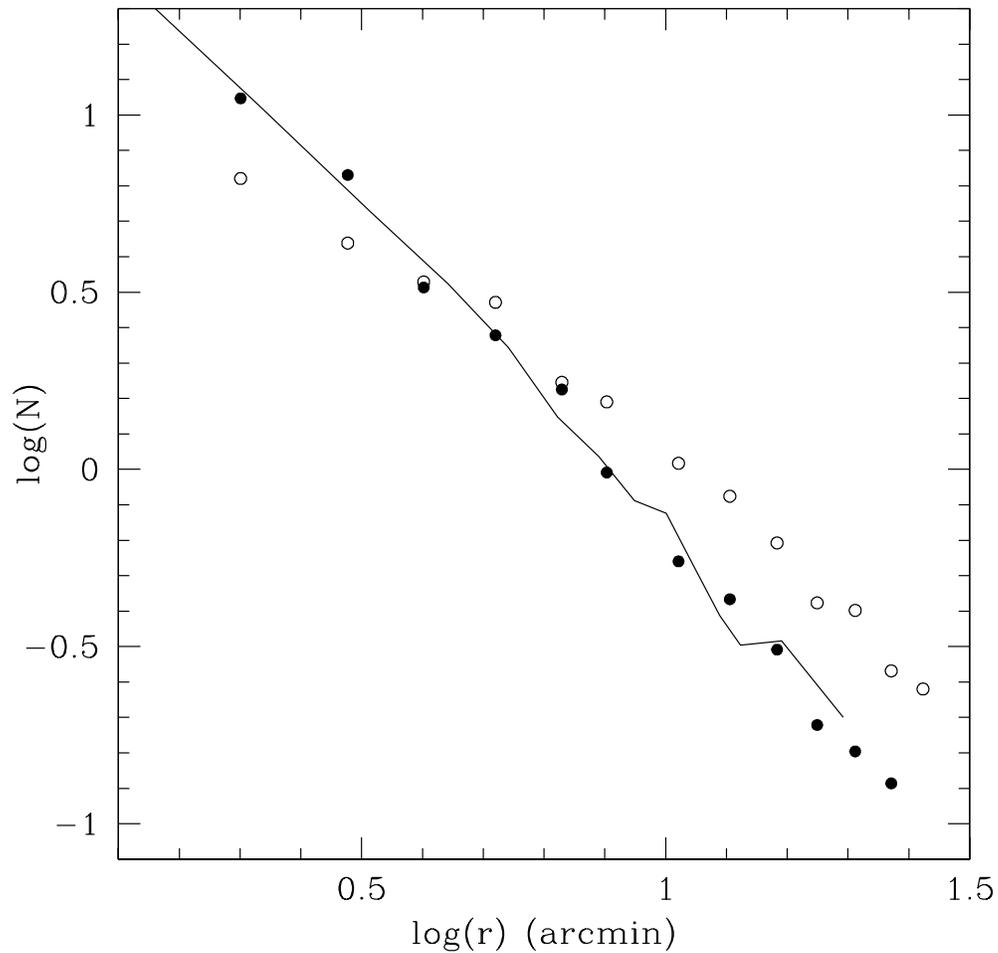}
\caption{Radial surface density distribution of metal-poor (open circles) and metal-rich (filled circles) GCs in the 
Fornax gE NGC 1399. The solid line is the scaled galaxy light profile in the $R$-band. The 
metal-rich GCs are more centrally 
concentrated, and closely follow the underlying galaxy light. The radial distribution of 
the metal-poor GCs
is flatter; they dominate the GC system at large radii (Bassino \etal~2006).}
\label{fig5}
\end{figure}

\begin{figure}
\epsscale{1}
\plotone{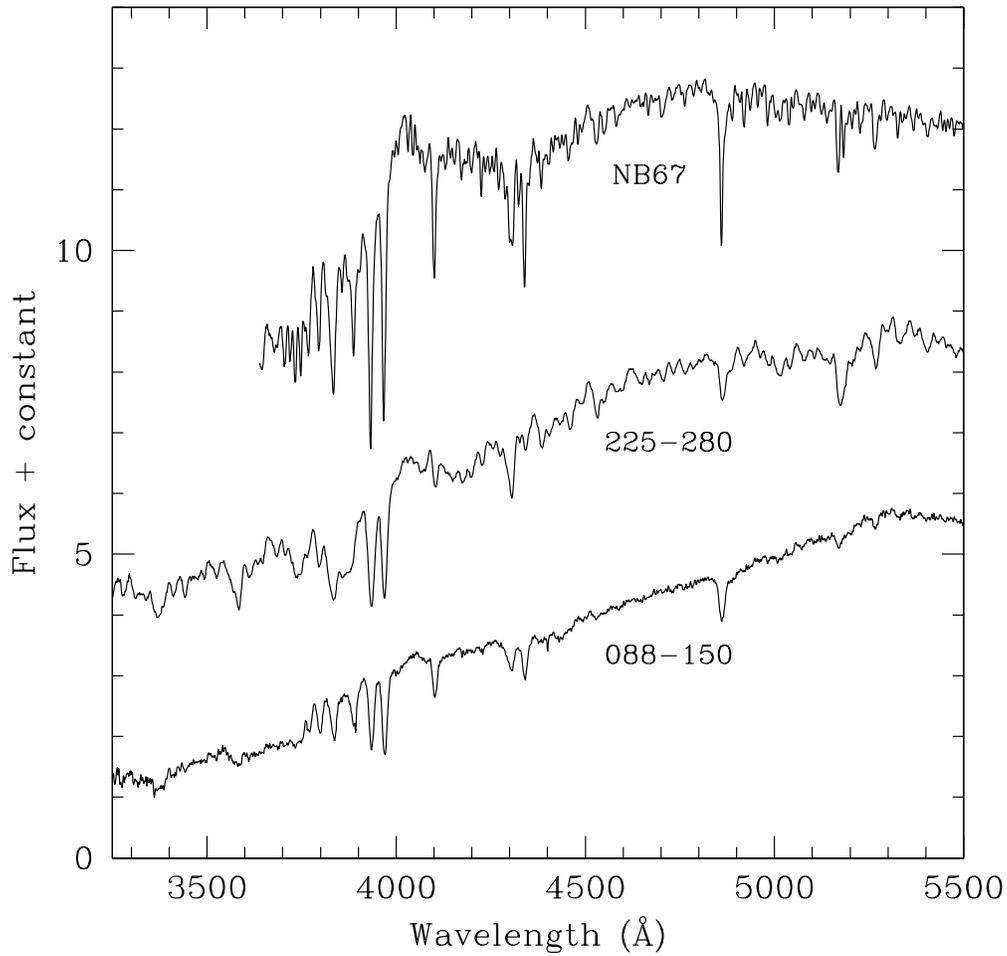}
\caption{Representative fluxed M31 GC spectra from Keck/LRIS: from bottom to top, an old 
metal-poor 
GC (088-150), an old metal-rich GC (225-280), and an intermediate-age, intermediate-metallicity GC 
(NB67). The spectrum of NB67 is from Beasley \etal~(2005), and has been smoothed with a 
5-pixel 
boxcar. The former two spectra are unpublished data of the authors. NB67 has a metallicity $\sim 
-1$ and an age of $\sim 2$ Gyr; its combination of strong Balmer and metal lines distinguishes it 
from the old GCs.}
\label{fig6}
\end{figure}  

\begin{figure}
\epsscale{0.8}
\plotone{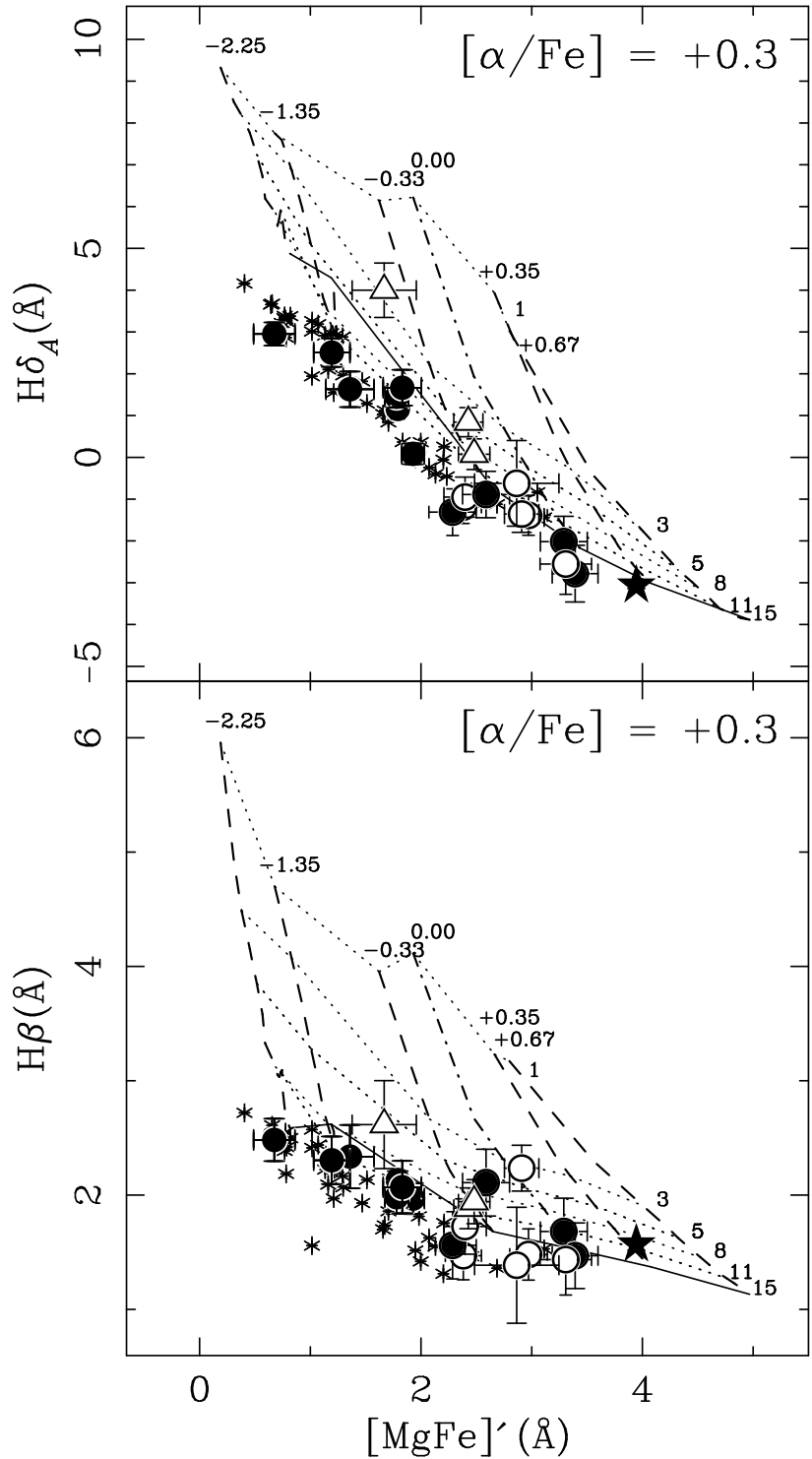}
\caption{H$\delta_{A}$ and H$\beta$ vs.~[MgFe]\arcmin age-metallicity index-index plots for 
GCs in NGC 1407 (circles and triangles; Cenarro \etal~2006) and Galactic GCs (stars, from 
Schiavon \etal~2004). The filled circles are ``normal" GCs, the open circles have anomalously 
high [Mg/Fe] and [C/Fe], and the open triangles are GCs with enhanced Balmer lines (probably 
due to blue horizontal branches, but younger ages are also a possibility). The large star is a 
central $r_{e}/8$ aperture for NGC 1407 itself. The overplotted grids are Thomas, Maraston, \& 
Korn (2004) models with [$\alpha$/Fe]=+0.3 and the ages and metallicities indicated. The model 
grid lines cross at old ages and low metallicities, making exact age estimates impossible.}
\label{fig7}
\end{figure}
 
\begin{figure}
\epsscale{1.0} 
\plotone{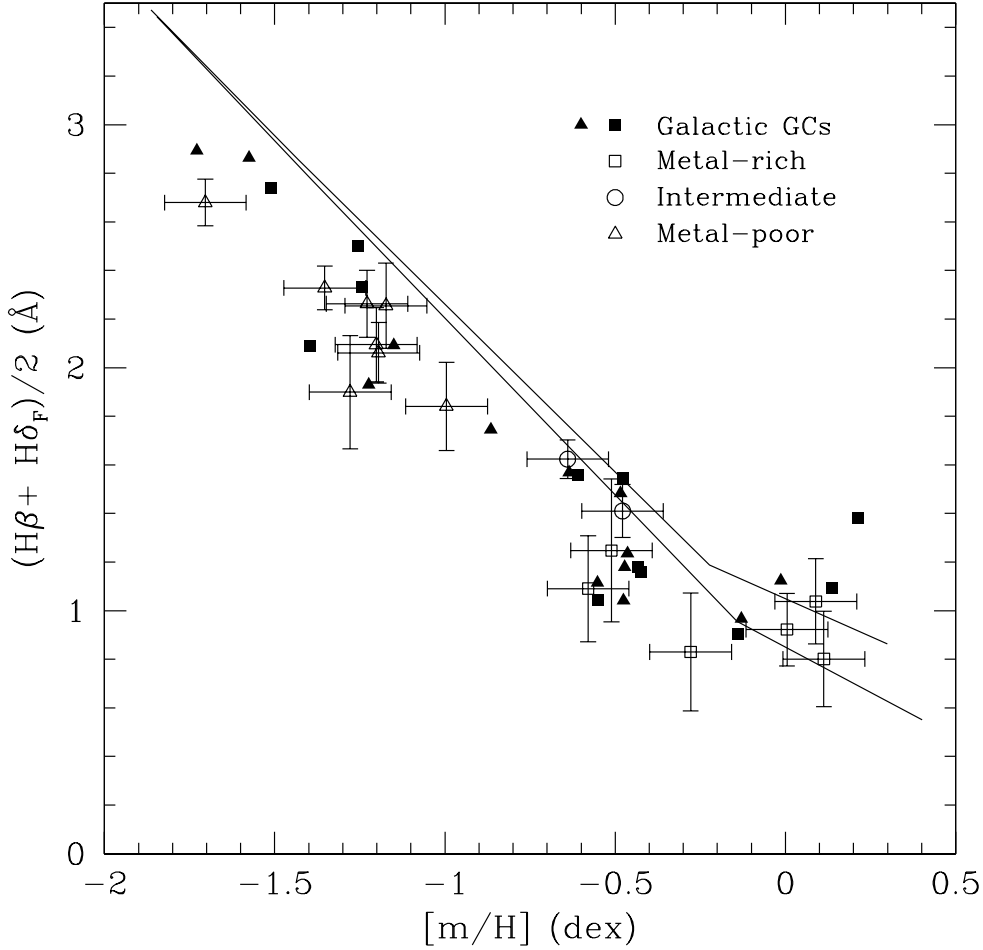}
\caption{Combined Balmer index vs.~metallicity for extragalactic GC subpopulations and Galactic GCs 
(Strader \etal~2005). The extragalactic subpopulations are plotted as open triangles (metal-poor), 
open circles (intermediate-metallicity), and open squares (metal-rich), with the subpopulations 
defined from broadband photometry of the GC systems. Individual Galactic GCs are 
plotted as filled triangles (data from Gregg 1994) and filled squares (data from Puzia 
\etal~2002b). At all metallicities, the extragalactic subpopulations appear coeval with (or older 
than) the comparison Galactic GCs. This suggests mean ages $> 10$--13 Gyr for the 
extragalactic GCs. 14 Gyr model lines from Thomas, Maraston, \& Korn (2004) are superposed 
([$Z$/H] = $-2.25$ to 0, with [$\alpha$/Fe] = 0 on bottom and +0.3 on top). These include a 
blue horizontal branch below [$Z$/H] = $-1.35$. At fixed metallicity, older ages lie at weaker 
Balmer line strength. The data are not fully calibrated to the models, so cannot be directly 
compared, and the offset between the data and the model lines is not significant.}
\label{fig8}
\end{figure}

\begin{figure}
\epsscale{1}
\plotone{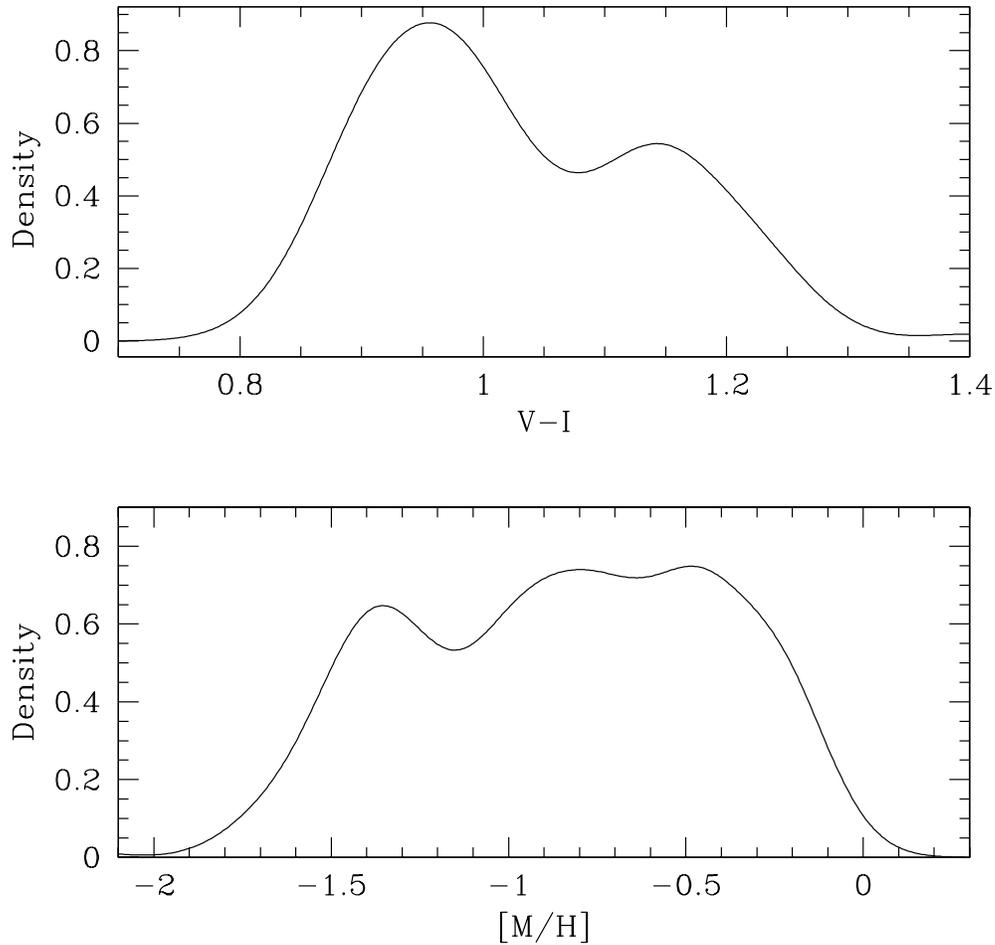}
\caption{Smoothed kernel density histograms in $V-I$ and [m/H] for GCs in NGC 5128. 
The kernels used are 0.03 mag and 0.1 dex, respectively. The 
metallicity histogram shows evidence for three distinct subpopulations of GCs, and a 
comparison of the two histograms indicates the clear nonlinearity of the 
color--metallicity relation (Beasley \etal~2006).}
\label{fig9}
\end{figure}

\begin{figure}
\epsscale{1}
\plotone{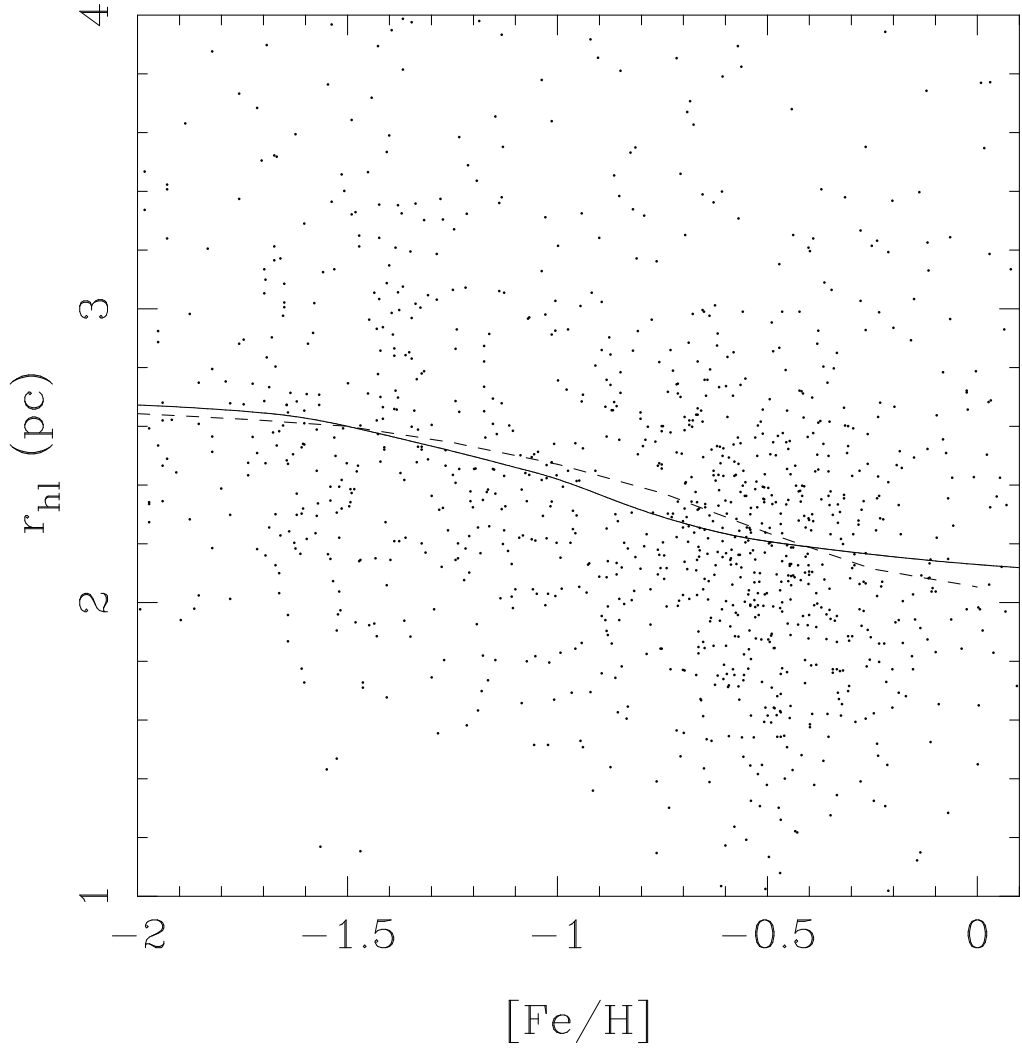}
\caption{Half-light radii of GCs in M87 vs.~[Fe/H] (derived from $g-z$ color). The overplotted 
solid and dashed lines are a running robust mean and a model fit (based upon King models; 
Jord{\'a}n 2004) normalized to the metal-poor GCs. As is typical in massive galaxies, the 
metal-rich GCs are smaller than the metal-poor GCs (figure courtesy A.~Jord{\'a}n). }
\label{fig11}
\end{figure}

\begin{figure}
\epsscale{1}
\plotone{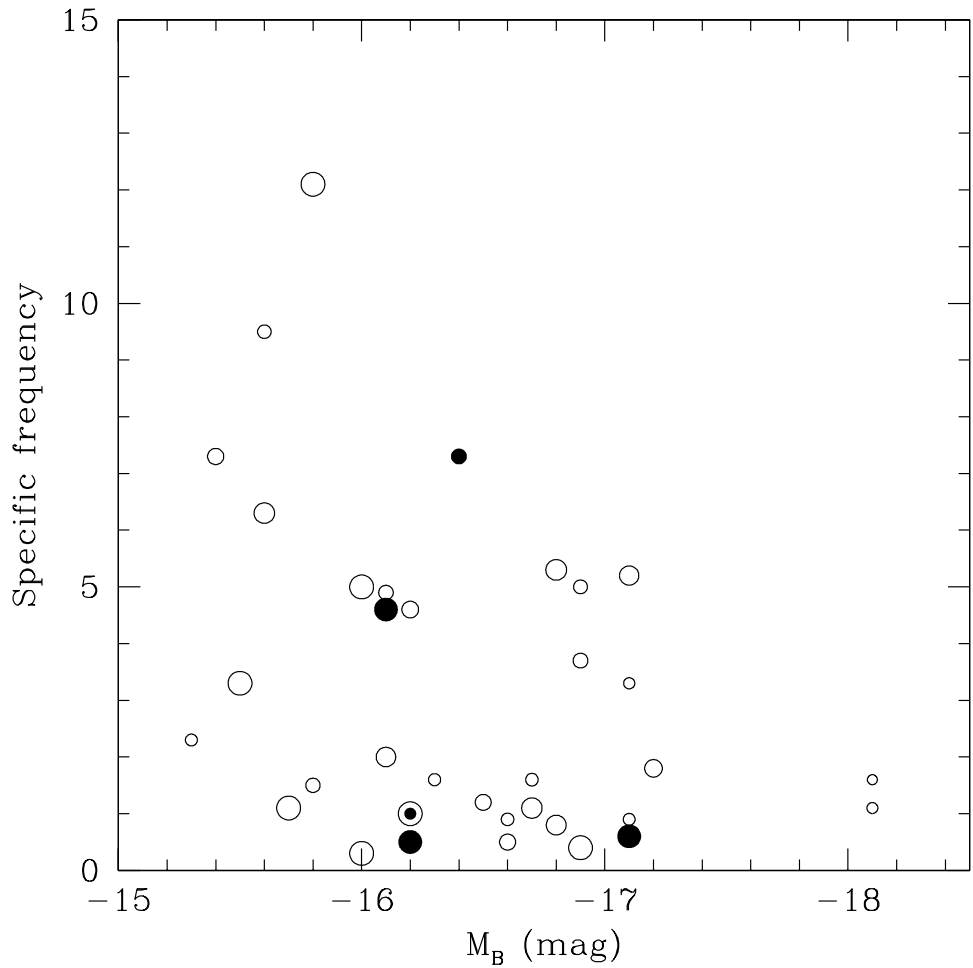}
\caption{Specific frequency ($S_N$) of dEs (open circles) and faint power-law Es (filled 
circles) vs.~parent galaxy $M_B$. The size of the points is proportional to the fraction 
of blue GCs. There is only a weak trend of increasing $S_N$ with decreasing $M_B$, and no 
substantial difference between the two galaxy classes. However, there is some evidence for 
a bimodal distribution of $S_N$ (Strader \etal~2006).}
\label{fig12}
\end{figure}

\begin{figure}
\epsscale{1}
\plotone{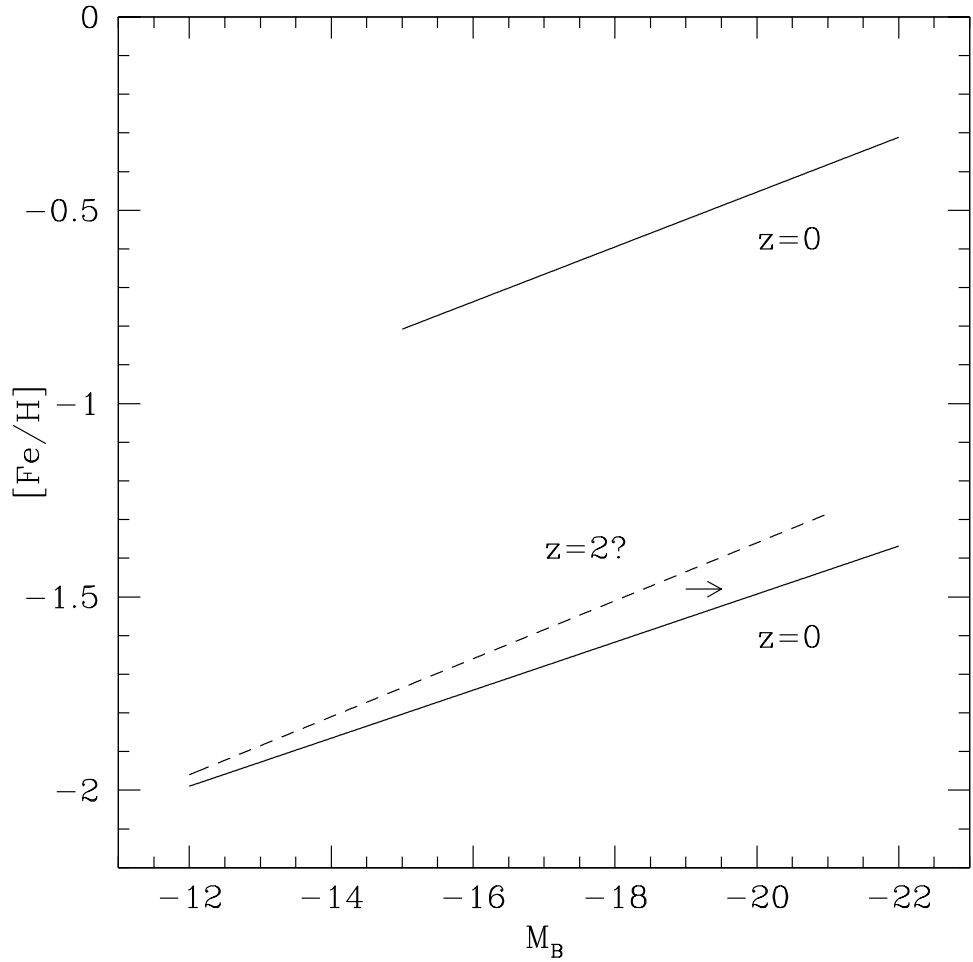}
\caption{A schematic plot of the evolution of the metal-poor GC metallicity--galaxy luminosity relation
due to biased galaxy merging. The solid lines show the $z=0$ relations for both subpopulations;
the dashed line shows a conceptual metal-poor relation at higher 
redshift.}
\label{fig12a}
\end{figure}

\begin{figure}
\epsscale{0.7}
\plotone{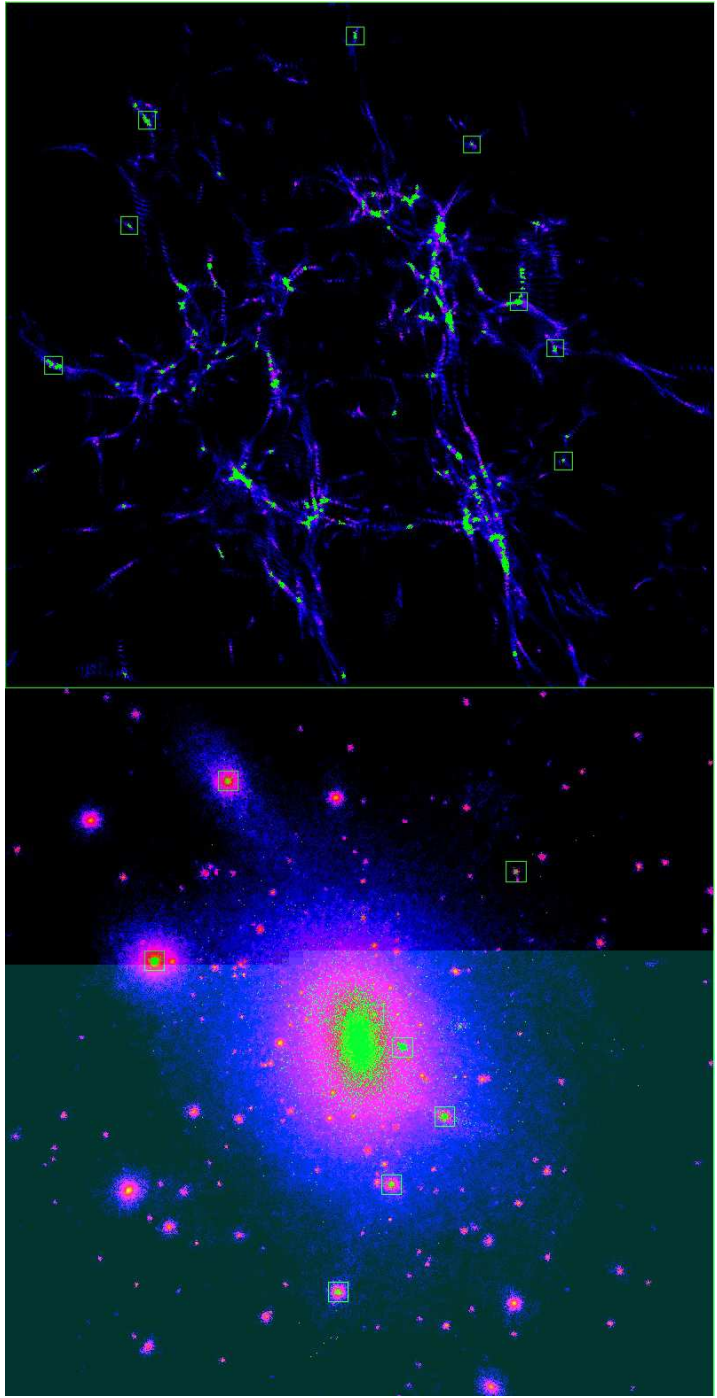}
\caption{Two snapshots of a high-resolution dark matter simulation of the formation of a 
$\sim 10^{12} M_{\odot}$ galaxy (Moore \etal~2006; Diemand \etal~2005). The top panel 
represents the simulation at $z=12$, and the bottom panel represents the present day. The 
blue to pink colors indicate dark matter of increasing density, while the green regions 
are those at $z=12$ with virial temperatures $> 10^4$ K (such that atomic line cooling is 
effective, and gas can cool to form stars). At $z=12$, these green regions represent halos 
with masses $10^8 - 10^{10} M_{\odot}$, and these same green particles are marked in the 
$z=0$ snapshot. These high-$\sigma$ peaks collapse in a filamentary structure at high $z$ 
but are concentrated toward the center of the final galaxy. The boxes can be identified as 
dwarf satellites of the final galaxy.}
\label{fig13}
\end{figure}

\begin{figure}
\epsscale{1}
\plotone{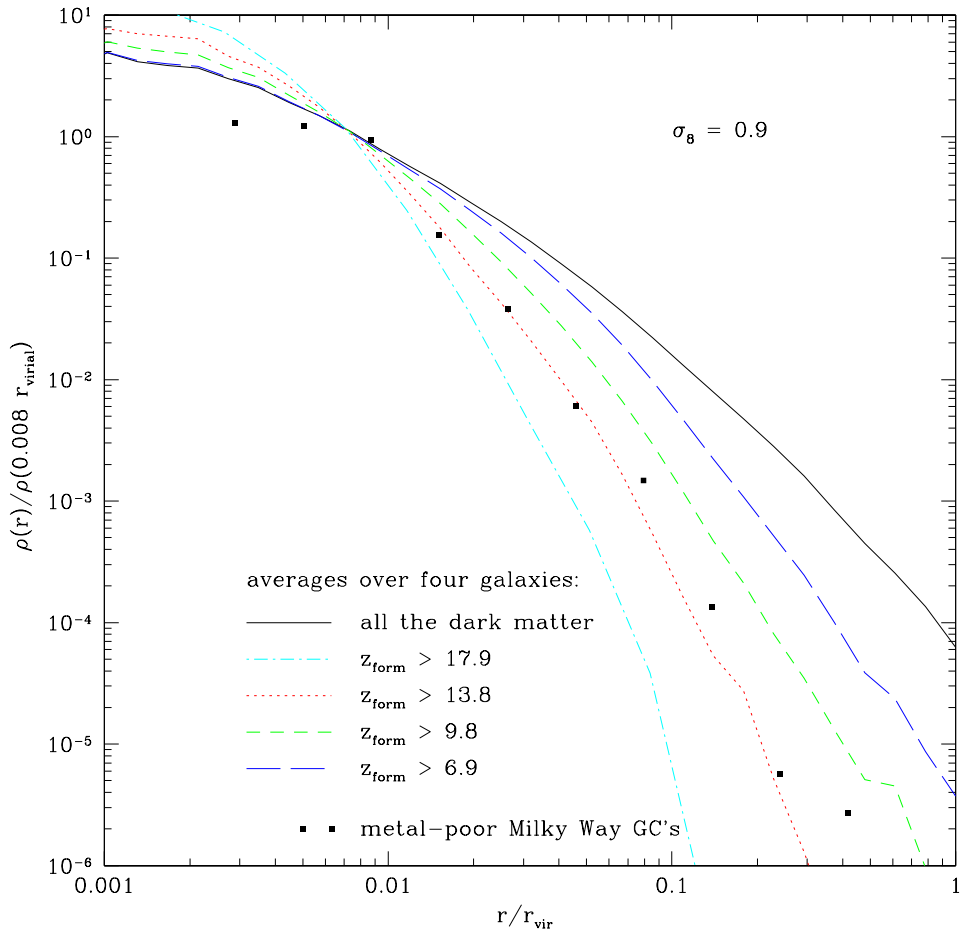}
\caption{The radial distribution of metal-poor GCs in the Galaxy compared to results from 
numerical simulations of the formation of a Galaxy-like dark matter halo in a $\Lambda$CDM 
cosmology (Moore \etal~2006). The lines represent the radial distribution of $2.2 \times 
10^8 M_{\odot}$ mass halos that collapsed before: $z=17.9$ (dot-dashed), $z=13.8$ 
(dotted), $z=9.8$ (short dashed), $z=6.9$ (long dashed), and the cumulative distribution 
(solid). The higher the redshift of collapse, the rarer the peak. Under the assumption 
that reionization truncates star formation in such halos, the comparison suggests 
reionization at $z \sim 12$. This redshift and mass combination corresponds to $\sim 2.5\sigma$ peaks.}
\label{fig14}
\end{figure}

\clearpage
\noindent
{\bf References}\\

\noindent
\nhi Abadi M, Navarro J, Steinmetz M. 2006. astro-ph/0506659\\
\nhi Armandroff TE, Zinn R. 1988. {\it Astron. J.} 96:92\\
\nhi Ashman KM. 1990. {\it  Mon. Not. R. Astron. Soc.} 247:662\\
\nhi Ashman KM, Bird CM. 1993. {\it Astron. J.} 106:2281\\
\nhi Ashman KM, Conti A, Zepf SE.\ 1995. {\it Astron. J.} 110:1164\\ 
\nhi Ashman KM, Zepf SE. 1998. Globular Cluster Systems. New York: Cambridge University Press\\
\nhi Ashman KM, Zepf SE. 1992. {\it Ap. J.} 384:50\\
\nhi Ashman KM, Zepf SE. 2001. {\it Astron. J.} 122:1888\\
\nhi Ashman KM, Walker S, Zepf SE. 2006. in preparation\\
\nhi Barmby P, Huchra JP, Brodie JP, Forbes DA, Schroder LL, Grillmair CJ. 2000. {\it Astron. J.} 119:727\\
\nhi Barmby P, Huchra JP. 2001. {\it Astron. J.} 122:2458\\
\nhi Bassino L, Richtler T, Dirsch B. 2006. {\it  Mon. Not. R. Astron. Soc.} in press\\
\nhi Baumgardt H, Makino J. 2003. {\it  Mon. Not. R. Astron. Soc.} 340:227\\ 
\nhi Beasley MA, Baugh CM, Forbes DA, Sharples RM, Frenk CS. 2002, {\it  Mon. Not. R. Astron. Soc.} 333:383 \\ 
\nhi Beasley MA, Baugh CM, Forbes DA, Sharples RM, Frenk CS. 2002. {\it Mon. Not. R. Astron. Soc.} 333:383\\
\nhi Beasley MA, Brodie JP, Strader J, Forbes DA, Proctor RN, et al. 2004. {\it Astron. J.} 128:1623\\
\nhi Beasley MA, Brodie JP, Strader J, Forbes DA, Proctor RN, et al. 2005. {\it Astron. J.} 129:1412\\
\nhi Bekki K, Beasley MA, Brodie JP, Forbes DA. 2005. {\it Mon. Not. R. Astron. Soc.} 363:1211\\
\nhi Bekki K, Couch WJ, Drinkwater MJ, Shioya Y. 2003. {\it Mon. Not. R. Astron. Soc.} 344:399\\
\nhi Bekki K, Forbes DA, Beasley MA, Couch WJ. 2002. {\it Mon. Not. R. Astron. Soc.} 335:1176\\
\nhi Bekki K, Forbes DA, Beasley MA, Couch WJ. 2003. {\it Mon. Not. R. Astron. Soc.} 344:1334\\
\nhi Bekki K, Forbes DA. 2005. {\it Astron. Astrophys.} in press\\
\nhi Bekki K, Harris WE, Harris GLH. 2003. {\it Mon. Not. R. Astron. Soc.} 338:587\\
\nhi Bell EF, et al.\ 2004. {\it Ap. J.} 608:752 \\
\nhi Bernstein R,  McWilliam A. 2005, astro-ph/0507042\\
\nhi Blakeslee JP, Tonry JL,  Metzger MR.\ 1997. {\it Astron. J.} 114:482 \\
\nhi Brodie JP, Hanes DA. 1986. {\it Ap. J.} 300:258\\
\nhi Brodie JP, Huchra JP. 1990. {\it Ap. J.} 362:503\\
\nhi Brodie JP, Huchra JP. 1991. {\it Ap. J.} 379:157\\
\nhi Brodie JP, Larsen SS, Kissler-Patig M. 2000. {\it Ap. J. Lett.} 543:19\\
\nhi Brodie JP, Larsen SS. 2002. {\it Astron. J.} 124:1410\\
\nhi Brodie JP, Schroder LL, Huchra JP, Phillips AC, Kissler-Patig M, Forbes DA. 1998. {\it Astron. J.} 116:691\\
\nhi Brodie JP, Strader J, Denicoló G, Beasley MA, Cenarro AJ, et al. 2005. {\it Astron. J.} 129:2643\\
\nhi Brodie JP. 1981. Ph.D. Thesis\\
\nhi Brown TM, Ferguson HC, Smith E, Kimble RA, Sweigart AV, Renzini A, Rich RM, VandenBerg DA.\ 2003. {\it Ap. J. Lett.} 592:17\\ 
\nhi Brown TM, Ferguson HC, Smith E, Kimble RA, Sweigart AV, Renzini A, Rich RM, VandenBerg DA.\ 2004. {\it Ap. J. Lett.} 613:125\\
\nhi Bruzual G, Charlot S. 2003. {\it Mon. Not. R. Astron. Soc.} 344:1000\\
\nhi Burgarella D, Kissler-Patig M,  Buat V.\ 2001. {\it Astron. J.} 121:2647\\ 
\nhi Burkert A, Smith GH. 2000. {\it Ap. J.} 542:95\\
\nhi Burkert A, Brodie J, Larsen S. 2005. {\it Ap. J.} 628:231\\
\nhi Burstein D, Faber SM, Gaskell CM, Krumm N. 1984. {\it Ap. J.} 287:586\\
\nhi Burstein D, Heiles C. 1984. {\it Ap. J. Suppl.} 54:33\\
\nhi Burstein D, Li Y, Freeman KC, Norris JE, Bessell MS, et al. 2004. {\it Ap. J.} 614:158\\
\nhi Burstein D. 1987. In Nearly Normal Galaxies: From the Planck Time to the Present; Proceedings of the Eighth Santa Cruz Summer Workshop in 
Astronomy and Astrophysics, ed. S Faber, p. 47 New York: Springer-Verlag\\
\nhi Carney BW, Latham DW, Laird JB. 1990. {\it Astron. J.} 99:572\\
\nhi Carney BW.\ 1996. {\it Publ. Astron. Soc. Pac.} 108:900\\ 
\nhi Carretta E, Gratton RG. 1997. {\it Astron. Ap. Suppl.} 121:95\\
\nhi Cenarro AJ, et al. 2006. in preparation\\
\nhi Chaboyer B, Demarque P, Sarajedini A. 1996. {\it Ap. J.} 459:558\\
\nhi Chandar R, Bianchi L, Ford HC, Sarajedini A. 2002. {\it Ap. J.} 564:712\\
\nhi Chandar R, Whitmore B,  Lee MG.\ 2004. {\it Ap. J.} 611:220 \\
\nhi Cohen JG, Blakeslee JP,  Ryzhov A.\ 1998. {\it Ap. J.} 496:808\\ 
\nhi Cohen JG, Blakeslee JP, Côté P. 2003. {\it Ap. J.} 592:866\\
\nhi Cohen JG, Blakeslee JP, Ryzhov A. 1998. {\it Ap. J.} 496:808\\
\nhi Cohen JG, Blakeslee JP. 1998. {\it Astron. J.} 115:2356\\
\nhi Cohen JG, Matthews K,  Cameron PB.\ 2005. {\it Ap. J. Lett.} 634:45\\ 
\nhi Couture J, Harris WE, Allwright JWB. 1991. {\it Ap. J.} 372:97\\
\nhi Cowie LL, Songaila A, Hu EM,  Cohen JG.\ 1996. {\it Astron. J.} 112:839 \\
\nhi C{\^o}t{\'e} P, Blakeslee JP, Ferrarese L, Jord{\'a}n A, Mei S, et al. 2004. {\it Ap. J. Suppl.} 153:223\\
\nhi C{\^o}t{\'e} P, Marzke RO, West MJ, Minniti D. 2000. {\it Ap. J.} 533:869\\
\nhi C{\^o}t{\'e} P, Marzke RO, West MJ. 1998. {\it Ap. J.} 501:554\\
\nhi C{\^o}t{\'e} P, McLaughlin DE, Cohen JG, Blakeslee JP. 2003. {\it Ap. J.} 591:850\\
\nhi C{\^o}t{\'e} P, McLaughlin DE, Hanes DA, Bridges TJ, Geisler D, et al. 2001. {\it Ap. J.} 559:828\\
\nhi C{\^o}t{\'e} P, West MJ, Marzke RO. 2002. {\it Ap. J.} 567:853\\
\nhi C{\^o}t{\'e} P. 1999. {\it Astron. J.} 118:406\\
\nhi Davies RL, et al.\ 2001. {\it Ap. J. Lett.} 548:33 \\
\nhi Dekel A,  Birnboim Y. 2006. astro-ph/0412300\\
\nhi Dekel A, Stoehr F, Mamon GA, Cox TJ, Novak GS, Primack JR. 2005. {\it Nature} 437:707\\
\nhi Diemand J, Madau P, Moore B. 2005. {\it Mon. Not. R. Astron. Soc.} 364:367\\
\nhi Dirsch B, Richtler T, Bassino LP. 2003. {\it Astron. Astrophys.} 408:929\\
\nhi Dirsch B, Richtler T, Geisler D, Forte JC, Bassino LP, Gieren WP. 2003. {\it Astron. J.} 125:1908\\
\nhi Dirsch B, Schuberth Y, Richtler T. 2005. {\it Astron. Astrophys.} 433:43\\
\nhi Dressler A, Oemler A Jr , Couch WJ, Smail I, Ellis RS, et al. 1997. {\it Ap. J.} 490:577\\
\nhi Durrell PR, Harris WE,  Pritchet CJ.\ 2001. {\it Astron. J.} 121:2557 \\
\nhi Durrell PR, Harris WE, Geisler D, Pudritz RE. 1996. {\it Astron. J.} 112:972\\
\nhi Durrell PR, Harris WE, Pritchet CJ. 1994. {\it Astron. J.} 108:2114\\
\nhi Durrell PR, McLaughlin DE, Harris WE, Hanes DA. 1996. {\it Ap. J.} 463:543\\
\nhi Elmegreen BG, Efremov YN.\ 1997. {\it Ap. J.} 480, 235 \\
\nhi Evans NW, Wilkinson MI. 2000. {\it Mon. Not. R. Astron. Soc.} 316:929\\
\nhi Faber SM, et al. 2005, astro-ph/0506044\\
\nhi Faber SM, Tremaine S, Ajhar EA, Byun Y-I, Dressler A, et al. 1997. {\it Astron. J.} 114:1771\\
\nhi Fall SM,  Rees MJ.\ 1985. {\it Ap. J.} 298:18 \\
\nhi Fall SM,  Rees MJ.\ 1988, IAU Symp126: The Harlow-Shapley Symposium on Globular Cluster Systems in Galaxies, 126:323 \\
\nhi Fall SM, Zhang Q. 2001. {\it Ap. J.} 561:751\\
\nhi Forbes DA, Beasley MA, Brodie JP, Kissler-Patig M. 2001. {\it Ap. J. Lett.} 563:143\\
\nhi Forbes DA, Brodie JP, Grillmair CJ. 1997. {\it Astron. J.} 113:1652\\
\nhi Forbes DA, Brodie JP, Huchra J. 1996. {\it Astron. J.} 112:2448\\
\nhi Forbes DA, Brodie JP, Larsen SS. 2001. {\it Ap. J. Lett.} 556:83\\
\nhi Forbes DA, Forte JC. 2001. {\it Mon. Not. R. Astron. Soc.} 322:257\\
\nhi Forbes DA, Franx M, Illingworth GD, Carollo CM. 1996. {\it Ap. J.} 467:126\\
\nhi Forbes DA, Georgakakis AE, Brodie JP. 2001. {\it Mon. Not. R. Astron. Soc.} 325:1431\\
\nhi Forbes DA, Masters KL, Minniti D, Barmby P. 2000. {\it Astron. Astrophys.} 358:471\\
\nhi Forbes DA, Strader J, Brodie JP. 2004. {\it Astron. J.} 127:3394\\
\nhi Forbes DA. 2005, astro-ph/0511291\\
\nhi Forte JC, Faifer F, Geisler D. 2005. {\it Mon. Not. R. Astron. Soc.} 357:56\\
\nhi Forte JC, Geisler D, Ostrov PG, Piatti AE, Gieren W. 2001. {\it Astron. J.} 121:1992\\
\nhi Fukugita, M, Hogan CJ,  Peebles PJE.\ 1998. {\it Ap. J.} 503:518\\ 
\nhi Fulbright J, McWilliam A,  Rich M. 2006. astro-ph/0510408\\
\nhi Fusi Pecci F, Bellazzini M, Buzzoni A, De Simone E, Federici L, Galleti S. 2005. {\it Astron. J.} 130:554\\
\nhi Gebhardt K, Kissler-Patig M. 1999. {\it Astron. J.} 118:1526\\
\nhi Geisler D, Lee MG, Kim E. 1996. {\it Astron. J.} 111:1529\\
\nhi Goerdt T, Moore B, Read JI, Stadel J, Zemp M. 2006, astro-ph/0601404\\
\nhi Gomez M, Richtler T. 2004. {\it Astron. Astrophys.} 415:499\\
\nhi Goudfrooij P, Alonso MV, Maraston C, Minniti D. 2001. {\it Mon. Not. R. Astron. Soc.} 328:237\\
\nhi Goudfrooij P, Gilmore D, Whitmore BC, Schweizer F. 2004. {\it Ap. J. Lett.} 613:121\\
\nhi Goudfrooij P, Mack J, Kissler-Patig M, Meylan G, Minniti D. 2001. {\it Mon. Not. R. Astron. Soc.} 322:643\\
\nhi Goudfrooij P, Strader J, Brenneman L, Kissler-Patig M, Minniti D, Edwin Huizinga J. 2003. {\it Mon. Not. R. Astron. Soc.} 343:665\\
\nhi Gratton R, Sneden C, Carretta E. 2004. Annu. Rev. {\it Astron. Astrophys.} 42:385\\
\nhi Grebel EK, Dolphin AE, Guhathakurta P. 2000. Astron. Gesellschaft Abstract Ser. 17:79 (Abstr.)\\
\nhi Gregg MD. 1994. {\it Astron. J.} 108:2164\\
\nhi Guhathakurta P, Ostheimer JC, Gilbert KM, Rich RM, Majewski SR, et al. 2005. astro-ph/0502366\\
\nhi Hansen J,  Moore B. 2006. astro-ph/0411473\\
\nhi Harris GLH, Geisler D, Harris HC,  Hesser JE.\ 1992. {\it Astron. J.} 104:613 \\
\nhi Harris GLH, Harris WE, Geisler D. 2004. {\it Astron. J.} 128:723\\
\nhi Harris GLH, Harris WE, Poole GB. 1999. {\it Astron. J.} 117:855\\
\nhi Harris WE, Harris GLH, Holland ST, McLaughlin DE. 2002. {\it Astron. J.} 124:1435\\
\nhi Harris WE, Harris GLH. 2001. {\it Astron. J.} 122:3065\\
\nhi Harris WE, Harris GLH. 2002. {\it Astron. J.} 123:3108\\
\nhi Harris WE, Kavelaars JJ, Hanes DA, Hesser JE, Pritchet CJ. 2000. {\it Ap. J.} 533:137\\
\nhi Harris WE, Pudritz RE. 1994. {\it Ap. J.} 429:177\\
\nhi Harris WE, van den Bergh S. 1981. {\it Astron. J.} 86:1627\\
\nhi Harris WE. 1986. {\it Astron. J.} 91:822\\
\nhi Harris WE. 1991. {\it Annu. Rev. Astron. Astrophys.} 29:543\\
\nhi Harris WE. 2001. Saas-Fee Advanced Course 28: Star Clusters 28:223\\
\nhi Harris WE.\ 2003, Extragalactic Globular Cluster Systems, 317\\ 
\nhi Hempel M, Hilker M, Kissler-Patig M, Puzia TH, Minniti D, Goudfrooij P. 2003. {\it Astron. Astrophys.} 405:487\\
\nhi Hempel M, Kissler-Patig M. 2004. {\it Astron. Astrophys.} 419:863\\
\nhi Hempel M, Kissler-Patig M. 2004. {\it Astron. Astrophys.} 428:459\\
\nhi Hilker M, Infante L, Richtler T. 1999. {\it Astron. Astrophy. Suppl.} 138:55\\
\nhi Hilker M. 1998. Ph.D. Thesis\\
\nhi Hodge PW, Dolphin AE, Smith TR, Mateo M. 1999. {\it Ap. J.} 521:577\\
\nhi Holtzman JA, et al.\ 1992. {\it Astron. J.} 103:691 \\
\nhi Huchra JP, Brodie JP, Kent SM. 1991. {\it Ap. J.} 370:495\\
\nhi Huxor AP, Tanvir NR, Irwin MJ, Ibata R, Collett JL, et al. 2005. {\it Mon. Not. R. Astron. Soc.} 360:1007\\
\nhi Hwang N, et al.\ 2005, IAU Colloq198: Near-fields cosmology with dwarf elliptical galaxies, 257 \\
\nhi Hwang N, Lee MG. 2006, astro-ph/0601280\\
\nhi Jord{\'a}n A. 2004. {\it Ap. J. Lett.} 613:117\\
\nhi Karachentsev ID, Karachentseva VE, Dolphin AE, Geisler D, Grebel EK, et al. 2000. {\it Astron. Astrophys.} 363:117\\
\nhi Karachentsev ID, Sharina ME, Grebel EK, Dolphin AE, Geisler D, et al. 2000. {\it Ap. J.} 542:128\\
\nhi Kent SM. 1992. {\it Ap. J.} 387:181\\
\nhi Kissler-Patig M, Ashman KM, Zepf SE,  Freeman KC.\ 1999. {\it Astron. J.} 118:197 \\
\nhi Kissler-Patig M, Forbes DA, Minniti D. 1998. {\it Mon. Not. R. Astron. Soc.} 298:1123\\
\nhi Kissler-Patig M. 2000. In Reviews in Modern Astronomy 13 : New Astrophysical Horizons, ed. RE Schielicke, p. 13. Hamburg, Germany: Astronomische Gesellschaft\\
\nhi Kormendy J. 1985. {\it Ap. J.} 295:73\\
\nhi Kormendy J. 1987. Nearly Normal Galaxies.~From the Planck Time to the Present, 163 \\
\nhi Kormendy J, Bender R. 1996. {\it Ap. J.} 464:119\\
\nhi Kormendy J, Kennicutt RC Jr. 2004. Annu. Rev. Astron. Astrophys. 42:603\\
\nhi Kormendy J, et al. 2006. in preparation\\
\nhi Korn AJ, Maraston C, Thomas D. 2005. {\it Astron. Astrophys.} 438:685\\
\nhi Kraft RP, Ivans II. 2003.  {\it Publ. Astron. Soc. Pac.} 115:804\\
\nhi Kravtsov AV,  Gnedin OY.\ 2005. {\it Ap. J.} 623:650 \\
\nhi Kundu A, Whitmore BC, Sparks WB, Macchetto FD, Zepf SE, Ashman KM. 1999. {\it Ap. J.} 513:733\\
\nhi Kundu A, Whitmore BC. 1998. {\it Astron. J.} 116:2841\\
\nhi Kundu A, Whitmore BC. 2001. {\it Astron. J.} 121:2950\\
\nhi Kundu A, Whitmore BC. 2001. {\it Astron. J.} 122:1251\\
\nhi Kundu A, et al.\ 2005. {\it Ap. J. Lett.} 634:41 \\
\nhi Kuntschner H, Ziegler BL, Sharples RM, Worthey G, Fricke KJ. 2002. {\it Astron. Astrophys.} 395:761\\
\nhi Larsen SS, Brodie JP,  Hunter DA.\ 2004. {\it Astron. J.} 128:2295 \\
\nhi Larsen SS, Brodie JP,  Strader J.\ 2005. {\it Astron. Astrophys.} 443:413\\
\nhi Larsen SS, Brodie JP, Beasley MA, Forbes DA, Kissler-Patig M, et al. 2003. {\it Ap. J.} 585:767\\
\nhi Larsen SS, Brodie JP, Beasley MA, Forbes DA. 2002. {\it Astron. J.} 124:828\\
\nhi Larsen SS, Brodie JP, Huchra JP, Forbes DA, Grillmair CJ. 2001. {\it Astron. J.} 121:2974\\
\nhi Larsen SS, Brodie JP, Sarajedini A, Huchra JP. 2002. {\it Astron. J.} 124:2615\\
\nhi Larsen SS, Brodie JP. 2000. {\it Astron. J.} 120:2938\\
\nhi Larsen SS, Brodie JP. 2002. {\it Astron. J.} 123:1488\\
\nhi Larsen SS, Brodie JP. 2003. {\it Ap. J.} 593:340\\
\nhi Larsen SS, Forbes DA, Brodie JP. 2001. {\it Mon. Not. R. Astron. Soc.} 327:1116\\
\nhi Larsen SS, Richtler T. 2000. {\it Astron. Astrophys.} 354:836\\
\nhi Law DR, Johnston KV, Majewski SR. 2005. {\it Ap. J.} 619:807\\
\nhi Layden AC, Sarajedini A. 2000. {\it Astron. J.} 119:1760\\
\nhi Lynden-Bell D.\ 1975. {\it Vistas in Astronomy} 19:299 \\
\nhi Majewski SR, Patterson RJ, Dinescu DI, Johnson WY, Ostheimer JC, et al. 2000. In Proceedings of the 35th Liege International Astrophysics Colloquium : The Galactic Halo; From Globular Cluster to Field Stars, eds. A Noels, P Magain, D Caro, E Jehin, G Parmentier, AA Thoul, pp. 61926. Liege, Belgium: Institut d'Astrophysique et de Geophysique\\
\nhi Maraston C, Bastian N, Saglia RP, Kissler-Patig M, Schweizer F,  Goudfrooij P.\ 2004. {\it Astron. Astrophys.} 416:467 \\
\nhi Maraston C. 2005. {\it Mon. Not. R. Astron. Soc.} 362:799\\
\nhi Martin NF, Ibata RA, Bellazzini M, Irwin MJ, Lewis GF,  Dehnen W.\ 2004. {\it Mon. Not. R. Astron. Soc.} 348:12 \\
\nhi Mashchenko S,  Sills A.\ 2004. {\it Ap. J. Lett.} 605:121 \\
\nhi Mashchenko S,  Sills A.\ 2005. {\it Ap. J.} 619:243 (2005a)\\
\nhi Mashchenko S,  Sills A.\ 2005. {\it Ap. J.} 619:258 (2005b)\\
\nhi Mashchenko S, Couchman H,  Sills A. astro-ph/0511361\\
\nhi Matteucci F. 1994. {\it Astron. Astrophys.} 288:57\\
\nhi McCrady N, Gilbert AM, Graham JR. 2003. {\it Ap. J.} 596:240\\
\nhi McCrady N, Graham JR, Vacca WD. 2005. {\it Ap. J.} 621:278\\
\nhi McLaughlin DE. 1994. {\it Publ. Astron. Soc. Pac.} 106:47\\
\nhi McLaughlin DE. 1999. {\it Astron. J.} 117:2398\\
\nhi Meylan G, Heggie DC. 1997. {\it Astron. Astrophys. Rev.} 8:1\\
\nhi Mighell KJ, Sarajedini A, French RS. 1998. {\it Astron. J.} 116:2395\\
\nhi Miller BW, Lotz JM, Ferguson HC, Stiavelli M,  Whitmore BC.\ 1998. {\it Ap. J. Lett.} 508:133 \\
\nhi Miller BW, Whitmore BC, Schweizer F,  Fall SM.\ 1997. {\it Astron. J.} 114:2381 \\
\nhi Minniti D, Meylan G, Kissler-Patig M. 1996. {\it Astron. Astrophys.} 312:49\\
\nhi Minniti D. 1995. {\it Astron. J.} 109:1663\\
\nhi Moore B, Katz N, Lake G, Dressler A, Oemler A Jr. 1996. {\it Nature} 379:613\\
\nhi Moore B, Lake G, Katz N. 1998. {\it Ap. J.} 495:139\\
\nhi Moore B.\ 1996. {\it Ap. J. Lett.} 461:13 \\
\nhi Moore, B.\ et al. 2006. astro-ph/0510370\\
\nhi Morrison HL, Harding P, Perrett K,  Hurley-Keller D.\ 2004. {\it Ap. J.} 603:87 \\
\nhi Mould J, Kristian J. 1986. {\it Ap. J.} 305:591\\
\nhi Muzzio JC. 1987. {\it Publ. Astron. Soc. Pac.} 99:245\\
\nhi Olsen KAG, Miller BW, Suntzeff NB, Schommer RA,  Bright J.\ 2004. {\it Astron. J.} 127:2674 \\
\nhi Ostrov P, Geisler D, Forte JC. 1993. {\it Astron. J.} 105:1762\\
\nhi Ostrov PG, Forte JC, Geisler D. 1998. {\it Astron. J.} 116:2854\\
\nhi Peng EW, Ford HC, Freeman KC. 2004. {\it Ap. J.} 602:685\\
\nhi Peng EW, Ford HC, Freeman KC. 2004. {\it Ap. J.} 602:705\\
\nhi Peng EW, Jord{\'a}n A, Côté P, Blakeslee JP, Ferrarese L, et al. 2006. astro-ph/0509654 \\
\nhi Peng EW, et al. 2006. astro-ph/0511251 \\
\nhi Perrett KM, Bridges TJ, Hanes DA, Irwin MJ, Brodie JP, et al. 2002. {\it Astron. J.} 123:249\\
\nhi Perrett KM, Stiff DA, Hanes DA, Bridges TJ. 2003. {\it Ap. J.} 589:790\\
\nhi Phillipps S, Drinkwater MJ, Gregg MD, Jones JB.\ 2001. {\it Ap. J.} 560:201 \\
\nhi Pierce M, Brodie JP, Forbes DA, Beasley MA, Proctor R, Strader J. 2005. {\it Mon. Not. R. Astron. Soc.} 358:419\\
\nhi Proctor RN, Forbes DA, Beasley MA. 2004. {\it Mon. Not. R. Astron. Soc.} 355:1327\\
\nhi Proctor RN, Sansom AE. 2002. {\it Mon. Not. R. Astron. Soc.} 333:517\\
\nhi Puzia TH, Kissler-Patig M, Brodie JP, Huchra JP. 1999. {\it Astron. J.} 118:2734\\
\nhi Puzia TH, Kissler-Patig M, Brodie JP, Schroder LL.\ 2000. {\it Astron. J.} 120:777 \\
\nhi Puzia TH. 2003, Ph.D. Thesis\\
\nhi Puzia TH, Kissler-Patig M, Thomas D, Maraston C, Saglia RP, et al. 2004. {\it Astron. Astrophys.} 415:123\\
\nhi Puzia TH, Kissler-Patig M, Thomas D, Maraston C, Saglia RP, et al. 2005. {\it Astron. Astrophys.} 439:997\\
\nhi Puzia TH, Perrett KM, Bridges TJ. 2005. {\it Astron. Astrophys.} 434:909\\
\nhi Puzia TH, Saglia RP, Kissler-Patig M, Maraston C, Greggio L, et al. 2002. {\it Astron. Astrophys.} 395:45\\
\nhi Puzia TH, Zepf SE, Kissler-Patig M, Hilker M, Minniti D, Goudfrooij P. 2002. {\it Astron. Astrophys.} 391:453\\
\nhi Read JI, Wilkinson MI, Evans NW, Gilmore G, Kleyna JT. 2006. astro-ph/0511759\\
\nhi Rejkuba M, Greggio L, Harris WE, Harris GLH, Peng EW. 2005. {\it Ap. J.} 631:262\\
\nhi Rejkuba M, Minniti D, Silva DR, Bedding TR. 2003. {\it Astron. Astrophys.} 411:351\\
\nhi Rhode KL, Zepf SE, Santos MR. 2005. {\it Ap. J. Lett.} 630:21\\
\nhi Rhode KL, Zepf SE. 2001. {\it Astron. J.} 121:210\\
\nhi Rhode KL, Zepf SE. 2003. {\it Astron. J.} 126:2307\\
\nhi Rhode KL, Zepf SE. 2004. {\it Astron. J.} 127:302\\
\nhi Rich RM, Corsi CE, Cacciari C, Federici L, Fusi Pecci F, Djorgovski SG, Freedman WL.\ 2005. {\it Astron. J.} 129:2670 \\
\nhi Richtler T. 2006. astro-ph/0512545 \\
\nhi Richtler T, Dirsch B, Gebhardt K, Geisler D, Hilker M, et al. 2004. {\it Astron. J.} 127:2094\\
\nhi Romanowsky AJ, Douglas NG, Arnaboldi M, Kuijken K, Merrifield MR, et al. 2003. {\it Science} 301:1696\\
\nhi Rose JA. 1985. {\it Astron. J.} 90:1927\\
\nhi Sadler EM, Rich RM, Terndrup DM. 1996. {\it Astron. J.} 112:171\\
\nhi Saitoh et al. 2005. {\it Astron. J.} in press\\
\nhi Santos MR.\ 2003. Extragalactic Globular Cluster Systems, 348 \\
\nhi Sarajedini A, Geisler D, Schommer R, Harding P. 2000. {\it Astron. J.} 120:2437\\
\nhi Sarajedini A, Jablonka P. 2005. {\it Astron. J.} 130:1627\\
\nhi Schiavon R. 2006. ApJS, submitted\\
\nhi Schiavon RP, Faber SM, Castilho BV, Rose JA. 2002. {\it Ap. J.} 580:850\\
\nhi Schiavon RP, Rose JA, Courteau S, MacArthur LA. 2004. {\it Ap. J. Lett.} 608:33\\
\nhi Schroder LL, Brodie JP, Kissler-Patig M, Huchra JP, Phillips AC.\ 2002. {\it Astron. J.} 123:2473 \\
\nhi Schuberth, et al. 2006. in preparation\\
\nhi Schommer RA, Olszewski EW, Suntzeff NB, Harris HC. 1992. {\it Astron. J.} 103:447\\
\nhi Schweizer F. 1987. In Nearly Normal Galaxies: From the Planck Time to the Present; Proceedings of the Eighth Santa Cruz Summer Workshop in Astronomy and Astrophysics, ed. S Faber, p. 18. New York: Springer-Verlag\\
\nhi Searle L, Wilkinson A, Bagnuolo WG. 1980. {\it Ap. J.} 239:803\\
\nhi Searle L, Zinn R. 1978. {\it Ap. J.} 225:357\\
\nhi Secker J, Geisler D, McLaughlin DE, Harris WE. 1995. {\it Astron. J.} 109:1019\\
\nhi Shapley H. 1939. {\it Proc. Nat. Acad. Sci.} 25:423\\
\nhi Shapley H. 1939. {\it Proc. Nat. Acad. Sci.} 25:565\\
\nhi Sharina ME, Sil'chenko OK, Burenkov AN. 2003. {\it Astron. Astrophys.} 397:831\\
\nhi Smith GH, Burkert A. 2002. {\it Ap. J. Lett.} 578:51\\
\nhi Smith LJ, Gallagher JS. 2001. {\it Mon. Not. R. Astron. Soc.} 326:1027\\
\nhi Sohn ST. et al. 2005. astro-ph/0510413\\
\nhi Soria R, Mould JR, Watson AM, Gallagher JS III, Ballester GE, et al. 1996. {\it Ap. J.} 465:79\\
\nhi Spitzer L. 1987. Dynamical Evolution of Globular Clusters. Princeton, NJ: Princeton University Press\\
\nhi Stetson PB, Vandenberg DA, Bolte M. 1996. {\it Publ. Astron. Soc. Pac.} 108:560\\
\nhi Strader J, Brodie JP, Schweizer F, Larsen SS, Seitzer P. 2003. {\it Astron. J.} 125:626\\
\nhi Strader J, Brodie JP, Forbes DA, Beasley MA, Huchra JP. 2003. {\it Astron. J.} 125:1291\\
\nhi Strader J, Brodie JP, Forbes DA. 2004. {\it Astron. J.} 127:295\\
\nhi Strader J, Brodie JP, Forbes DA. 2004. {\it Astron. J.} 127:3431\\
\nhi Strader J, Brodie JP. 2004. {\it Astron. J.} 128:1671\\
\nhi Strader J, Brodie JP, Cenarro AJ, Beasley MA, Forbes DA. 2005. {\it Astron. J.} 130:1315\\
\nhi Strader J, Brodie J, Beasley M, Spitler L. 2006. astro-ph/0508001 \\
\nhi Tantalo R, Chiosi C, Bressan A. 1998. {\it Astron. Astrophys.} 333:419\\
\nhi Thomas D, Maraston C, Bender R. 2003. {\it Mon. Not. R. Astron. Soc.} 339:897\\
\nhi Thomas D, Maraston C, Bender R, de Oliveira CM. 2005, {\it Ap. J.} 621:673\\
\nhi Toomre A, Toomre J. 1972. {\it Ap. J.} 178:623\\
\nhi Toomre A. 1977. In Evolution of Galaxies and Stellar Populations, Proceedings of a Conference at Yale University, eds. BM Tinsley, RB Larson, p. 40. New Haven: Yale University Observatory\\
\nhi Trager SC, Faber SM, Worthey G, González JJ. 2000. {\it Astron. J.} 120:165\\
\nhi Trager SC, Worthey G, Faber SM, Burstein D, Gonzalez JJ. 1998. {\it Ap. J. Suppl.} 116:1\\
\nhi Trager SC.\ 2004. {\it Origin and Evolution of the Elements} 391 \\
\nhi Tripicco MJ, Bell RA. 1995. {\it Astron. J.} 110:3035\\
\nhi van den Bergh S, Mackey AD. 2004. {\it Mon. Not. R. Astron. Soc.} 354:713\\
\nhi van den Bergh S, Morbey C, Pazder J. 1991. {\it Ap. J.} 375:594\\
\nhi van den Bergh S. 1975. {\it Annu. Rev. Astron. Astrophys.} 13:217\\
\nhi van den Bergh S. 1982. {\it Publ. Astron. Soc. Pac.} 94:459\\
\nhi van den Bergh S. 1994, In: Smith R.C., Storm J. (eds.) The Local Group. ESO: Garching\\
\nhi van den Bergh S. 1999. {\it Astron. Astrophys.} Rev. 9:273\\
\nhi van den Bergh S. 2004. {\it Astron. J.} 127:897\\
\nhi Vesperini E,  Heggie DC.\ 1997. {\it Mon. Not. R. Astron. Soc.} 289:898\\ 
\nhi Vesperini E, Zepf SE, Kundu A, Ashman KM. 2003. {\it Ap. J.} 593:760\\
\nhi Vesperini E, Zepf SE. 2003. {\it Ap. J. Lett.} 587:97\\
\nhi Vesperini E. 2000. {\it Mon. Not. R. Astron. Soc.} 318:841\\
\nhi Vesperini E. 2001. {\it Mon. Not. R. Astron. Soc.} 322:247\\
\nhi Walcher CJ, Fried JW, Burkert A, Klessen RS. 2003. {\it Astron. Astrophys.} 406:847\\
\nhi West MJ, C{\^o}t{\'e} P, Marzke RO, Jord{\'a}n A.\ 2004. {\it Nature}. 427:31 \\
\nhi West MJ.\ 1993. {\it Mon. Not. R. Astron. Soc.} 265:755  \\
\nhi Whitmore BC, Schweizer F, Kundu A, Miller BW. 2002. {\it Astron. J.} 124:147\\
\nhi Whitmore BC, Sparks WB, Lucas RA, Macchett FD, Biretta JA. 1995. {\it Ap. J. Lett.} 454:73\\
\nhi Whitmore BC, Schweizer F. 1995. {\it Astron. J.} 109:960\\
\nhi Whitmore BC, Zhang Q. 2002. {\it Astron. J.} 124:1418\\
\nhi Williams BF, Hodge PW. 2001. {\it Ap. J.} 548:190\\
\nhi Wolf M. et al. 2006. {\it Ap. J.} submitted\\
\nhi Woodley KA, Harris WE,  Harris GLH.\ 2005. {\it Astron. J.} 129:2654\\ 
\nhi Woodworth SC, Harris WE. 2000. {\it Astron. J.} 119:2699\\
\nhi Worthey G, Faber SM, González JJ. 1992. {\it Ap. J.} 398:69\\
\nhi Worthey G, Faber SM, González JJ, Burstein D. 1994. {\it Ap. J. Suppl.} 94:687\\
\nhi Worthey G, España A, MacArthur LA, Courteau S. 2005. {\it Ap. J.} 631:820\\
\nhi Worthey G. 1994. {\it Ap. J. Suppl.} 95:107\\
\nhi Wu X, Tremaine S. 2006. astro-ph/0508463\\
\nhi Yoon S, Yi SK, Lee Y. 2006. astro-ph/0601526\\
\nhi Zepf SE, Ashman KM. 1993. {\it Mon. Not. R. Astron. Soc.} 264:611\\
\nhi Zepf SE, Beasley MA, Bridges TJ, Hanes DA, Sharples RM, et al. 2000. {\it Astron. J.} 120:2928\\
\nhi Zinn R, West MJ. 1984. {\it Ap. J. Suppl.} 55:45\\
\nhi Zinn R. 1985. {\it Ap. J.} 293:424\\

\end{document}